\documentclass[11pt,twoside,a4paper,cmspaper,final,collab]{cms-tdr}
\def\svnVersion{5e09472}\def\svnDate{2026/03/23}\def\cmsCernNoTag{CERN-EP-2025-170}\def\cmsCernDate{\today}\def\cmsMessage{Published in the Journal of High Energy Physics as \href{http://dx.doi.org/10.1007/JHEP03(2026)083}{\doi{10.1007/JHEP03(2026)083}.}}
\begin{document}\cmsNoteHeader{TOP-24-003}

\newlength\cmsTabSkip\setlength{\cmsTabSkip}{1ex}
\providecommand{\cmsTable}[2][1]{\resizebox{#1\textwidth}{!}{#2}}

\newcommand{\abseta}{\ensuremath{\abs{\eta}}\xspace}
\newcommand{\deepjet}{\textsc{DeepJet}\xspace}

\newcommand{\pp}{\ensuremath{\Pp\Pp}\xspace}
\newcommand{\tW}{\ensuremath{\PQt\PW}\xspace}
\newcommand{\ttW}{\ensuremath{\ttbar\PW}\xspace}
\newcommand{\ttWp}{\ensuremath{\ttbar\PWp}\xspace}
\newcommand{\ttWm}{\ensuremath{\ttbar\PWm}\xspace}
\newcommand{\ttH}{\ensuremath{\ttbar\PH}\xspace}
\newcommand{\ttZ}{\ensuremath{\ttbar\PZ}\xspace}
\newcommand{\ttX}{\ensuremath{\ttbar\PX}\xspace}
\newcommand{\tZq}{\ensuremath{\PQt\PQq\PZ}\xspace}
\newcommand{\tWZ}{\ensuremath{\PQt\PW\PZ}\xspace}
\newcommand{\WZ}{\ensuremath{\PW\PZ}\xspace}
\newcommand{\tttt}{\ensuremath{\ttbar\ttbar}\xspace}
\newcommand{\ZZ}{\ensuremath{\PZ\PZ}\xspace}
\newcommand{\Zgamma}{\ensuremath{\PZ\PGg}\xspace}
\newcommand{\ttgamma}{\ensuremath{\ttbar\PGg}\xspace}
\newcommand{\tgamma}{\ensuremath{\PQt\PGg}\xspace}

\newcommand{\relIso}{\ensuremath{I_\text{rel}}\xspace}

\newcommand{\twoLeptonss}{\ensuremath{2\Pell\mkern1mu\text{SS}}\xspace}
\newcommand{\threeLepton}{\ensuremath{3\Pell}\xspace}

\newcommand{\vectheta}{\ensuremath{\boldsymbol{\theta}}\xspace}
\newcommand{\vecsigma}{\ensuremath{\boldsymbol{\sigma}}\xspace}

\newcommand{\Nreco}{\ensuremath{N_{\text{reco}}}\xspace}
\newcommand{\Ngen}{\ensuremath{N_{\text{gen}}}\xspace}
\newcommand{\Acl}{\ensuremath{A_\text{c}^{\Pell}}\xspace}

\newcommand{\ttjets}{\ensuremath{\ttbar\text{+jets}}\xspace}
\newcommand{\gammajets}{\ensuremath{\PGg\text{+jets}}\xspace}
\newcommand{\DYjets}{\ensuremath{\text{DY+jets}}\xspace}
\newcommand{\mZ}{\ensuremath{m_{\PZ}}\xspace}
\newcommand{\Deta}{\ensuremath{\Delta\eta}\xspace}
\newcommand{\ZGst}{\ensuremath{\PZ/\PGg^\ast}\xspace}

\newcommand{\Dyl}{\ensuremath{\Delta y^{\Pell}}\xspace}
\newcommand{\Dylreco}{\ensuremath{\Dyl_{\text{reco}}}\xspace}

\newcommand{\Detaloneltwo}{\ensuremath{\abs{\Deta(\Pell_1,\Pell_2)}}\xspace}
\newcommand{\DRloneltwo}{\ensuremath{\DR(\Pell_1,\Pell_2)}\xspace}
\newcommand{\DRlonejetmin}{\ensuremath{\DR(\Pell_1,\text{jet})_{\text{min}}}\xspace}
\newcommand{\mleptons}{\ensuremath{m(\text{leptons})}\xspace}

\newcommand{\ratiodescription}{%
    The lower panels show the ratio of the data to the sum of the postfit predictions (points) and the ratio of the data to the prefit predictions (red line).
    The vertical lines on the data points represent the statistical uncertainty in the data and the hatched (filled) band corresponds to the total uncertainty in the postfit (prefit) prediction.%
\xspace}
\newcommand{\overflow}{%
    Events that exceed the range of the plot are included in the last bin.%
\xspace}
\newcommand{\binwidth}{%
    The bin contents are divided by bin width.%
\xspace}
\newcommand{\diffextracaption}{%
    The upper panels show the results of the measurement together with the theoretical predictions.
    The blue band shows the uncertainty of the prediction from Ref.~\cite{Frederix:2021agh}.
    The lower panels show the ratio between the predictions and the measurement.
    The vertical lines on the unfolded data points represent the total experimental uncertainty of the unfolded cross section, while the horizontal bars show the statistical component of the uncertainty.%
\xspace}
\newcommand{\diffextracaptionTogether}{%
    The upper panels show the results of the two measurements together with the theoretical predictions.
    The blue band shows the uncertainty of the prediction from Ref.~\cite{Frederix:2021agh}.
    Each two lower panels shows the ratio between the predictions and each of the measurements.
    The vertical lines on the unfolded data points represent the total experimental uncertainty of the unfolded cross section, while the horizontal bars show the statistical component of the uncertainty.%
\xspace}

\cmsNoteHeader{TOP-24-003}
\title{Measurements of \texorpdfstring{\ttW}{ttW} differential cross sections and the leptonic charge asymmetry at \texorpdfstring{$\sqrt{s}=13\TeV$}{sqrt(s)=13 TeV}}
\date{\today}

\abstract{Measurements of properties of top quark-antiquark pair production in association with a \PW boson in proton-proton collisions at a center-of-mass energy of 13\TeV are presented, using a data sample corresponding to an integrated luminosity of 138\fbinv, recorded by the CMS experiment at the CERN LHC. Events are selected based on the presence of either two leptons with the same electric charge or three leptons, and multiple jets and \PQb-tagged jets. We present measurements of differential production cross sections as a function of kinematic variables sensitive to different aspects of the process modeling, using a multivariate discriminator in the two-lepton selection region and a simple selection-based method in the three-lepton region. The normalized cross section measurements are generally consistent with the standard model expectations, while we observe larger values compared to the expectations in the absolute cross section measurements, consistent with previous inclusive cross section measurements. In addition, we measure the leptonic charge asymmetry of this process, obtaining an observed value of $\Acl=-0.19\,^{+0.16}_{-0.18}$, consistent with the expectation of $-0.085\pm0.006$ predicted by next-to-leading order simulations.}

\hypersetup{%
pdfauthor={CMS Collaboration},%
pdftitle={Measurements of ttW differential cross sections and the leptonic charge asymmetry at sqrt{s}=13 TeV},%
pdfsubject={CMS},%
pdfkeywords={CMS, ttW, top quark, W boson, charge asymmetry}}

\maketitle

\section{Introduction}

\begin{figure}[!h]
\centering
\includegraphics[width=0.5\textwidth]{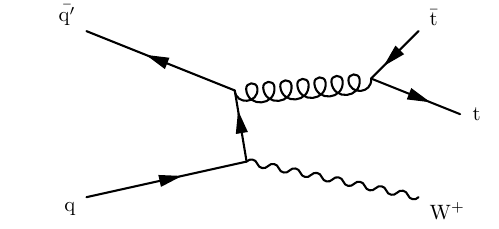}%
\includegraphics[width=0.5\textwidth]{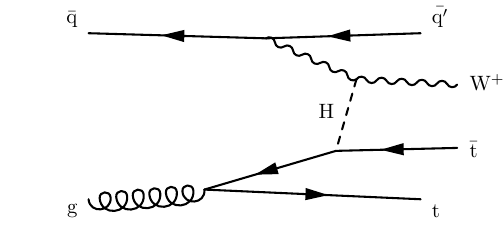}
\caption{%
    Examples of Feynman diagrams for \ttW production: at LO (left) and one of the NLO diagrams introducing sizeable electroweak corrections (right).
}
\label{fig:feynman_diagrams}
\end{figure}

The production of a top quark pair in association with a \PW boson (\ttW) is sensitive to the interaction of the top quark with the electroweak bosons, which makes it one of the most intriguing processes studied recently at the LHC in proton-proton (\pp) collisions.
At leading order (LO), this process is induced by the $\qqbar^\prime$ partonic initial state and receives contributions from the $\Pg\PQq$ channel only at next-to-LO (NLO), while the $\Pg\Pg$ contribution is suppressed and only starts to appear at next-to-NLO (NNLO).
Because of this, perturbative calculations for the \ttW production introduce large corrections at higher orders in the quantum chromodynamics (QCD) expansion~\cite{LHCHiggsCrossSectionWorkingGroup:2016ypw,Buonocore:2023ljm} and sizable electroweak corrections~\cite{Dror:2015nkp,Frederix:2017wme}.
Figure~\ref{fig:feynman_diagrams} shows representative Feynman diagrams for this process.

In inclusive top quark-antiquark (\ttbar) production, top quarks are produced with a wider rapidity distribution than top antiquarks due to interference effects present at NLO accuracy, which break the symmetry under charge conjugation, giving rise to an asymmetry in the \ttbar system~\cite{Kuhn:1998jr,Kuhn:1998kw}.
In the case of \ttW production, this asymmetry is enhanced due to the absence of the gluon-gluon partonic initial state and to the \PW boson acting as a polarizer for the incoming quark in LO diagrams, such as the one in Fig.~\ref{fig:feynman_diagrams}~\cite{Maltoni:2014zpa}.
This feature can be used as an experimentally robust probe of physics beyond the standard model (BSM) that could modify these asymmetries.
In addition, the \ttW process is a unique probe of four-quark contact interactions~\cite{CMS:2023xyc}, which may introduce a dominant contribution to the production cross section, in contrast to other processes that are dominated by gluon-fusion production, such as the inclusive production of \ttbar pairs.
Finally, being one of the few processes that give rise to pairs of leptons (electrons or muons, \Pell) with the same electric charge (same-sign (SS) lepton pairs), this process constitutes the leading irreducible background to studies that profit from this topology, including measurements of the simultaneous production of four top quarks (\tttt)~\cite{CMS:2023ftu,ATLAS:2023ajo} or the production of a top quark pair in association with a Higgs boson (\ttH)~\cite{CMS:2020mpn}.

Considering the production of on-shell top quarks without their subsequent decay, the state-of-the-art in the calculation of the \ttW cross section is an NNLO computation~\cite{Buonocore:2023ljm}, which estimates the contribution from two-loop amplitudes by means of approximate methods.
Calculations at NLO in QCD and electroweak corrections including resummation at next-to-next-to-leading logarithm~\cite{Kulesza:2018tqz,Broggio:2019ewu,Kulesza:2020nfh} are also available.
In addition, recent advances~\cite{Bevilacqua:2020pzy,Bevilacqua:2020srb} show that this process receives contributions from off-shell top quark effects in low and high transverse momentum (\pt) phase space regions.
Finally, several event generators~\cite{Frederix:2021agh,FebresCordero:2021kcc,Ferencz:2023fso} are available to simulate this process at NLO matched to parton showers, including up to two extra partons in the matrix element calculation.

One additional motivation for precision studies of \ttW production is the continuing tension between previous measurements and the set of standard model (SM) predictions.
The inclusive production cross section for this process has been measured 17--29\% larger than the SM expectation, by both the ATLAS~\cite{ATLAS:2024moy} and CMS~\cite{CMS:2020mpn,CMS:2022tkv} Collaborations, in measurements that achieve precisions of the order of 10\%.
The ATLAS Collaboration has in addition performed a measurement of its differential production cross section~\cite{ATLAS:2024moy} and of the charge asymmetry of the \ttbar system in this process~\cite{ATLAS:2023xay}.
The measurements of the normalized distributions and the charge asymmetry are in agreement with the SM predictions.

In this paper, we report measurements of the \ttW production cross section as a function of several kinematic variables, and a measurement of the leptonic charge asymmetry in the \ttbar system, in events with at least one SS lepton pair and hadronic jets recorded by the CMS experiment.
One of the challenges in the differential cross section measurement is the contribution of processes with leptons not produced in the prompt decays of \PW or \PZ bosons, that can contaminate the signal selection.
Two different strategies are pursued to mitigate the effect of this contribution.
In the first method, denoted as ``multivariate analysis (MVA)-based method'', we employ a loose lepton selection, and use an MVA discriminator with information on the event kinematic properties to discriminate the signal against the main background processes, which are events with nonprompt or charge misidentified leptons and associated \ttbar production processes.
An alternative method, dubbed as ``counting method'', that instead relies on a more stringent lepton selection achieving a larger signal purity is also used.
Both methods use a likelihood-based unfolding strategy, performing a fit to the signal region (SR) and several control regions (CRs).
We use the MVA-based method to measure the differential cross section in events with two leptons. Both measurements attain a similar sensitivity in most bins, but the MVA-based method is more precise in all measured bins, achieving up to a 40\% better precision in some bins of this region. The counting method is used to measure the differential cross section and charge asymmetry in events with three leptons.
We also provide the results of the counting method in the two-lepton region in the appendix of the paper to complement the results from the MVA-based method.
Tabulated results are provided in the HEPData record for this analysis~\cite{hepdata}.

This manuscript is organized as follows.
Section~\ref{sec:cms} provides an overview of the CMS detector and Section~\ref{sec:samples} describes the collision data and simulated samples used in this study.
The object and event selections are described in Section~\ref{sec:objects_events}, the estimation of background events contributing to the measurements in Section~\ref{sec:background}, and the systematic uncertainties affecting the measurements in Section~\ref{sec:systematics}.
In Section~\ref{sec:differential}, we discuss the differential cross section measurements performed with the MVA-based and counting methods.
The measurement of the charge asymmetry is shown in Section~\ref{sec:charge_asym} and a summary is given in Section~\ref{sec:summary}.
Finally, the results of the differential cross section measurements are shown in appendix~\ref{app:countingmethod_2lss_results}.

\section{The CMS detector and event reconstruction}
\label{sec:cms}

The central feature of the CMS apparatus is a superconducting solenoid of 6\unit{m} internal diameter, providing a magnetic field of 3.8\unit{T}.
Within the solenoid volume are a silicon pixel and strip tracker, a lead tungstate crystal electromagnetic calorimeter (ECAL), and a brass and scintillator hadron calorimeter (HCAL), each composed of a barrel and two endcap sections.
Forward calorimeters extend the pseudorapidity ($\eta$) coverage provided by the barrel and endcap detectors.
Muons are measured in gas-ionization detectors embedded in the steel flux-return yoke outside the solenoid.
More detailed descriptions of the CMS detector, together with a definition of the coordinate system used and the relevant kinematic variables, can be found in Refs.~\cite{CMS:2008xjf,CMS:PRF-21-001}.

Events of interest are selected using a two-tiered trigger system.
The first level, composed of custom hardware processors, uses information from the calorimeters and muon detectors to select events at a rate of around 100\unit{kHz} within a fixed latency of about 4\mus~\cite{CMS:2020cmk}.
The second level, known as the high-level trigger, consists of a farm of processors running a version of the full event reconstruction software optimized for fast processing, and reduces the event rate to around 1\unit{kHz} before data storage~\cite{CMS:2016ngn,CMS:2024aqx}.

The CMS particle-flow (PF) algorithm~\cite{CMS:2017yfk} provides a global event description combining optimally the information from all subdetectors, to reconstruct and identify all individual particles in the event.
The particles are subsequently classified into five mutually exclusive categories: photons, electrons, muons, and charged and neutral hadrons.

In this work, we consider particles produced at the primary vertex (PV), which is taken to be the vertex corresponding to the hardest scattering in the event, evaluated using tracking information alone, as described in Section~9.4.1 of Ref.~\cite{CMS-TDR-15-02}.
Photons are identified from clusters in the ECAL not compatible with tracks in the tracker, which are also used to determine the photon energy.
Electrons are reconstructed via a combination of tracker and ECAL measurements~\cite{CMS:2020uim,CMS:prefire2} in the range $\abseta<2.5$.
Their energy is determined from a combination of the electron momentum at the PV as determined by the tracker, the energy of the corresponding ECAL cluster, and the energy sum of all bremsstrahlung photons spatially compatible with originating from the electron track.
Muons are reconstructed in the range $\abseta<2.4$ by combining information from the tracker and the muon spectrometers in a global fit~\cite{CMS:2018rym}.
The energy of muons is obtained from the curvature of the corresponding track.
The energy of charged hadrons is determined from a combination of their momentum measured in the tracker and the matching ECAL and HCAL energy deposits, corrected for the response function of the calorimeters to hadronic showers.
Finally, the energy of neutral hadrons is obtained from the corresponding corrected ECAL and HCAL energies.

Hadronic jets are clustered from the PF objects using the infrared- and collinear-safe anti-\kt algorithm~\cite{Cacciari:2008gp,Cacciari:2011ma} with a distance parameter of 0.4.
We only consider in the analysis jets that have a $\pt>25\GeV$ and $\abseta<2.4$, and are separated by $\DR=\sqrt{\smash[b]{(\Deta)^2+(\Delta\phi)^2}}>0.4$ from any identified lepton (following the criteria given in Section~\ref{sec:objects_events}), where $\phi$ is the azimuthal angle with respect to the beam axis.
Jet momentum is determined as the vectorial sum of all particle momenta in the jet, and is found from simulation to be, on average, within 5 to 10\% of the true momentum over the whole \pt spectrum and detector acceptance.
Additional \pp interactions within the same or nearby bunch crossings (pileup) can contribute additional tracks and calorimetric energy depositions, increasing the apparent jet momentum.
To mitigate this effect, tracks identified to be originating from vertices in the same bunch crossing but different from the PV are discarded and an offset correction is applied to correct for remaining contributions~\cite{CMS:2020ebo}.
Jet energy corrections are derived from simulation studies so that the average measured energy of jets becomes identical to that of particle level jets.
In situ measurements of the momentum balance in dijet, photon+jets, {\PZ}+jets, and multijet events are used to determine any residual differences between the jet energy scale in data and in simulation, and appropriate corrections are made.
The jet energy resolution (JER) amounts typically to 15--20\% at 30\GeV, 10\% at 100\GeV, and 5\% at 1\TeV.
A smearing procedure is applied to match the JER in simulation to that in data~\cite{CMS:2016lmd}.
Additional selection criteria are applied to each jet to remove jets potentially dominated by instrumental effects or reconstruction failures~\cite{CMS:2020ebo}.

The missing transverse momentum vector \ptvecmiss is computed as the negative vector sum of the \pt of all the PF candidates in an event, and its magnitude is denoted as \ptmiss~\cite{CMS:2019ctu}.
The \ptvecmiss is modified to account for corrections to the energy scale of the reconstructed jets in the event.
Anomalous high-\ptmiss events can be due to a variety of reconstruction failures, detector malfunctions or noncollision backgrounds.
Such events are rejected by event filters that are designed to identify more than 85--90\% of the spurious high-\ptmiss events with a mistagging rate less than 0.1\%~\cite{CMS:2019ctu}.

Jets originating from \PQb quarks are identified with the \deepjet algorithm~\cite{Sirunyan:2017ezt, Bols:2020bkb, CMS-DP-2018-058}, with two defined working points (WPs) labeled ``loose'' and ``medium''.
The loose (medium) WP has a selection efficiency for \PQb quark jets of about 90 (76)\%, and a misidentification rate of about 49 (17)\% for \PQc quark jets and 18 (3)\% for light quark and gluon jets.
Unless specified otherwise, the term ``\PQb-tagged jets'' is used to refer to jets that pass the loose WP requirements.

\section{Data and simulated samples}
\label{sec:samples}

We use \pp collision events at $\sqrt{s}=13\TeV$ recorded at the LHC between 2016 and 2018, with a total integrated luminosity of 138\fbinv.
The events used in this study were collected using a set of triggers requiring the presence of one, two, or three reconstructed leptons above a certain \pt threshold, which varies depending on the data-taking era and the lepton flavor and multiplicity.

A suite of simulated samples produced via Monte Carlo (MC) generators are used to train the MVA discriminators and to estimate the efficiency of the reconstruction and identification techniques in signal and background events.
These samples use various matrix-element generators, that are described below, and the NNPDF3.1NNLO parton distribution functions (PDFs)~\cite{NNPDF31}.
Unless specified otherwise, these generators are interfaced to \PYTHIA v8.240~\cite{Sjostrand:2014zea} to simulate the parton shower, hadronization, decay, and underlying event using the CP5 tune~\cite{CMS:2022awf}.
We use the MLM~\cite{Alwall:2007fs} (FxFx~\cite{Frederix:2012ps}) matching scheme for \MGvATNLO~\cite{Alwall:2014hca} samples that are produced at LO (NLO) using matrix elements with additional partons.
Generated hard-scattering events are overlaid with simulated pileup collisions, with a number of events per bunch crossing matching the distribution inferred from data.
The response of the CMS detector is fully simulated using a model of the detector based on the \GEANTfour toolkit~\cite{GEANT4:2002zbu}.

The nominal model for the \ttW signal is obtained using the \MGvATNLO v.2.6.5 generator, accounting for terms proportional to $\alpS^3\alpha$ (QCD production) and $\alpS\alpha^3$ (electroweak production) and higher order contributions in the QCD expansion, where \alpS and $\alpha$ are the strong coupling and fine-structure constants, respectively.
For the QCD production, we consider NLO corrections to both \ttW production with and without an additional parton.
In addition to the nominal sample, the results are also compared with an alternative sample produced with \MGvATNLO v3.3.1 interfaced with \PYTHIA v8.306, which uses the improved FxFx matching setup described in Ref.~\cite{Frederix:2021agh} with a merging scale of 42\GeV.

To assess potential biases in the measurements due to deviations of the signal predictions from the nominal model, we consider a BSM scenario that can be parametrized by an effective field theory (EFT) that introduces additional terms to the SM Lagrangian.
This contribution is modeled using LO signal simulations using \MGvATNLO, introducing the EFT contribution using the \texttt{dim6top} model~\cite{Aguilar-Saavedra:2018ksv}, considering non-zero values of 22 different operators.
Samples are reweighted according to the method described in Ref.~\cite{Mattelaer:2016gcx} and as used in Ref.~\cite{CMS:2023xyc}.

The contributions from backgrounds such as \ttH, the production of a top quark pair in association with a \Z boson (\ttZ), the production of a top quark pair in association with a photon (\ttgamma), the production of a single top quark in association with a \Z boson (\tZq), the production of a single top quark in association with a photon (\tgamma), the production of a single top quark in association with a \Z and \PW (\tWZ) and \ZGst production in association with jets (\DYjets) are estimated using \MGvATNLO at NLO.
We use the diagram-removal scheme~\cite{Frixione:2019fxg} to remove the diagrams contributing both to \ttZ and \tWZ production.
The associated production of a \PW and \Z (\WZ) or two \Z bosons (\ZZ) are both estimated at NLO using the \POWHEG generator.
We use NLO predictions with electroweak corrections~\cite{LHCHiggsCrossSectionWorkingGroup:2016ypw} to normalize the \ttH samples, and the NLO computations from the generators for \tZq and \tWZ productions.
The normalizations of the \ttZ and diboson processes are determined in data, so we do not assume any specific reference cross section.
Simulations of \ttbar production, \DYjets, and \tW production are used to train the multivariate discriminators described later in the paper and for validating the estimation of the nonprompt-lepton background, as described below.
We simulate \ttbar and \tW production using the \POWHEG generator~\cite{EXT:Powheg-2004,EXT:Powheg-2007,EXT:Powheg-2010,EXT:Powheg-singlet-2009,EXT:Powheg-ttbar-2007}.
The \ttgamma process is normalized to the cross section measured by CMS~\cite{CMS:2022lmh}.
The grouping of the \ttZ and \ttH processes is referred to as \ttX, while \ZZ and \WZ are grouped into ``Diboson''.
Rare top quark related backgrounds and multiboson productions are grouped into ``Other''.

We account for differences between data and simulation arising in lepton reconstruction, identification, and triggering efficiencies, the energy scale of jets and JER, the response of the \deepjet algorithm, and the resolution in \ptmiss.
These corrections are typically at the level of a few percent~\cite{CMS:2010svw,CMS:2016lmd,Sirunyan:2017ezt,CMS:2019ctu} and are measured using as standard candles a variety of SM processes, such as $\PZ\to\Pe\Pe$, $\PZ\to\PGm\PGm$, \ttjets, and \gammajets production.

\section{Object and event selection}
\label{sec:objects_events}

Muons and electrons are considered in this analysis if they are within the detector acceptance, removing electrons in the region between the barrel and endcap of the ECAL ($1.44<\abseta<1.57$).
Both types of leptons are subjected to a set of loose quality requirements referred to as baseline selection.
They must be compatible with originating from the PV and fulfill a set of pre-defined selections~\cite{CMS:2020uim, CMS:2018rym}.
This includes requirements on their relative isolation, \relIso, which is defined as the scalar \pt sum of all particles within a certain distance \DR around the lepton, divided by the lepton \pt.
We use a \pt-dependent distance~\cite{Rehermann_2011} of $\DR<0.2$ for leptons with $\pt<50\GeV$, $\DR<10\GeV/\pt$ for $50<\pt<200\GeV$, and $\DR<0.05$ for $\pt>200\GeV$, and require all leptons to pass a loose requirement of $\relIso<0.4$.
Specifically for electrons, the number of missing hits, defined as the number of positions in the pixel tracker where a hit by the passing electron is expected but not found, is required to be less than two.
In addition, those electrons that are identified as likely to come from photons converting in the detector material are rejected~\cite{CMS:2015xaf}.

We denote leptons originating from the decays of electroweak bosons or \PGt leptons produced in the primary interaction as prompt leptons, while those originating from other sources, \eg, hadron decays or misidentified jets or hadrons are called nonprompt leptons.
The presence of several prompt leptons in the final state is one of the main signatures of our signal, and hence the discrimination between both types of leptons is a crucial part of the analysis.
We rely on an MVA classifier, described in more detail in Refs.~\cite{CMS:2023dsu,CMS:2020mpn,CMS:2023ftu}, which is denoted as prompt-MVA.
This algorithm uses a number of variables, including the relative isolation of the lepton with both fixed and variable cone sizes, transverse and longitudinal impact parameters, and several variables related to the jet with the smallest \DR with respect to the lepton, including its \PQb tagging score.
We consider two WPs for the prompt-MVA, one for each of the strategies followed.
We define a loose WP with 89 (93)\% efficiency for prompt electrons (muons) and 2 (3)\% acceptance rate for nonprompt electrons (muons), and a tight WP with 75 (88)\% efficiency for prompt electrons (muons) and 1 (2)\% acceptance rate for nonprompt electrons (muons).
The cross section measurement with the counting strategy and the charge asymmetry measurement use the tighter WP to reduce the contribution from nonprompt leptons to the SR, while the MVA-based approach to measure the cross section instead uses the looser WP, and relies on another MVA to discriminate \ttW from background events.

We classify events in two regions, depending on the number of reconstructed leptons and charge.
For the SS dilepton (\twoLeptonss) SR, we select events with exactly two leptons, with \pt of at least 25 and 15\GeV for the highest \pt (leading) and second-highest \pt (subleading) lepton, respectively.
We require that the two leptons have the same electric charge to suppress backgrounds from \ttbar and \DYjets production, and we reject events with a lepton pair whose invariant mass is below 30\GeV to reduce contributions from low-mass resonances.
For electrons in the \twoLeptonss SR, we apply additional criteria that demand consistency among three measurements of the charge sign, described as ``selective algorithm'' in Ref.~\cite{CMS:2015xaf}.
Two of these measurements are performed using two different parametrizations of the electron track, while the other exploits the relative position between the electron track and the calorimeter energy deposit.
This approach further reduces the electron charge mismeasurement rate by a factor of five with an efficiency of about 97\%.
We also reject the events in which the invariant mass of the two selected electrons is closer than 10\GeV to the \PZ boson mass, $\mZ=91.19\GeV$~\cite{ParticleDataGroup:2022pth}.
In the MVA-based strategy, we additionally require $\ptmiss>30\GeV$, allowing us to create one of the CRs enriched in nonprompt leptons with $\ptmiss<30\GeV$ (described in Section~\ref{sec:background}).
Finally, we require events to have at least three jets, of which at least two are \PQb tagged.

For the three lepton (\threeLepton) SR, we select events with exactly three leptons with $\pt>25$, 15, and 15\GeV, respectively, among which there is at least one pair with opposite electric charge.
We reject events with a lepton pair whose invariant mass is below 30\GeV or with an opposite-charge same-flavor (OSSF) lepton pair with an invariant mass within 10\GeV of \mZ.
We select only events with at least two jets, two of which are required to be \PQb tagged.

\begin{figure}[!ht]
\centering
\includegraphics[width=0.8\textwidth]{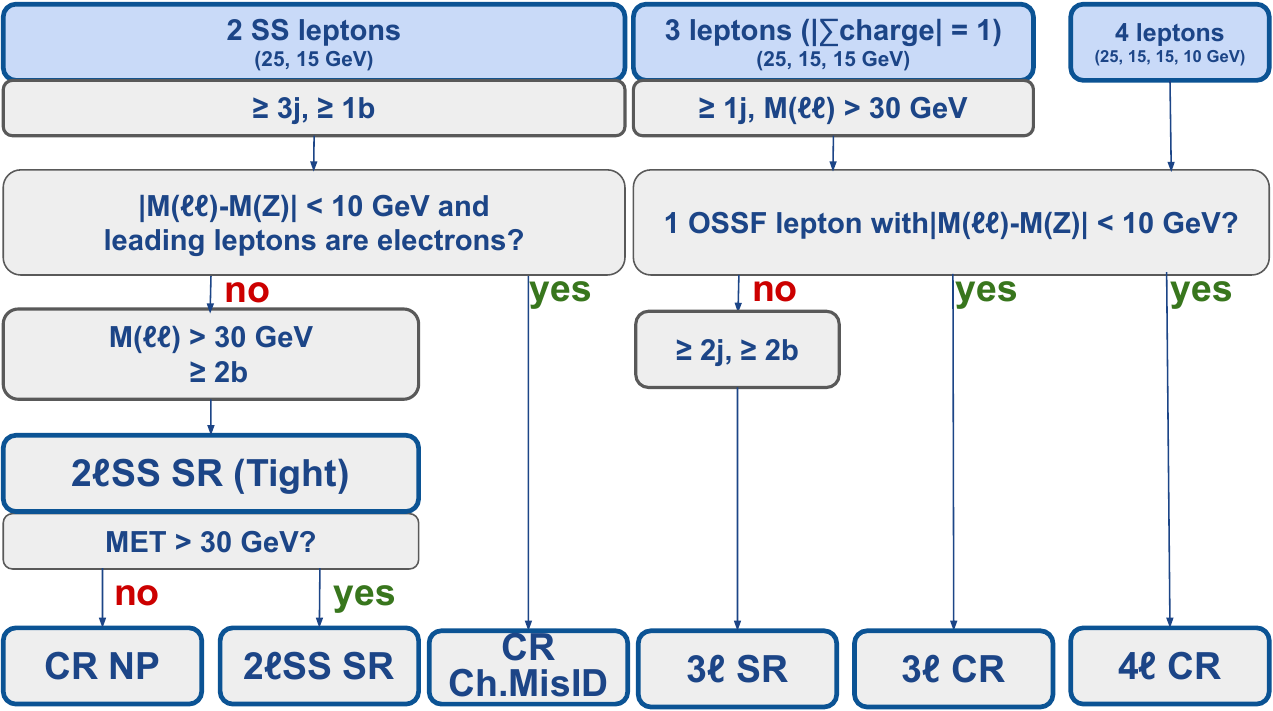}
\caption{%
    An overview of all control regions (CR) and signal regions (SR) used in the analysis.
}
\label{fig:regions:overview}
\end{figure}

Several event regions, which are discussed in Section~\ref{sec:background}, are defined to validate the modeling of the main background processes.
The yields of some of these regions are included in the likelihood fit described in Section~\ref{sec:differential} to constrain the relevant backgrounds.
We denote these regions as CRs, while we denote the regions that are not included in the fit as validation regions (VRs).
We note that predictions in the VRs are affected by the same sources of uncertainties as the SRs and CRs and are therefore modified by the result of the likelihood fit, even if they do not directly affect it.
An overview of all analysis regions is shown in Fig.~\ref{fig:regions:overview}.

\section{Background estimation}
\label{sec:background}

The background contributions to the SRs are divided into four categories: processes with at least one nonprompt lepton, processes with at least one charge-misidentified lepton, processes with one electron produced by a photon conversion, or irreducible backgrounds that contribute to the SR with prompt leptons whose charge has been correctly identified.
The latter two categories are estimated from simulated samples (described in Section~\ref{sec:samples}) that are normalized according to the measured luminosity and predicted cross sections, while the nonprompt and charge-misidentified lepton backgrounds are estimated from sideband regions in data.

\subsection{Nonprompt leptons}

Events with at least one nonprompt lepton that pass the signal selection are mainly coming from \ttbar and, to a smaller extent, \DYjets where one of the selected electrons and muons is either a secondary lepton coming from the decay of a \PQb or a \PQc hadron, or a light jet wrongly identified as a lepton.

The nonprompt-lepton background is estimated with a ``tight-to-loose'' ratio method~\cite{CMS:2020mpn,CMS:2023xyc,CMS:2023ftu,CMS:2021ugv}, performed separately for electrons and muons passing the tight and loose selection.
The main feature of the method is the usage of a sideband region with events that pass the same selection as the SRs, but where one or more of the selected leptons fail the required identification criteria while passing a loosened set of identification criteria specifically designed for this method.
This region is enriched in events containing nonprompt leptons.
After subtracting the residual contribution from prompt leptons using predictions from simulation, we weight these events by a misidentification rate to estimate the nonprompt-lepton contribution in the SRs.
This misidentification rate is measured in a data sample enriched in nonprompt leptons produced in SM events composed uniquely of jets produced through the strong interaction, referred to as multijet events, that is collected by a set of dedicated triggers.

The method is validated in simulation separately with \ttbar and \DYjets events, which have different sources of nonprompt leptons, and also separately for nonprompt electrons and muons.
We compare the event yields predicted by the simulations in the SRs with the predictions from nonprompt leptons that pass the loose but fail the tight ID and are reweighted with the tight-to-loose ratio method.
We observe an overall good agreement between the two methods, with the largest deviations being smaller than 30\%.

\begin{figure}[!ht]
\centering
\includegraphics[width=0.4\textwidth]{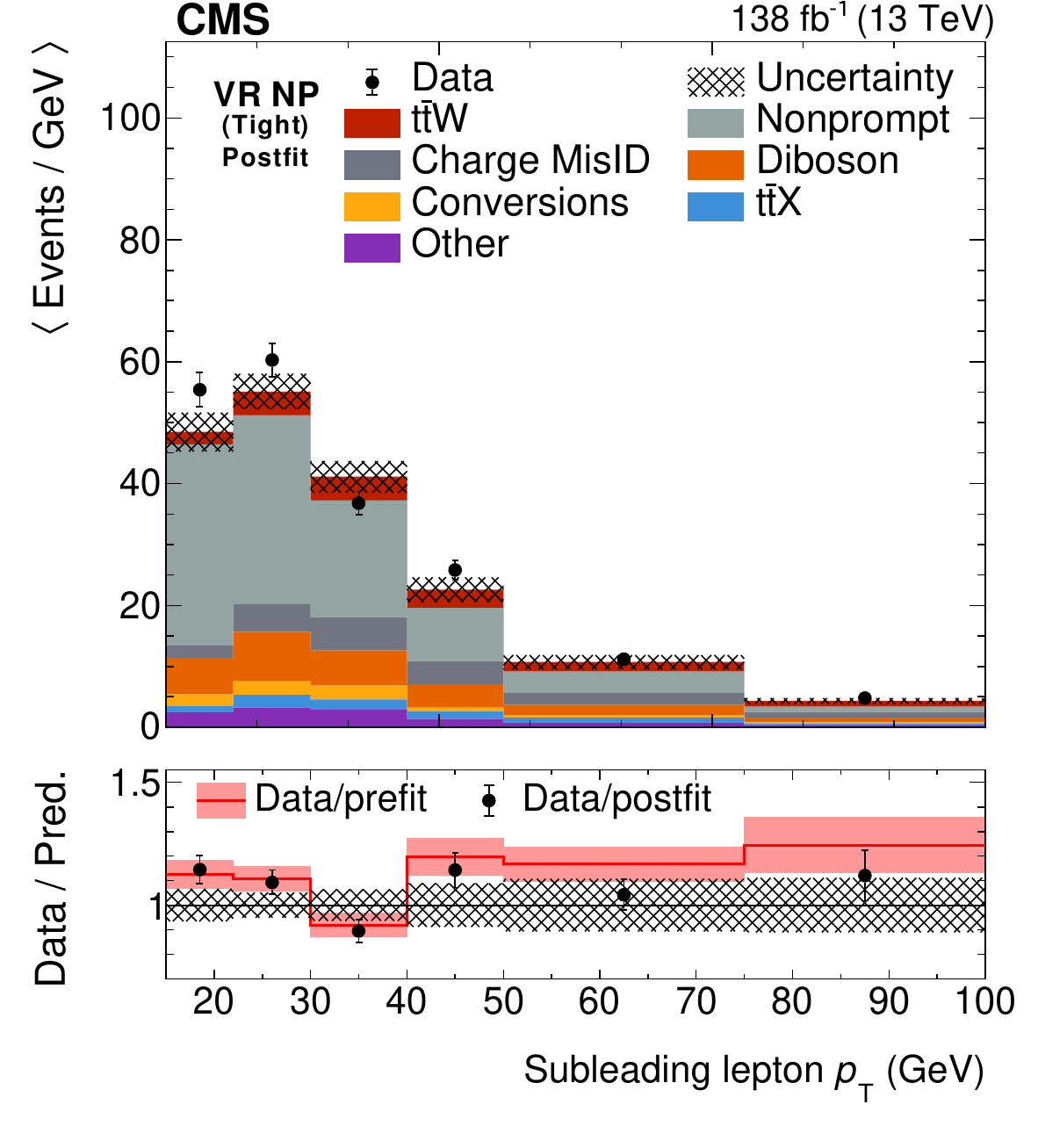}%
\hspace*{0.05\textwidth}%
\includegraphics[width=0.4\textwidth]{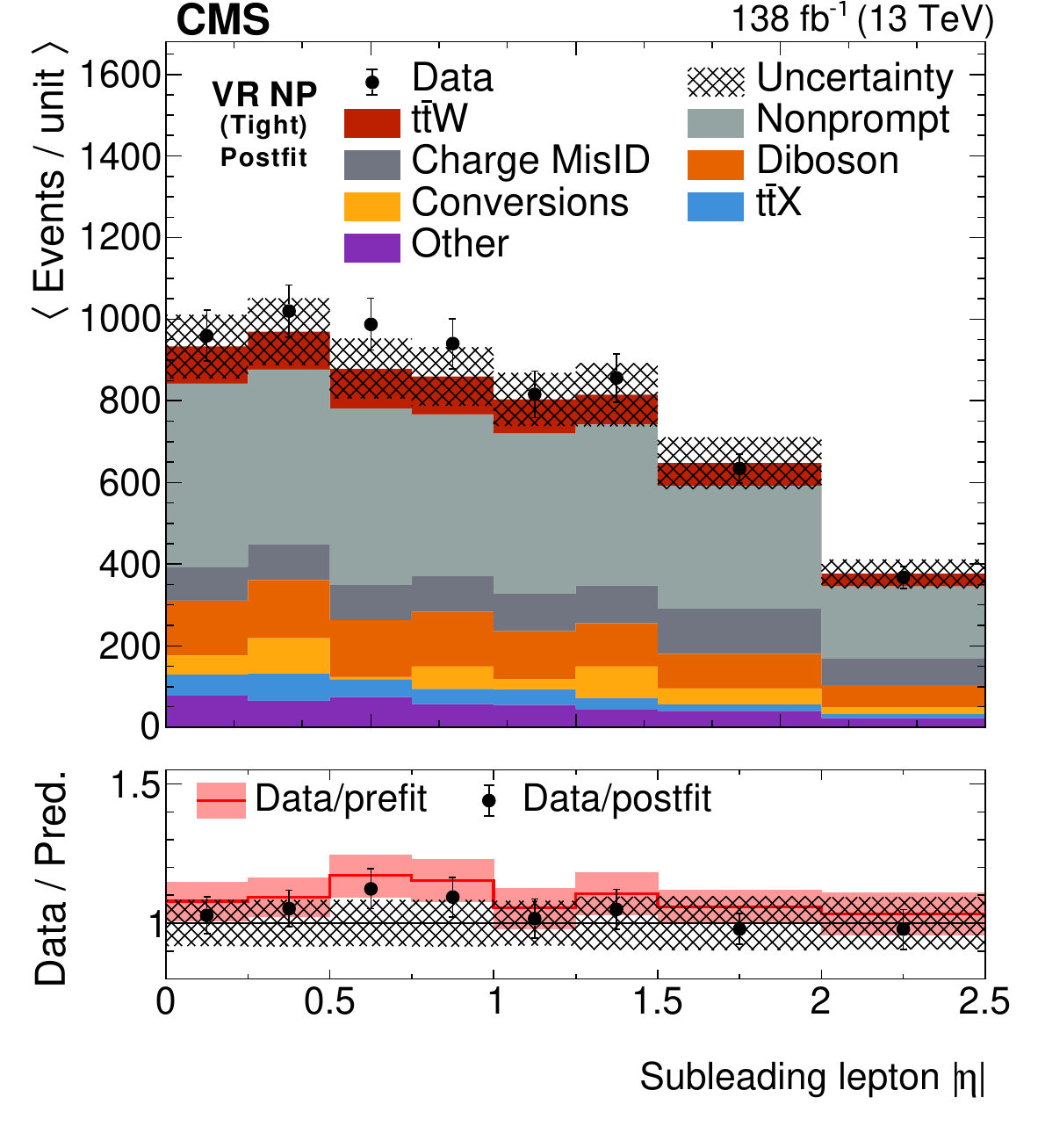}
\caption{%
    \pt of the sub-leading lepton (left) and \abseta of the subleading lepton (right), in a VR enriched with nonprompt leptons by applying requirements on the number of (\PQb-tagged) jets orthogonal to the SR, for data (points) and predictions (filled histograms) after the fit to the data in the SR and CR as described in Section~\ref{sec:differential} for the tight lepton selection.
    \ratiodescription
    \overflow
    \binwidth
}
\label{fig:backgrounds:nonprompt_jets}
\end{figure}

\begin{figure}[!ht]
\centering
\includegraphics[width=0.4\textwidth]{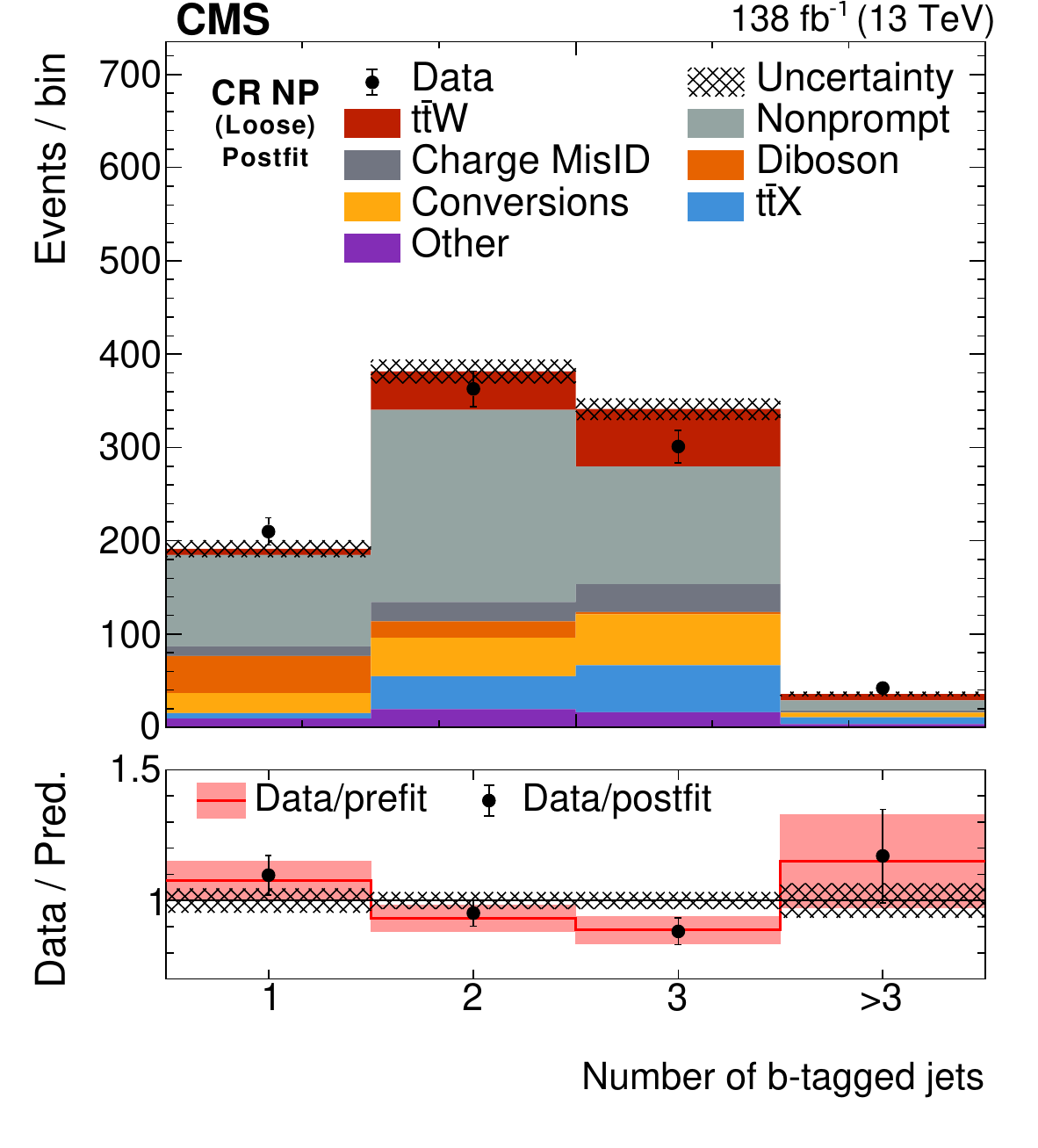}%
\hspace*{0.05\textwidth}%
\includegraphics[width=0.4\textwidth]{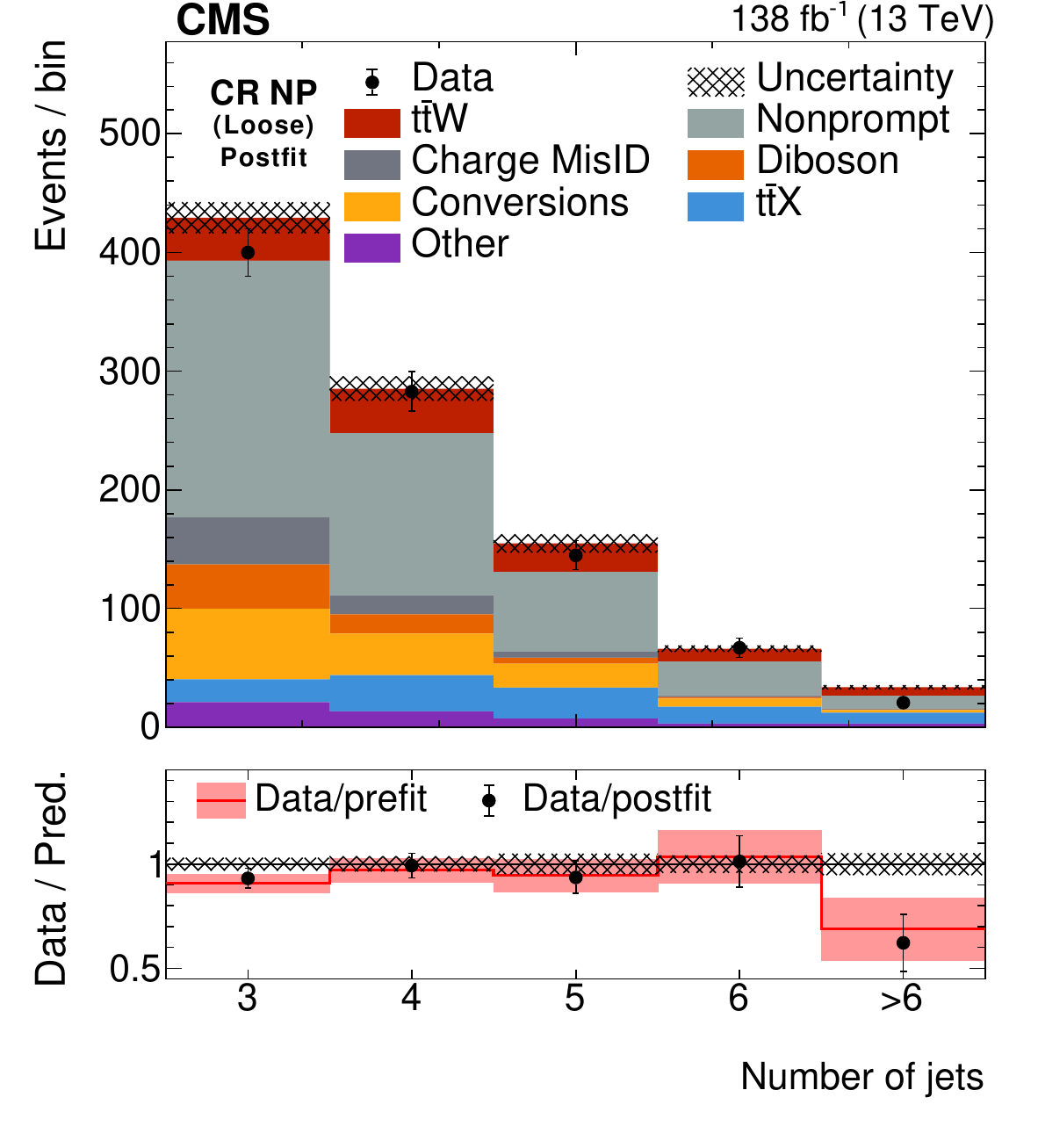} \\
\includegraphics[width=0.4\textwidth]{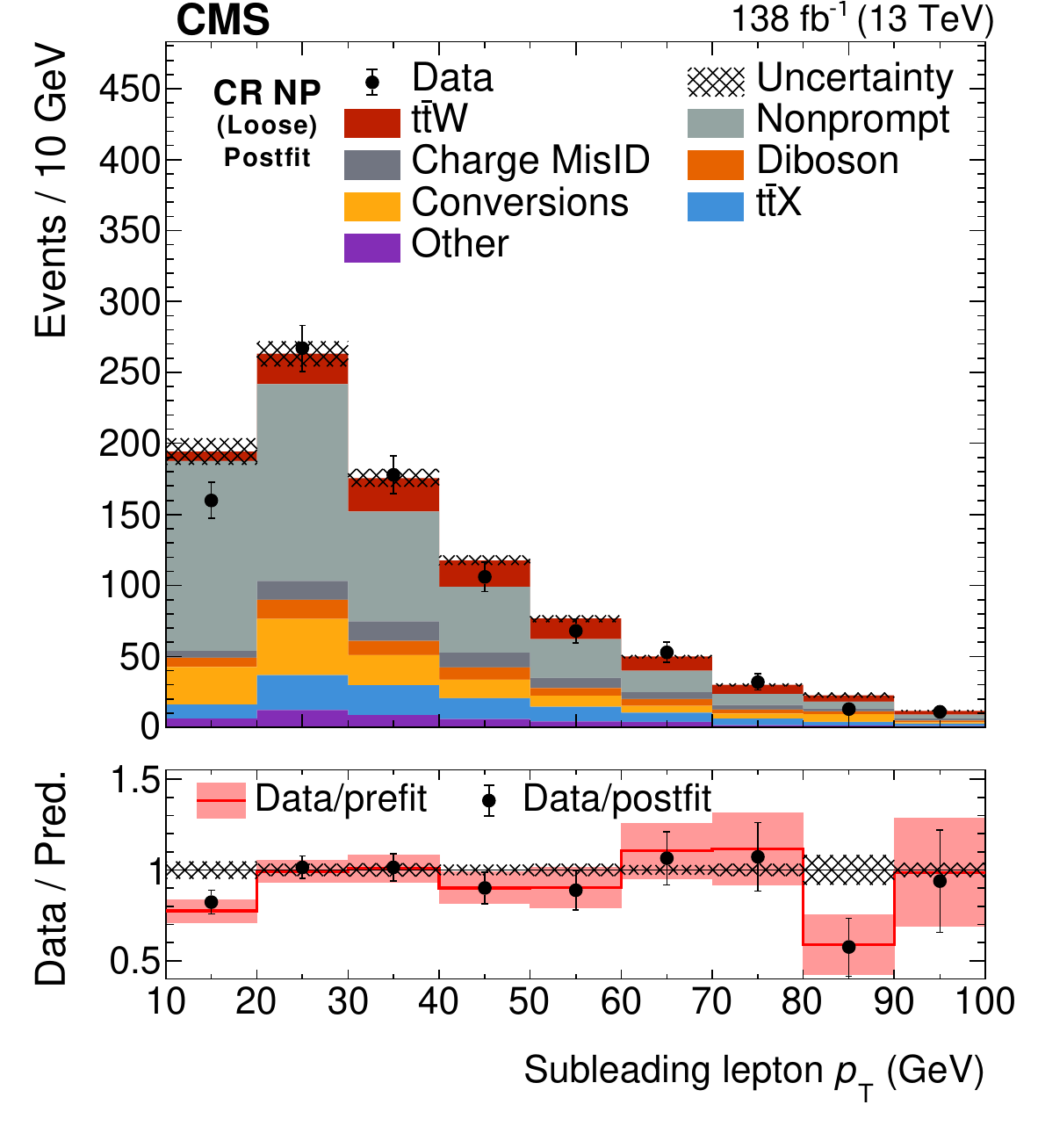}%
\hspace*{0.05\textwidth}%
\includegraphics[width=0.4\textwidth]{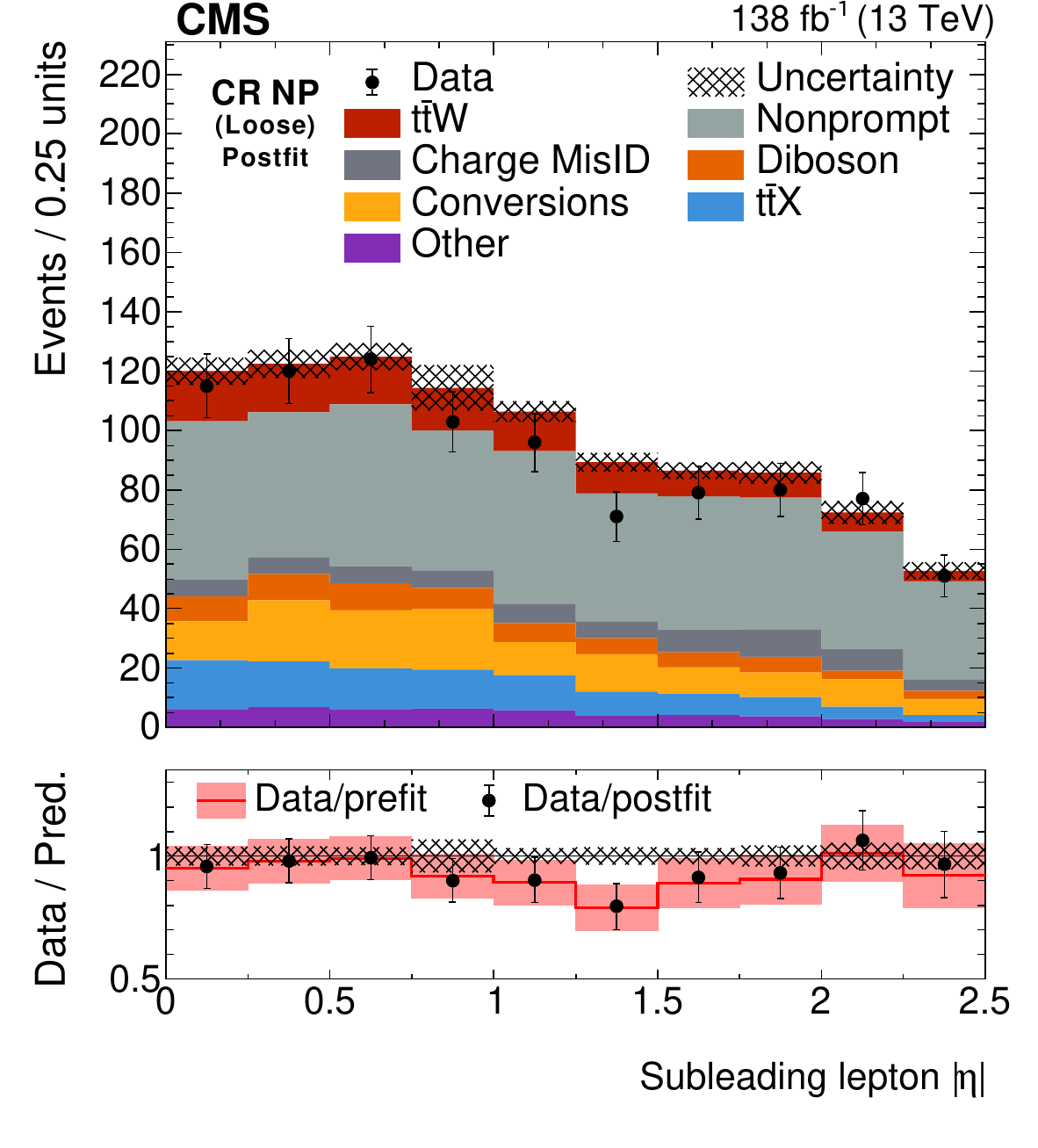}
\caption{%
    Number of selected \PQb-tagged jets in the event (upper left), number of selected jets in the event (upper right), \pt of the sub-leading lepton (lower left), and \abseta of the sub-leading lepton (lower right), in a CR enriched with nonprompt leptons by inverting the \ptmiss requirement, for data (points) and predictions (filled histograms) after the fit to the data in the SR and CR as described in Section~\ref{sec:differential} for the loose lepton selection.
    \ratiodescription
    \overflow
}
\label{fig:backgrounds:nonprompt_met}
\end{figure}

The method is then checked in one data CR and one data VR enriched in the nonprompt-lepton background.
For the VR (referred to as VR NP), we select events that pass similar requirements as in the \twoLeptonss SR, but modify the jet and \PQb-tagged jet multiplicity requirements: we require events to have exactly two jets, of which one is a \PQb-tagged jet using the medium WP.
For the CR (referred to as CR NP and only used with the loose lepton selection), we subdivide events in the \twoLeptonss SR in two categories.
The region with $\ptmiss<30\GeV$, which is a region depleted in signal and rich in nonprompt leptons, is used as the CR NP.
The region with $\ptmiss>30\GeV$ remains as the \twoLeptonss SR.
The prediction matches the data accurately in both regions within the uncertainties, as shown in Figs.~\ref{fig:backgrounds:nonprompt_jets}--\ref{fig:backgrounds:nonprompt_met} for leptons passing the tight and loose lepton selections, respectively.

\subsection{Charge-misidentified leptons}
\label{sub:charge_flips}

The \twoLeptonss SR contains a non-negligible contribution of events where the electric charge of one of the leptons has been wrongly measured.
This contribution is dominated by electrons, which undergo Bremsstrahlung as they traverse the detector, making the measurement of the track curvature more challenging.
The charge-misidentification probability is negligible for muons.

To determine the charge-misidentified background, which we denote as ``Charge MisID'', the electron charge-misidentification probability is determined in simulated \DYjets samples and parameterized as a function of \pt and \abseta of the lepton.
We then estimate the contribution from charge-misidentified leptons in the SR by weighting data events in a sideband, defined with the same requirement as the SR but where the electric charges of the two leptons have the opposite sign, by the charge-misidentification probability we estimated.

\begin{figure}[!ht]
\centering
\includegraphics[width=0.4\textwidth]{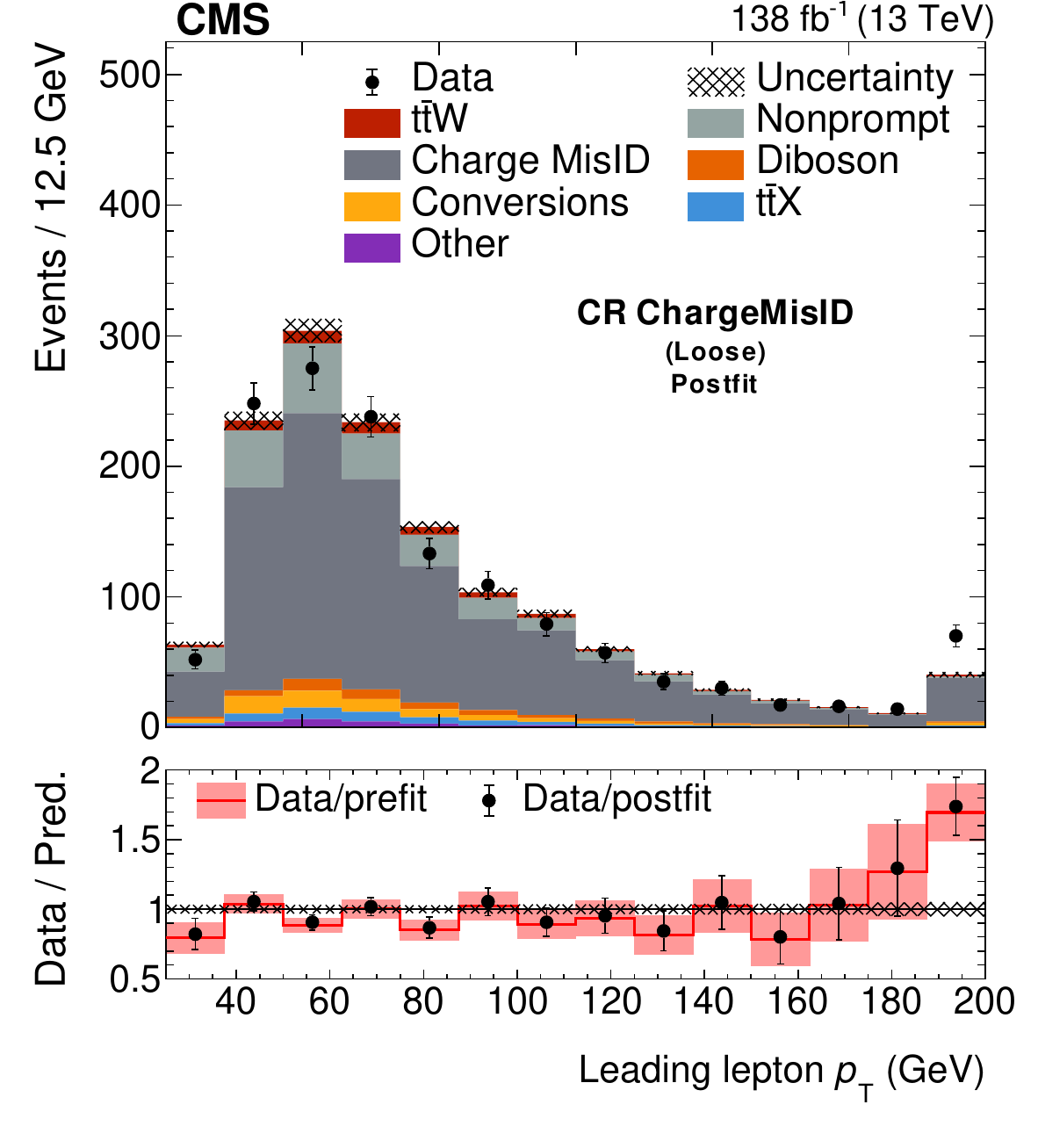}%
\hspace*{0.05\textwidth}%
\includegraphics[width=0.4\textwidth]{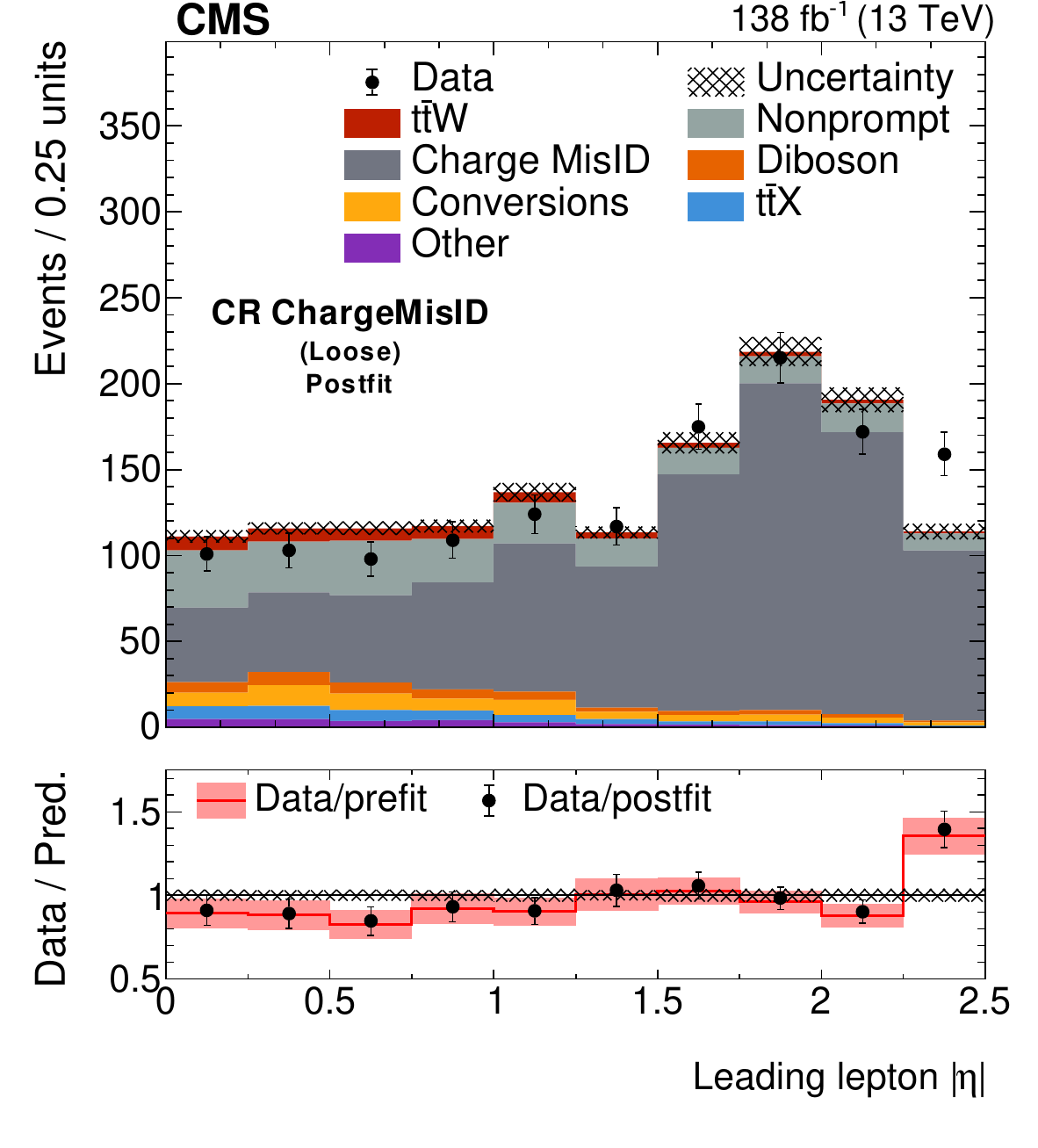}
\caption{\pt of the leading lepton (left), \abseta of the leading lepton (right), in a CR enriched with charge-misidentified leptons and additional jets for data (points) and predictions (filled histograms) after the fit to the data in the SR and CR as described in Section~\ref{sec:differential} for the loose lepton selection. \ratiodescription
\overflow}
\label{fig:backgrounds:chargeflips_jets}
\end{figure}

To account for residual differences of the charge-misidentification probability between data and simulations, we select events with an SS electron pair whose mass is within 10\GeV of \mZ, which is enriched in events with charge-misidentified leptons.
We compare the data yield with the yield predicted by this method.
The two agree between 10 and 40\%, depending on the data-taking year and the lepton selection, with residual differences due to the complexity of modeling misreconstruction effects.
We use this per-year factor to correct the charge-misidentification probability predicted by the simulations.
After this correction, data agrees well with the predictions.
A CR, labeled as CR ChargeMisId, is defined, requiring in addition the presence of at least three jet and at least one \PQb-tagged jet, which represents more closely the phase space of the SR.
The good agreement between data and prediction after applying these scale factors is preserved in this region, as shown in Fig.~\ref{fig:backgrounds:chargeflips_jets}.
Slight deviations can be observed in the high lepton \pt and lepton $\eta$ phase space regions, which are related to the binning in which the charge-misidentification probability is determined.
These effects however do not influence the measurement as the binning of the differential measurement is much coarser.

\subsection{Photon conversions}

An additional source of events in the SR is the conversion of photons into electrons in their interaction with the detector material, which we refer to as ``conversions'', which is also suppressed by means of dedicated identification algorithms.
Nevertheless, a significant contribution from \ttgamma, \tgamma, or \Zgamma is present, amounting to roughly 9\% of the events in the SRs, which is estimated using samples of simulated events.
The simulation of photon conversion is validated in a data sample with three-lepton events where the invariant mass of the three leptons is required to be consistent with the Z boson mass within 15\GeV.
Good agreement between data and simulation is observed in this VR.

\subsection{Irreducible backgrounds}

Irreducible backgrounds consist of events in which all selected leptons are prompt leptons with their electric charge correctly identified.
The major contributions to this background category are various production modes of top quark pairs in association with bosons, notably \ttH and \ttZ.
Additional contributions are coming from diboson processes such as \WZ and \ZZ or multiboson processes and the associated production of a single top quark with bosons, \eg, \tZq, \tWZ.
These background contributions are estimated from simulated samples, as detailed in Section~\ref{sec:samples}.

\begin{figure}[!ht]
\centering
\includegraphics[width=0.4\textwidth]{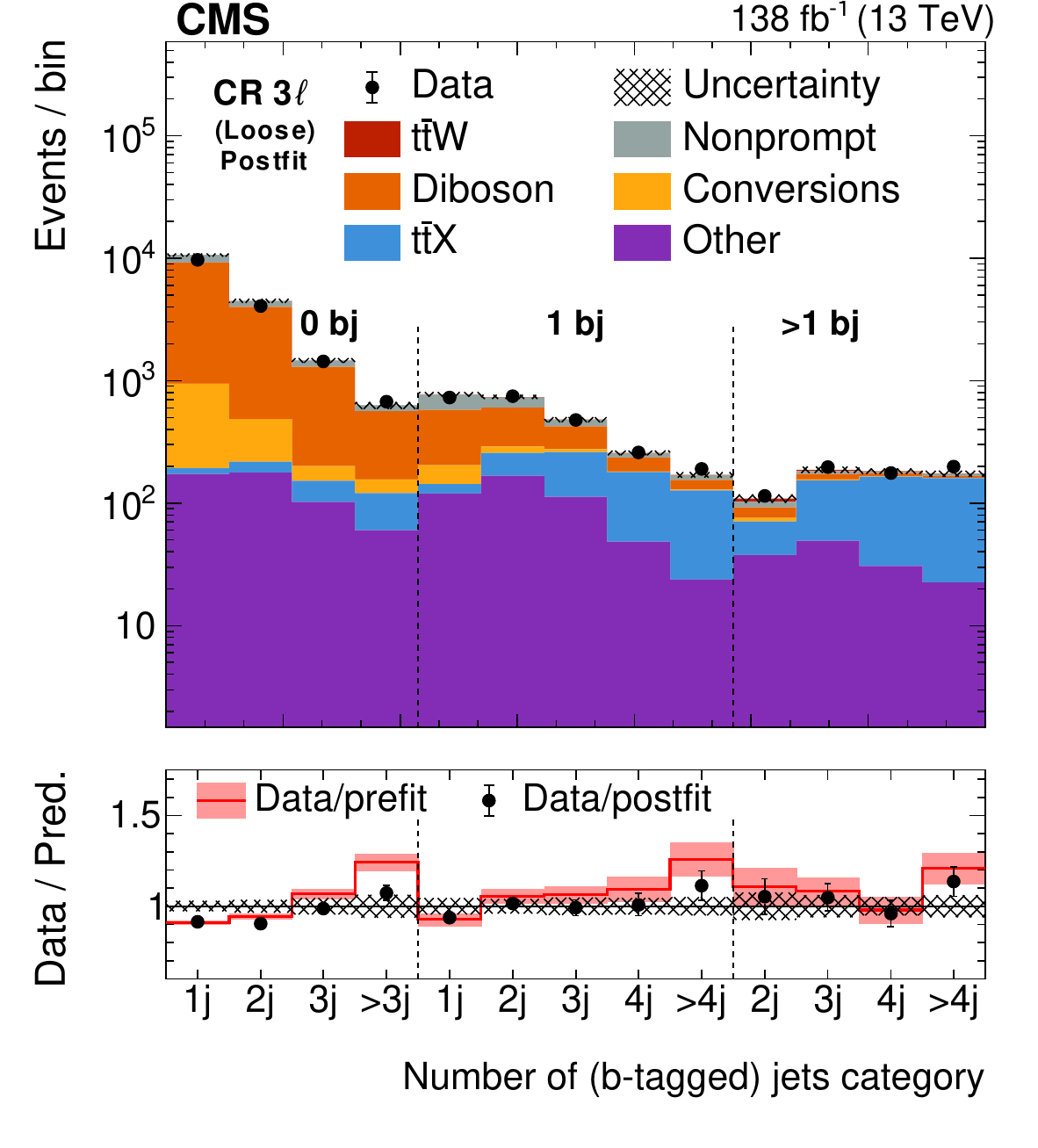}%
\hspace*{0.05\textwidth}
\includegraphics[width=0.4\textwidth]{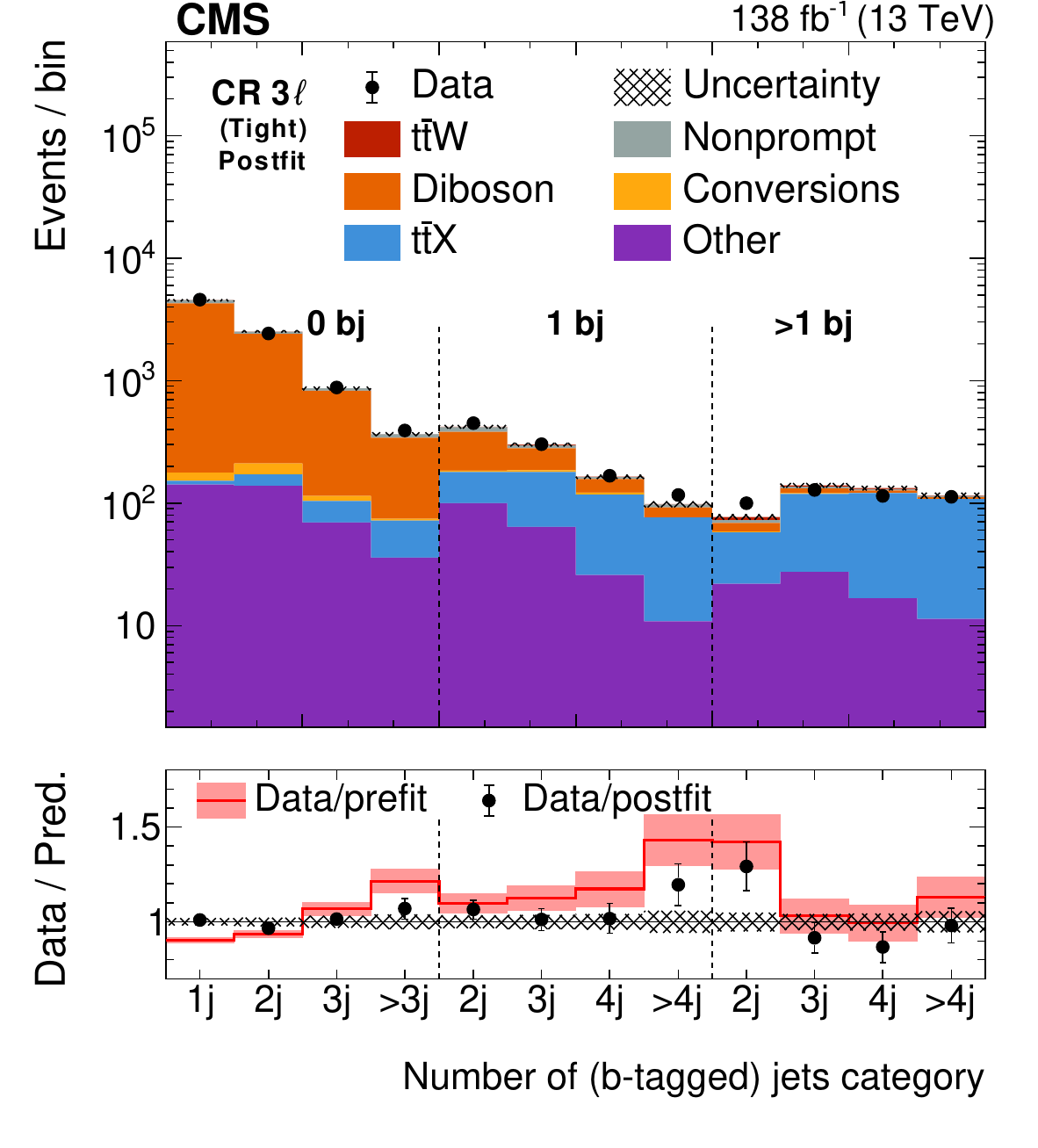}
\caption{%
    Number of selected (\PQb-tagged) jets in the event for events with three leptons passing the loose (left) and tight (right) selection after the fit to the data in the SR and CR as described in Section~\ref{sec:differential}.
    Events with no \PQb-tagged jets are classified in bins 1 to 4, events with one \PQb-tagged jet are classified in bins 5 to 9 (8) and events with more than one \PQb tag are classified in bins 10 (9) to 13 (12) for the loose (tight) lepton selection.
    \ratiodescription
}
\label{fig:backgrounds:threelepton}
\end{figure}

The modeling of irreducible backgrounds is studied in a number of VRs and two CRs: the three-lepton CR and four-lepton CR.
We build the three-lepton CR by selecting events in the \threeLepton region, but require that the events contain an OSSF lepton pair with an invariant mass within 10\GeV from \mZ.
We furthermore do not put the same requirements on the jet and \PQb-tagged jet multiplicities described in Section~\ref{sec:objects_events}, selecting instead events with at least one jet.
This CR is enriched mainly in \WZ and \ttZ, showing good agreement between data and prediction.
Events in this selection are classified depending on the number of jets and \PQb-tagged jets as shown in Fig.~\ref{fig:backgrounds:threelepton}, which helps to discriminate between \WZ and \ttZ backgrounds.

The four-lepton CR is defined by selecting events with exactly four leptons with $\pt>25$, 15, 15, and 10\GeV, respectively.
The presence of an OSSF lepton pair with an invariant mass within 10\GeV from \mZ is required.
This CR is enriched mainly in \ZZ and \ttZ events, and features a good agreement between data and predictions in all relevant variables.

\section{Systematic uncertainties}
\label{sec:systematics}

The precision on the differential cross section measurements is limited by various sources of experimental and theoretical uncertainties, the nature and estimation of which is described in this section.
The total uncertainty in the measurements is however dominated by the limited statistical precision.

\subsection{Experimental uncertainties}

The integrated luminosity of each data-taking period is used to normalize the predictions obtained from simulation.
Its measured value has an associated systematic uncertainty of 1.2--2.5\%, partially correlated across the data-taking years~\cite{cms-lumi-2016, CMS-PAS-LUM-17-004, CMS-PAS-LUM-18-002}, with an overall uncertainty of 1.6\%.

The uncertainty related to the amount of pileup interactions is taken into account by varying the total \pp inelastic cross section in all simulated samples by $\pm$4.6\% with respect to its nominal value of 69.2\unit{mb}, affecting their distributions as well as their normalization.
The pileup uncertainty is fully correlated between all data-taking years and processes.

Uncertainty in the efficiency with which events passing the signal selection are triggered in data is taken into account as well.
The trigger efficiency is measured in simulation, as well as in a data sample selected with a set of reference triggers uncorrelated with any of the lepton triggers used in this analysis.
Differences in trigger efficiency between simulation and data are taken into account for the SS dilepton selection applying weights to simulated events.
For events with three or four leptons, the trigger efficiency is found to be higher than 98\% in data at the confidence level of one standard deviation.
For these events, an uncertainty of 2\% is applied to take into account the limited statistical precision of the efficiency measurement.
These uncertainties are considered to be uncorrelated across data-taking years, as they are mainly statistical in nature.

Simulated samples are corrected for the so-called \textit{\Lone trigger prefiring}, \ie, a time jitter of the trigger signal in the muon~\cite{CMS:prefire1} and ECAL~\cite{CMS:2020cmk} systems, which led to a loss of events at the level of 0.5\% or less.

We take into account the jet energy scale (JES) and JER uncertainties by varying the jet energy scales up and down within their uncertainties~\cite{CMS:2020ebo} for all jets in the event.
We consider 14 JES uncertainty sources, some of which are independent per data-taking period, making a total of 29 independent nuisance parameters, and 6 JER sources, all independent per data-taking period.
These variations are propagated to \ptvecmiss and the \PQb tagging efficiency scale factors~\cite{CMS:2016lmd}.
They are considered to be partially correlated across all data-taking years and processes, depending on the origin of the uncertainty.
An additional uncertainty in the unclustered energy is taken into account by varying all unclustered contributions to \ptvecmiss within their respective resolutions and propagating these changes to \ptvecmiss~\cite{CMS:2019ctu}.
This uncertainty is considered to be uncorrelated among the data-taking years.

As the event selection relies on \PQb tagging, the associated uncertainty is taken into account by varying the data-to-simulation scale factors for the \PQb tagging efficiency up and down within their uncertainties.
These uncertainties are related to the purity of heavy and light flavor jets in their respective measurement regions and to the statistical fluctuations.
They are considered to be fully correlated across the processes and correlated (uncorrelated) across the years for the uncertainty sources of systematic (statistical) origin.

The lepton reconstruction and identification come with a set of scale factors, determined using the tag-and-probe method~\cite{CMS:2010svw} in \ZGst events decaying to two leptons.
Uncertainties in the resulting scale factors stem from the limited sizes of the data and simulated samples and from different sources of systematic effects, such as the choice of the dilepton invariant mass line-shape, the dilepton invariant mass window, or the phase space of the measurement.
The effect of these uncertainties is evaluated by varying the scale factors up and down within their uncertainties.

For the nonprompt background prediction, we apply different uncertainty sources to account for potential biases introduced by the method.
We assign uncertainties based on the non-closure observed in the simulations (between 5 and 30\%, depending on the lepton flavor and identification WP), which is found to be independent from the lepton kinematic properties within the statistical uncertainty of the samples.
We also assign different sources of uncertainty coming from the limited size of the multijet data samples that were used to measure the nonprompt-lepton rate, from the subtraction of the prompt lepton contamination in the measurement region, and from the differences in the background composition between measurement region (dominated by multijet background) and application region (dominated by {\ttbar}+jets background).
These uncertainties sum up to between 10 and 40\% depending on the lepton \pt, $\eta$, flavor, and data-taking period.
These uncertainties are parametrized as several independent sources that vary the integrated nonprompt-lepton rate, as well as its dependence as a function of \pt and \abseta.
The uncertainty in the charge misidentification prediction obtained from data is set to 30\%, independently from electron kinematics, which is large enough to modify the scale factor used to cover the disagreement between data and simulations that is described in Section~\ref{sub:charge_flips}, and is considered to be independent in the different data-taking periods.

Two sources of uncertainty are added to account for the mismodeling of the \WZ and \ZZ processes as a function of number of jets and \PQb-tagged jets, that is observed in the three-lepton and four-lepton CRs, as can be seen in Fig.~\ref{fig:backgrounds:threelepton}.
The first of these uncertainties adds a 10\% uncertainty per selected jet in each event up to a maximum of 50\%; the second adds a 10\% uncertainty in the event yields for events with zero or one \PQb-tagged jets and 40\% for events with two or more \PQb-tagged jets.
Both sources of uncertainty are correlated across the data-taking periods.

\subsection{Theoretical uncertainties}

We assign a normalization uncertainty to the background processes estimated with simulation.
Dedicated studies have been performed in CMS of \WZ, \ZZ, \ttZ, \ttH, \Zgamma, and \ttgamma, measuring their cross sections with a precision between 4 and 8\%~\cite{CMS:2019too,CMS:2019efc,CMS:2017dzg,CMS:2020mpn,CMS:2021klw,CMS:2020ioi}.
However, since the phase space of these dedicated measurements does not necessarily correspond to our signal selection, we apply a larger a-priori normalization uncertainty of 30\% around their normalizations for the \Zgamma and \ttgamma~\cite{CMS:2022lmh} processes.
The normalization of the \WZ, \ZZ, and \ttZ processes are free parameters in the likelihood fit described in Section~\ref{sec:differential} and are determined by the observed yields.
The \ttH process takes a 10\% normalization uncertainty according to a recent prediction~\cite{Catani:2022mfv}.
The remaining processes are of minor importance in the SR, thus we assign conservatively a 50\% normalization uncertainty around their latest theory predictions, in agreement with Ref.~\cite{CMS:2022tkv}.

Imperfect knowledge of the PDFs for the colliding protons yields another source of systematic uncertainty.
The PDF uncertainty is taken into account using NNPDF replicas~\cite{NNPDF31} and applying the procedure described in Ref.~\cite{EXT:Butterworth_2016}, resulting in 100 individual variations that are evaluated simultaneously for all processes, in addition to two variations of the strong coupling constant \alpS.

Uncertainty in the renormalization and factorization scale in the matrix element is taken into account in a similar way.
All simulated processes are reweighted using both independent and correlated up and down variations by a factor of 2 of both scales (excluding the two extreme variations).
These sources of uncertainty are correlated between the data-taking years but are treated uncorrelated between different processes.

Uncertainties related to the initial-state and final-state radiation (ISR and FSR, respectively) are evaluated similarly by variations in the renormalization scale for emissions in ISR and FSR of a factor two (up and down).
These variations are considered to be fully correlated across the data-taking years, while only FSR is correlated between different processes.

Some of these sources of uncertainty can affect both the production cross section and the acceptance of all simulated processes.
The systematic effect in the cross section is not taken into account for processes that are assigned a normalization uncertainty based on dedicated measurements or that are left freely floating in the fit.
For those processes, only the acceptance effects are considered.

\section{Differential cross section measurements}
\label{sec:differential}

We measure the \ttW differential cross sections using two complementary methods that share the same statistical model and only differ on the definition of the reconstruction-level bins used in the signal extraction fit.
In the \twoLeptonss region we use an event-level MVA discriminator and the variable of interest to define the reconstruction-level bins.
The other method only uses a reconstruction-level binning in the variable of interest and requires the tighter lepton selection described in Section~\ref{sec:objects_events}.
This method is used in the \threeLepton region as the main method and in the \twoLeptonss region as a complementary measurement to the MVA-based approach.

\subsection{Definition of measured observables}

Particle-level object definitions~\cite{Collaboration:2267573} are used to define the fiducial phase space and to construct the measured observables.
We consider particle-level leptons by clustering prompt electrons and muons with photons not arising from hadron decays, using the anti-\kt algorithm with a cone size of $R=0.1$.
Only particle-level leptons with $\pt>15\GeV$ and $\abseta<2.5$ (electrons) or $\abseta<2.4$ (muons) are selected for further processing.
Jets are constructed by clustering all stable particles excluding prompt selected leptons and neutrinos using the anti-\kt algorithm with a cone size of $R=0.4$.
Only particle-level jets with $\pt>25\GeV$ and $\abseta<2.4$ are selected in the analysis.
To determine the jet flavor, we use the ``ghost clustering'' procedure described in Ref.~\cite{Collaboration:2267573}, in which jets are re-clustered including the decayed \PQb hadrons with their momentum scaled to an arbitrary small value (ghosts), in addition to the set of particles included in the jet definition.
We define \PQb jets as those that contain a ghost when this reclustering is performed.

Using these objects, we define a fiducial region that aligns with the categories used to perform the measurement.
We define a \twoLeptonss fiducial region taking events with exactly two leptons with $\pt>25$ and 15\GeV that have the same electric charge.
Additionally, events are required to have at least 3 jets, out of which at least one is a \PQb jet.
For the \threeLepton region we select events having at least 3 leptons with $\pt>25$, 15, and 15\GeV respectively, and at least two jets, out of which at least one is a \PQb jet.
Signal events that do not pass this fiducial selection are taken into account as background events.

We also use these objects to construct the variables as a function of which the cross section is measured, which are listed in Table~\ref{tab:variables}.
Many of these distributions, such as the jet multiplicity or the scalar sum of jet \pt, \HT, are useful to assess the \ttW modeling as a background to measurements of other processes with larger jet multiplicity, like \ttH or \tttt production~\cite{Ferencz:2023fso}.
Other variables, such as the jet and lepton kinematic distributions, are generic probes of the \ttW modeling.

\begin{table}[!ht]
\centering
\topcaption{%
      Kinematic variables studied in the differential cross section measurement of the signal. 
}
\label{tab:variables}
\renewcommand{\arraystretch}{1.1}
\cmsTable{\begin{tabular}{ll}
    Variable & Description \\
    \hline
    \multicolumn{2}{c}{Distances between objects} \\
    \Detaloneltwo & Absolute difference in $\eta$ between the two leptons with the highest \pt \\
    \DRloneltwo & Separation in $(\eta,\phi)$ space between the two leptons with the highest \pt \\
    \DRlonejetmin & Separation in $(\eta,\phi)$ space between the lepton with the highest \pt and its closest jet \\[\cmsTabSkip]
    \multicolumn{2}{c}{Jet-related variables} \\
    \HT & Scalar sum of jet \pt \\
    Leading jet \abseta & \abseta of the jet with highest \pt \\
    Leading jet \pt & \pt of the jet with highest \pt \\
    Subleading jet \abseta & \abseta of the jet with next-to-highest \pt \\
    Subleading jet \pt & \pt of the jet with next-to-highest \pt \\[\cmsTabSkip]
    \multicolumn{2}{c}{Lepton-related variables} \\
    Leading lepton \pt & \pt of the lepton with highest \pt \\
    Subleading lepton \pt & \pt of the lepton with next-to-highest \pt \\
    Lepton maximum \abseta & Maximum \abseta among the leptons \\
    $\sum$ lepton \pt & Sum of the lepton \pt's \\
    \mleptons & Invariant mass of the system of the two leptons with the highest \pt \\[\cmsTabSkip]
    \multicolumn{2}{c}{Object multiplicity} \\
    Number of jets & Jet multiplicity \\
\end{tabular}}
\end{table}

\subsection{Statistical model}

We determine the differential cross section as a function of a given variable in \Ngen particle-level bins using the number of observed events $n_i$ in \Nreco reconstructed-level bins, using a likelihood-based unfolding method.
We fit data to a statistical model implemented in the \textsc{combine} software~\cite{CMS:2024onh} and described by the likelihood function $\mathcal{L}$, given by
\begin{equation}\label{eq:likelihood}
    \mathcal{L}\big(\text{data};\,\vecsigma,\vectheta\big)=\prod_{i}^{\Nreco}\mathcal{P}\Bigg(n_{i}\,\Bigg\vert\sum_j^{\Ngen} \sigma_j\,\epsilon_{ij}(\vectheta)L(\vectheta)+b_i(\vectheta)\Bigg)\prod_{k}p\big(\widetilde{\theta}_{k}\big\vert\theta_{k}\big).
\end{equation}
The symbol $\mathcal{P}$ denotes the Poisson probability for observing $n_i$ events in a given reconstruction-level bin with an expectation given by the expected background events and signal events outside the fiducial region in that bin $b_i(\vectheta)$, the luminosity $L(\vectheta)$, the fiducial cross section in each particle-level bin $\sigma_j$, which are the parameters of interest in the fit, and the elements of the response matrix $\epsilon_{ij}(\theta)$.
This matrix parametrizes the detector response and is defined by
\begin{equation}
    \epsilon_{ij}=\frac{N\big(\text{signal events generated in bin }j\text{ and reconstructed in bin }i\big)}{N\big(\text{signal events generated in bin }j\big)}.
\end{equation}
The likelihood model also includes the explicit dependence of the different quantities as a function of the nuisance parameters \vectheta that model the impact of the systematic uncertainties and that are constrained by their prior probability density functions, $p$.

In this kind of measurement, additional regularization terms are sometimes added to the likelihood to penalize nonphysical fluctuations, effectively reducing the uncertainties at the price of introducing a controlled bias.
In our case, we have not included such terms since the response matrices used in this analysis are sufficiently well-conditioned.
We quantified that using the condition number, defined as the ratio of the largest to the smallest value in a singular value decomposition of each matrix.
The condition of all the matrices if below 4, which indicates that the inverse of the matrix can be computed with high accuracy.

\subsection{Signal extraction in the \texorpdfstring{\twoLeptonss}{2lSS} region}

The method we use for the differential cross section measurement in the \twoLeptonss region is based on separating the signal and the backgrounds using an MVA discriminator that takes event-level variables as input.
This strategy uses the loose lepton identification criterion described in Section~\ref{sec:objects_events} to ensure a high signal efficiency.
In addition, a requirement of $\ptmiss>30\GeV$ is applied to reject the contribution from nonprompt leptons.

\begin{table}[!pt]
\centering
\topcaption{%
    Overview of input variables to the BDT and their prefit $p$-values, obtained in a $\chi^2$ goodness-of-fit test.
}
\label{tab:bdt_input_variables}
\renewcommand{\arraystretch}{1.1}
\cmsTable[0.8]{\begin{tabular}{l@{}r}
    MVA input variables & Prefit $p$-value \\
    \hline
    Leptons & \\
    Leading lepton \pt & 0.17 \\
    Subleading lepton \pt & 0.33 \\
    Leading lepton $\eta$ & 0.85 \\
    Subleading lepton $\eta$ & 0.29 \\
    Leading lepton energy & 0.65 \\
    Subleading lepton energy & 0.82 \\[\cmsTabSkip]
    Jets & \\
    Leading jet \pt & 0.61 \\
    Subleading jet \pt & 0.29 \\
    Leading jet $\eta$ & 0.37 \\
    Subleading jet $\eta$ & 0.98 \\
    Leading jet mass & 0.47 \\
    Subleading jet mass & 0.87 \\
    Number of jets & 0.36 \\
    Maximum \abseta among all jets & 0.80 \\[\cmsTabSkip]
    \PQb-tagged jets & \\
    Leading \PQb-tagged jet \pt & 0.47 \\
    Leading \PQb-tagged jet $\eta$ & 0.75 \\
    Highest \deepjet score among all jets & 0.78 \\
    \deepjet score of leading jet & 0.26 \\
    \deepjet score of subleading jet & 0.98 \\
    Number of loose \PQb-tagged jets & 0.72 \\
    Number of medium \PQb-tagged jets & 0.69 \\
    Number of tight \PQb-tagged jets & 0.59 \\[\cmsTabSkip]
    Angular separation & \\
    Minimum \DR between a lepton and a \PQb-tagged jet & 0.40 \\
    \DR between the leading and subleading lepton & 0.23 \\
    \Deta between the leading and subleading lepton & 0.31 \\
    Minimum \DR between the leading lepton and a jet & 0.96 \\
    Minimum \DR between the leading lepton and a \PQb-tagged jet & 0.15 \\
    Minimum \DR between the leading lepton and a non-\PQb-tagged jet & 0.96 \\[\cmsTabSkip]
    Dijet and dilepton properties & \\
    Maximum dijet mass & 0.43 \\
    Maximum dijet \pt & 0.81 \\
    Invariant mass of the leading and subleading lepton system & 0.31 \\[\cmsTabSkip]
    Global variables & \\
    Scalar sum of the \pt of the jets & 0.34 \\
    Scalar sum of the \pt of the leptons and \ptmiss & 0.16 \\
    Lepton flavor category ($\Pe\Pe$, $\Pe\PGm$, $\PGm\Pe$, or $\PGm\PGm$, with \pt ordering) & 0.21 \\
    Sum of lepton charges & 0.24 \\
    \ptmiss & 0.18 \\
    Data-taking year & \NA \\
\end{tabular}}
\end{table}

\begin{figure}[!ht]
\centering
\includegraphics[width=0.4\textwidth]{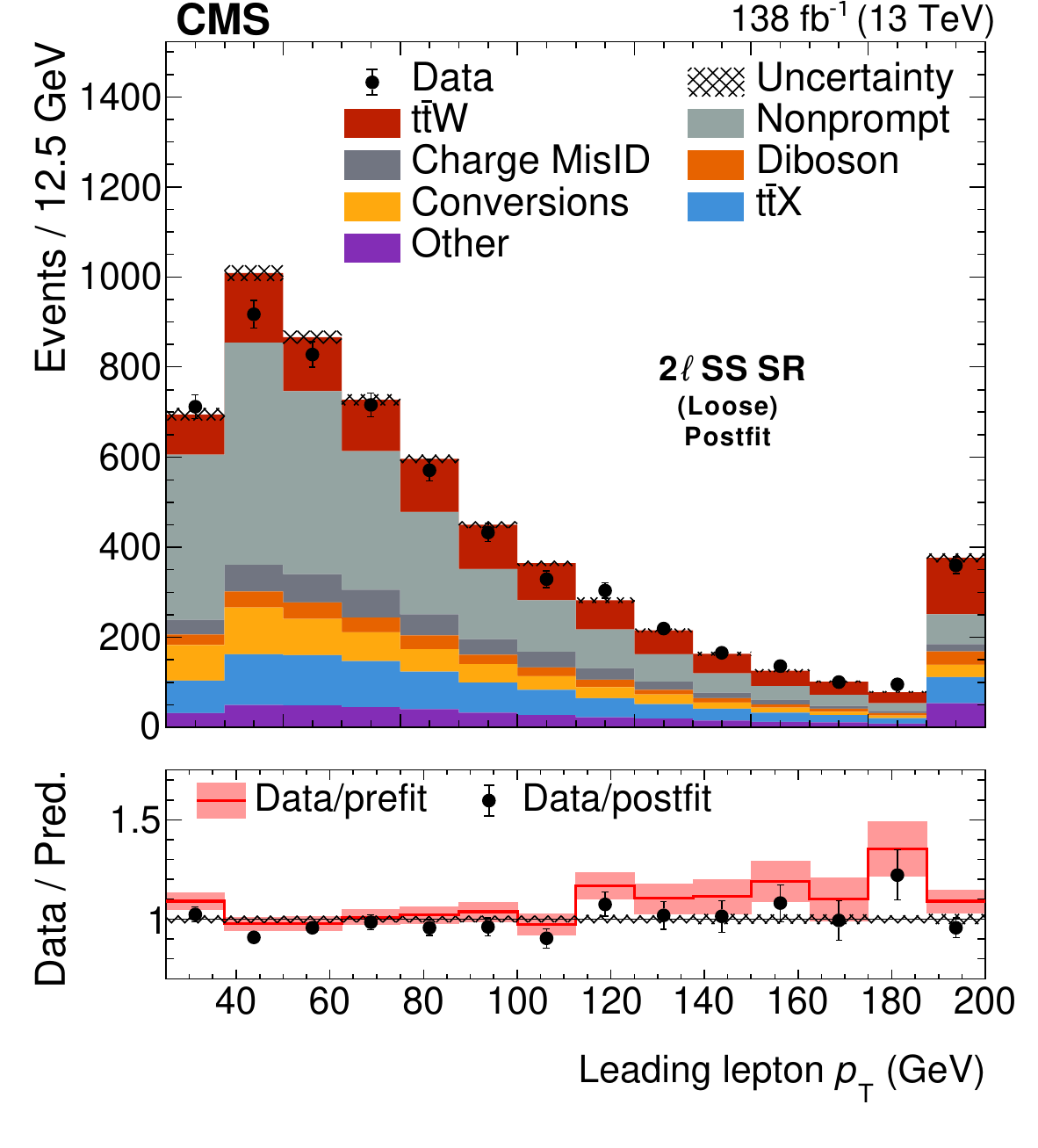}%
\hspace*{0.05\textwidth}%
\includegraphics[width=0.4\textwidth]{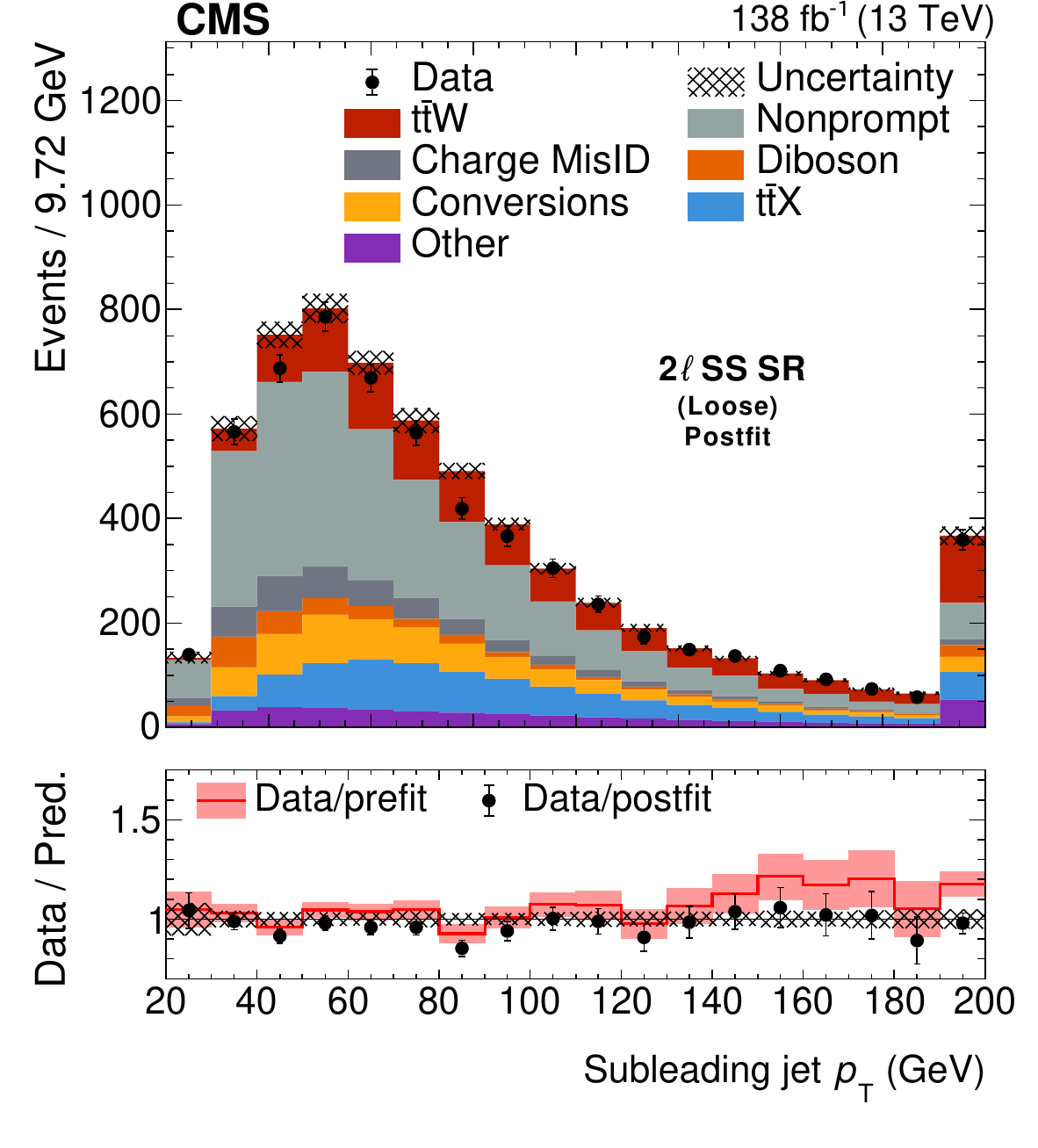} \\
\includegraphics[width=0.4\textwidth]{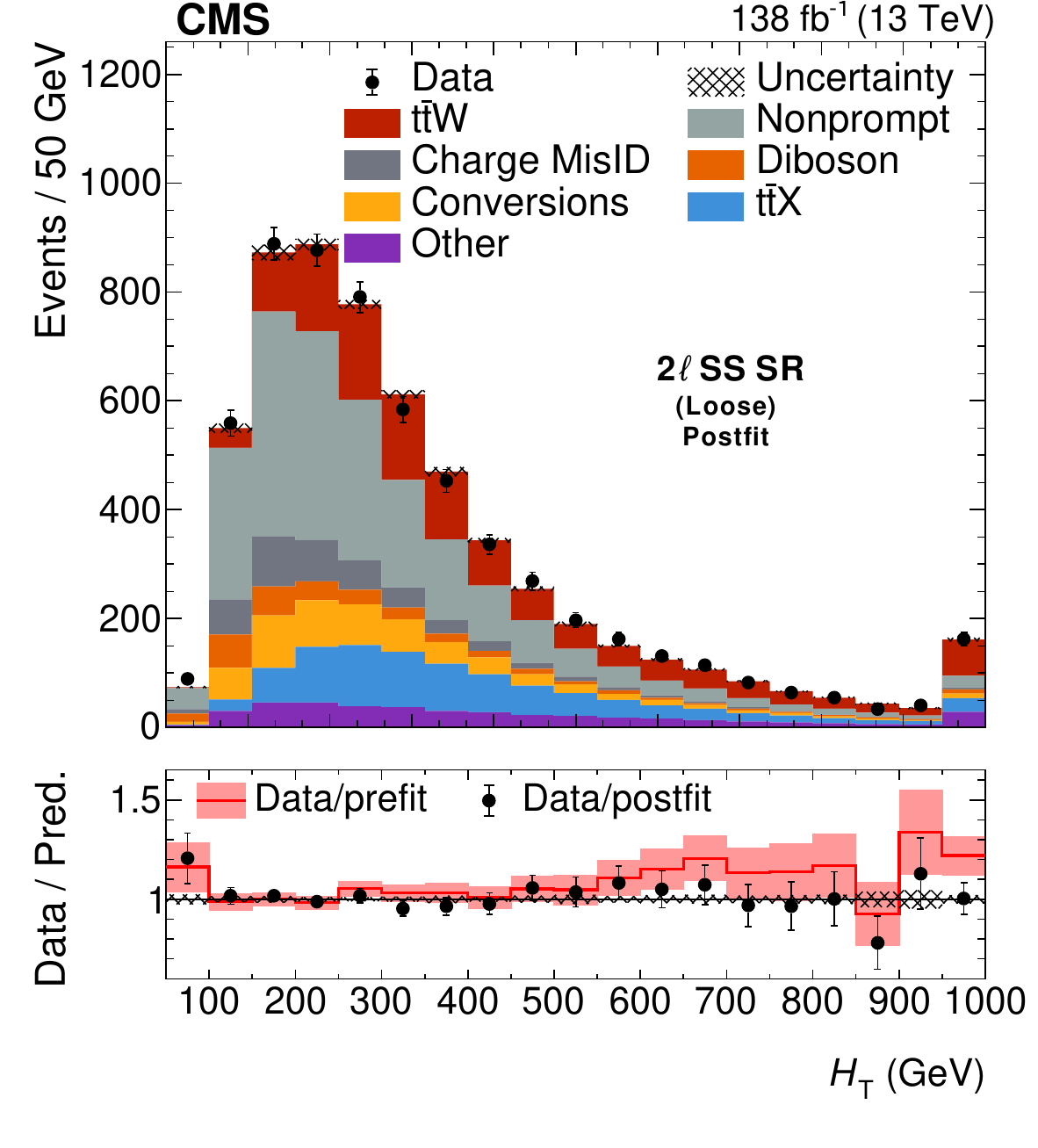}%
\hspace*{0.05\textwidth}%
\includegraphics[width=0.4\textwidth]{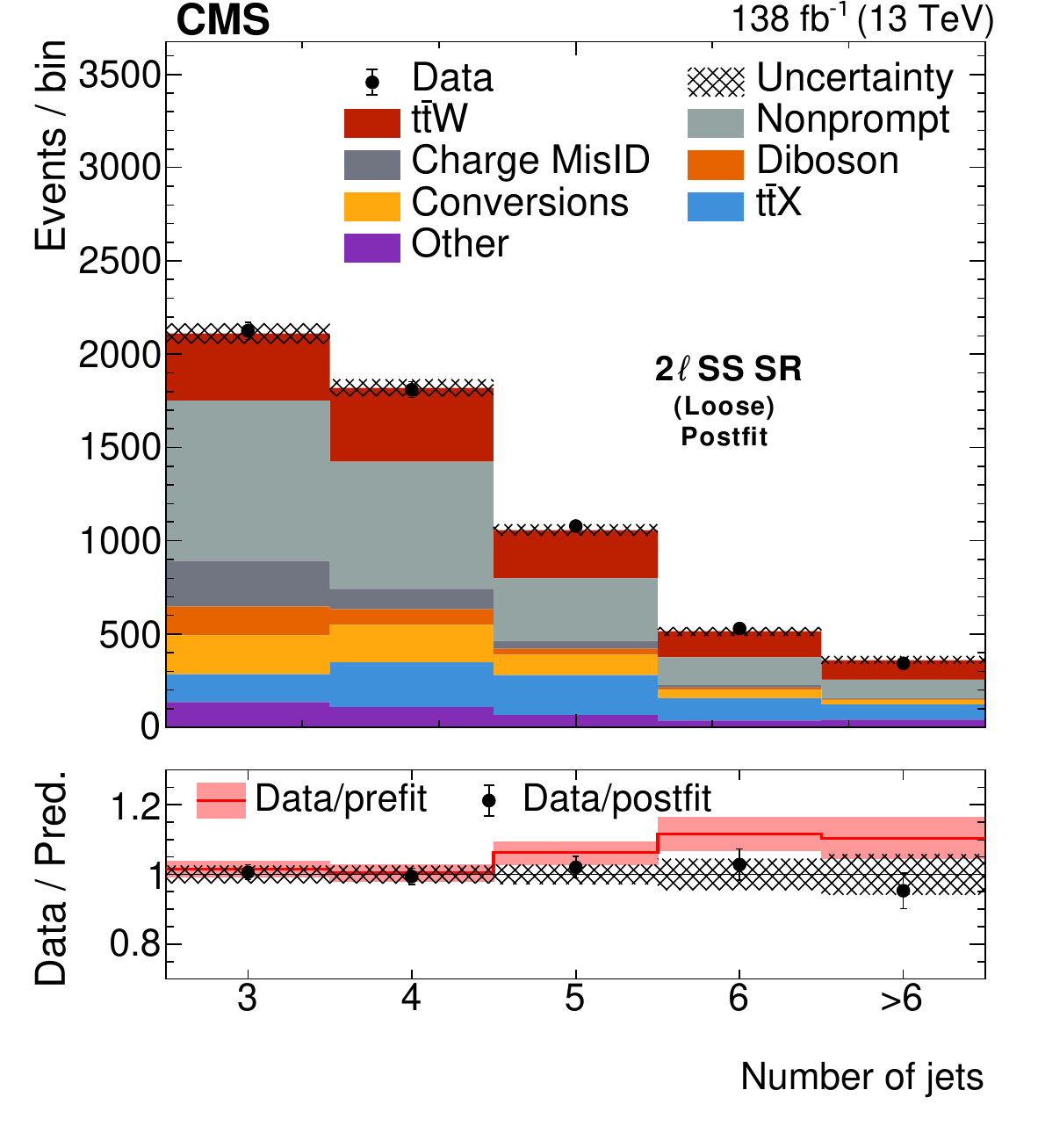}
\caption{%
    \pt of the leading lepton (upper left), \pt of the subleading jet (upper right),  scalar \pt sum of selected jets in the event (lower left), and the number of selected jets in the event (lower right), in the two-lepton signal selection for the loose lepton selection for data (points) and predictions (filled histograms) after the fit to the data in the SR and CR.
    \ratiodescription
    \overflow
}
\label{fig:signalregion:dilepton_1}
\end{figure}

\begin{figure}[!ht]
\centering
\includegraphics[width=0.4\textwidth]{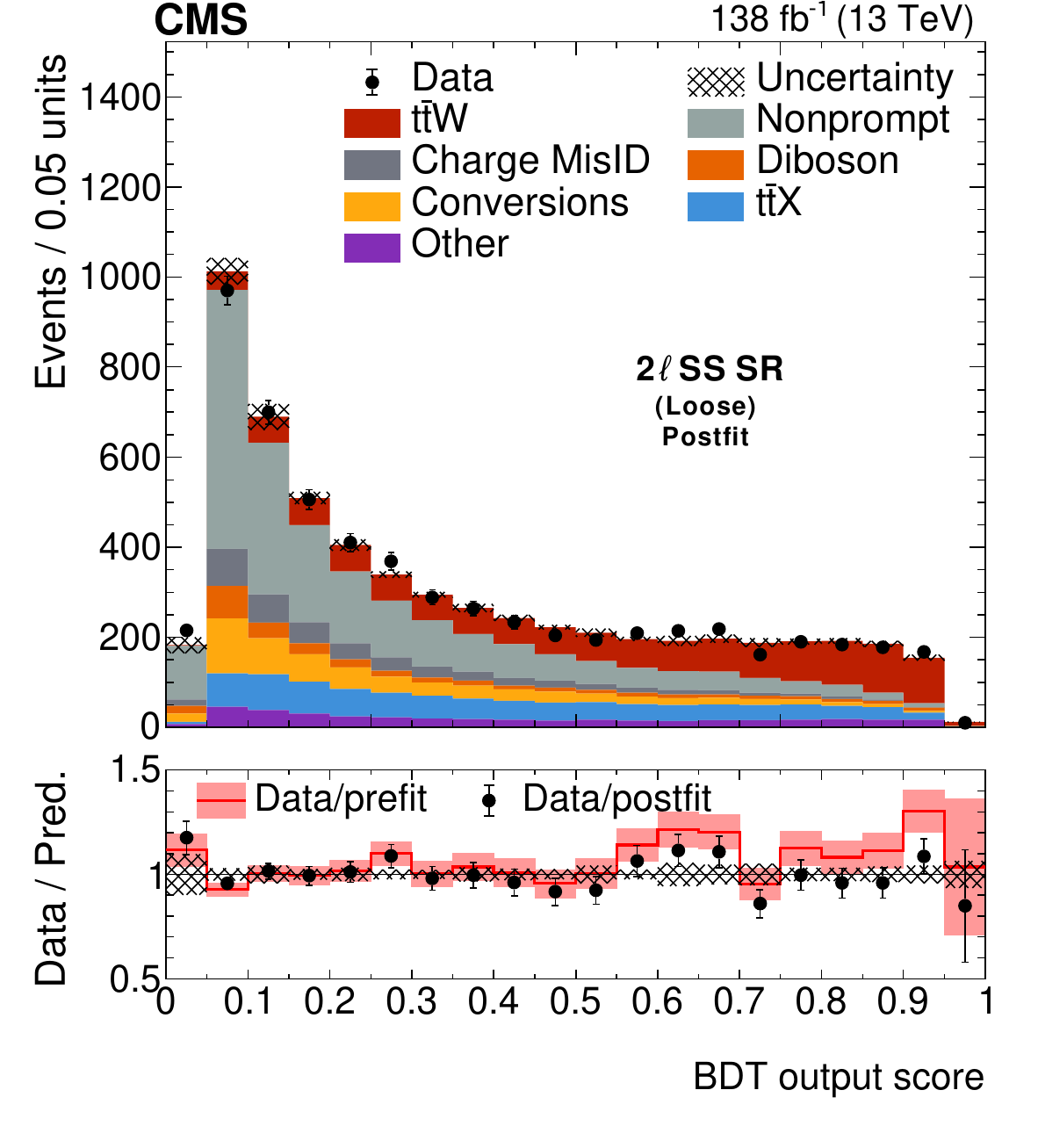}
\caption{%
    Distribution of the BDT output node in the two-lepton signal selection for the loose lepton selection for data (points) and predictions (filled histograms) after the fit to the data in the SR and CR.
    \ratiodescription
}
\label{fig:signalregion:dilepton_2}
\end{figure}

The MVA algorithm of choice is gradient boosted decision tree (gBDT), trained with the \textsc{XGBoost} library~\cite{xgboost}.
For the training and validation of the gBDT, 20\% of the simulated events are used, and this part of the samples is excluded from the rest of the analysis to avoid potential overtraining biases.
In total, 37 variables are selected as input features.
It was checked that their distributions are well described by our predictions and their mutual correlation matrix agrees well between simulation and data.
The input features include basic kinematic properties of the leptons and jets in the event, the \PQb tagging scores of the jets, the directional separation between various combinations of leptons and jets, the invariant mass and \pt of various dilepton and dijet systems, and some global event variables such as the scalar \pt sum of all selected jets, the lepton flavor category, the sum of lepton charges, and the \ptmiss.
The data-taking year is also included as an input variable to allow the algorithm to account for changing experimental conditions.
A full overview of the input variables in the \twoLeptonss region is provided in Table~\ref{tab:bdt_input_variables}, together with their corresponding prefit $p$-values, obtained in a $\chi^2$ goodness-of-fit test, which show no indication of significant tensions between data and the prediction for these variables.
The training was performed in the \twoLeptonss region, including the $\ptmiss>30\GeV$ requirement.
The learning rate, number of estimators, and maximum depth per estimator were optimized simultaneously in a grid search.
The good postfit agreement between data and prediction in this selection is shown for a number of variables of interest and the BDT output score in Figs.~\ref{fig:signalregion:dilepton_1} and~\ref{fig:signalregion:dilepton_2}, respectively.

\begin{figure}[!tp]
\centering
\includegraphics[width=0.4\textwidth]{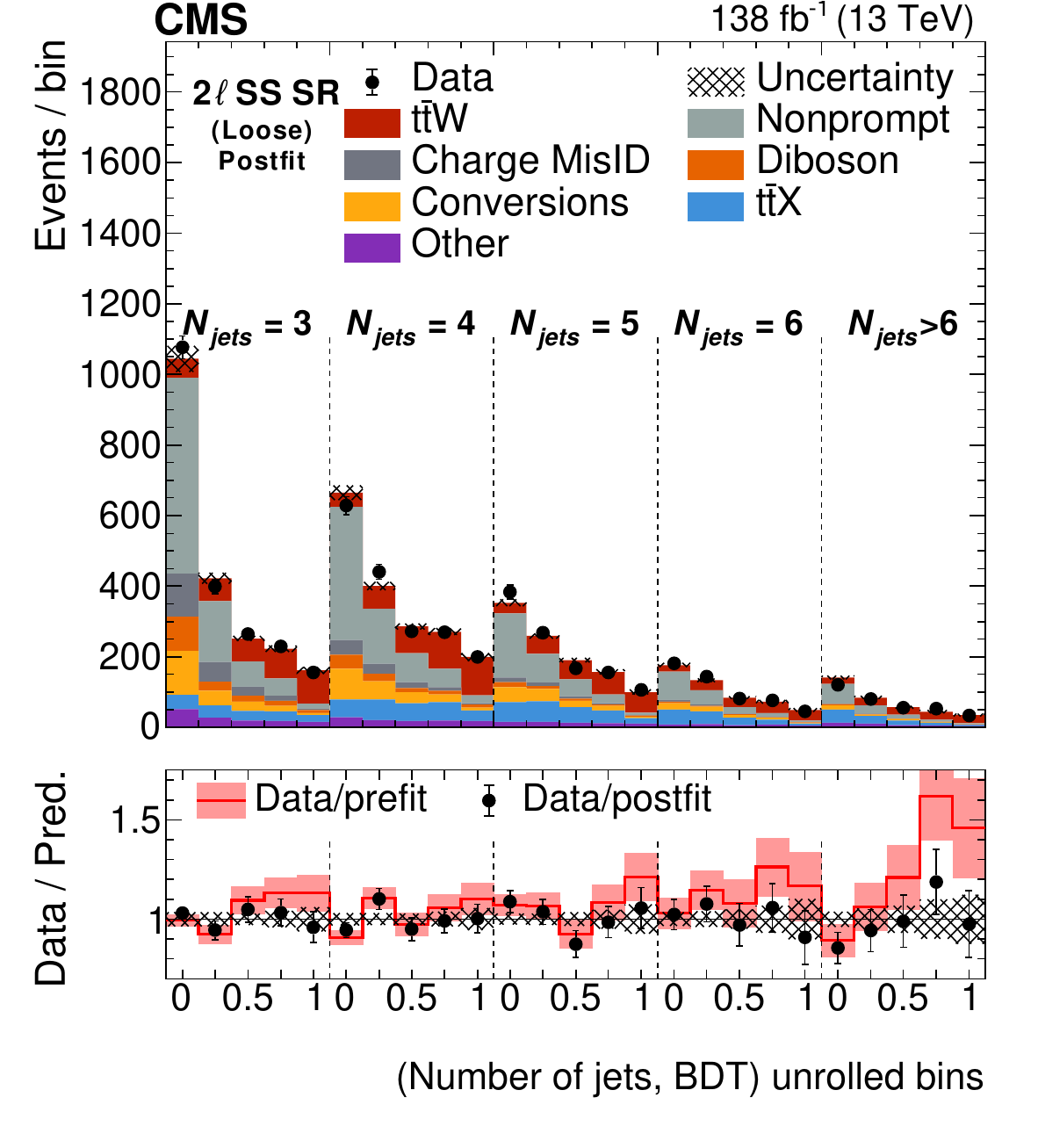}%
\hspace*{0.05\textwidth}%
\includegraphics[width=0.4\textwidth]{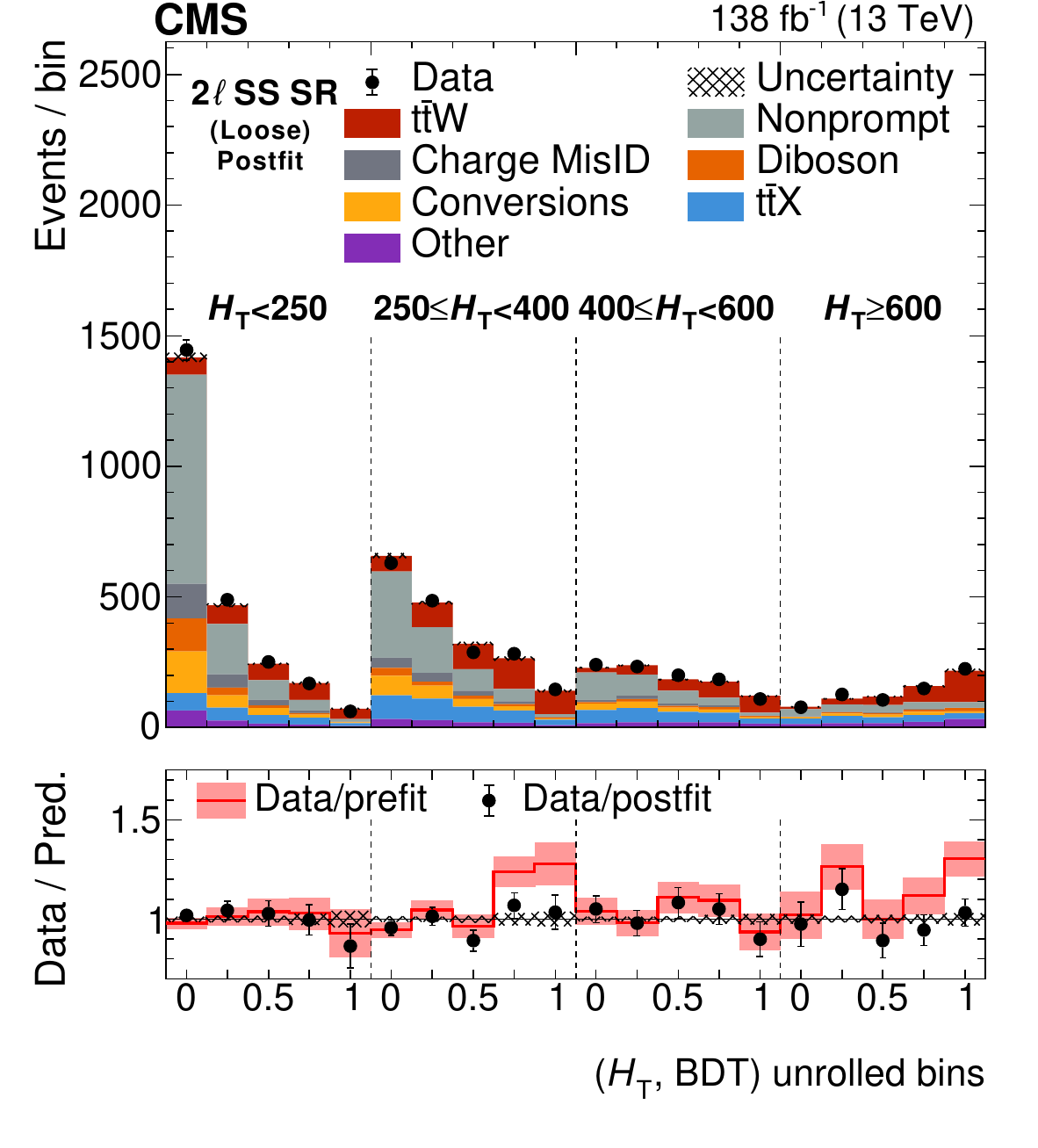} \\
\includegraphics[width=0.4\textwidth]{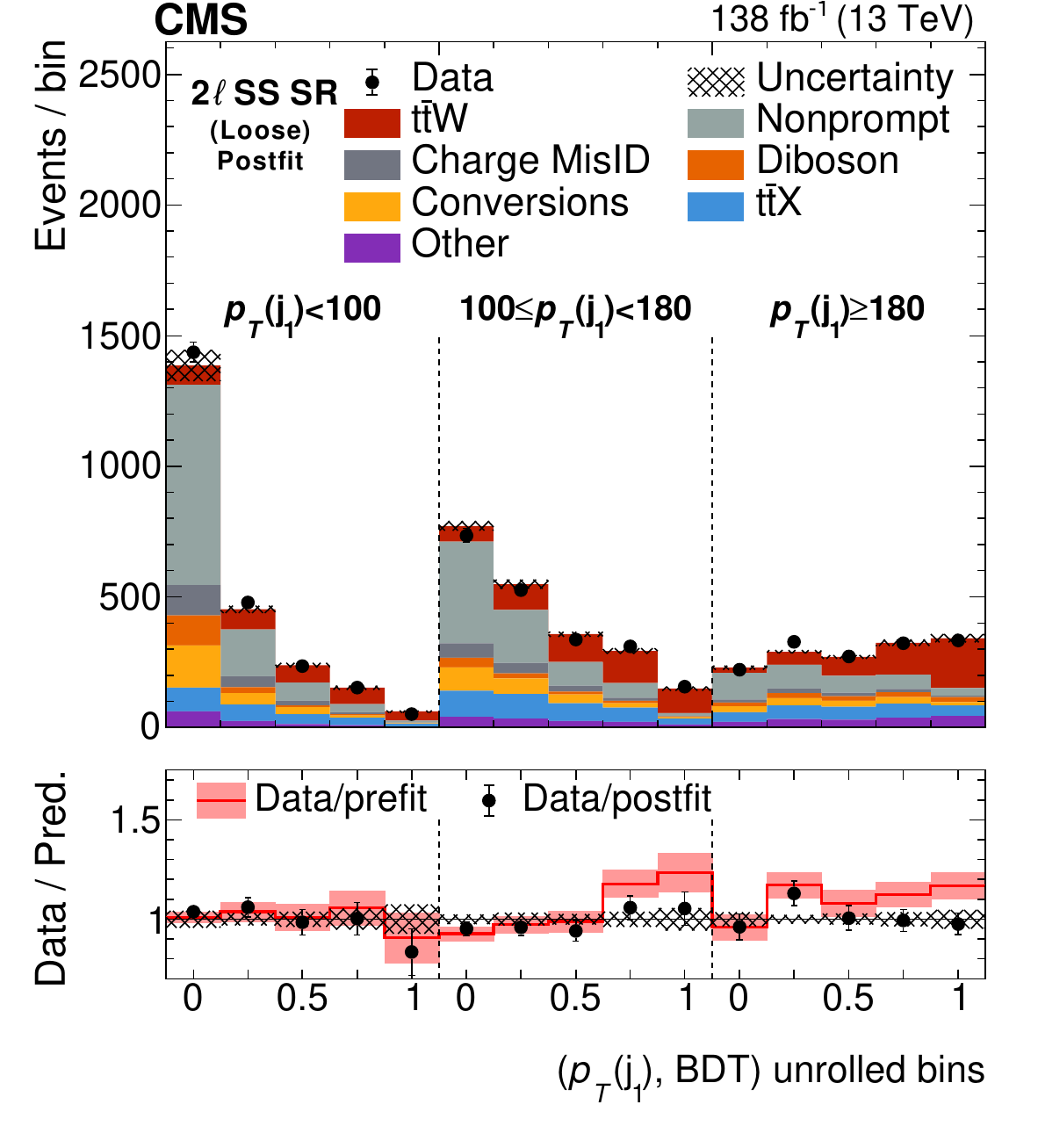}%
\hspace*{0.05\textwidth}%
\includegraphics[width=0.4\textwidth]{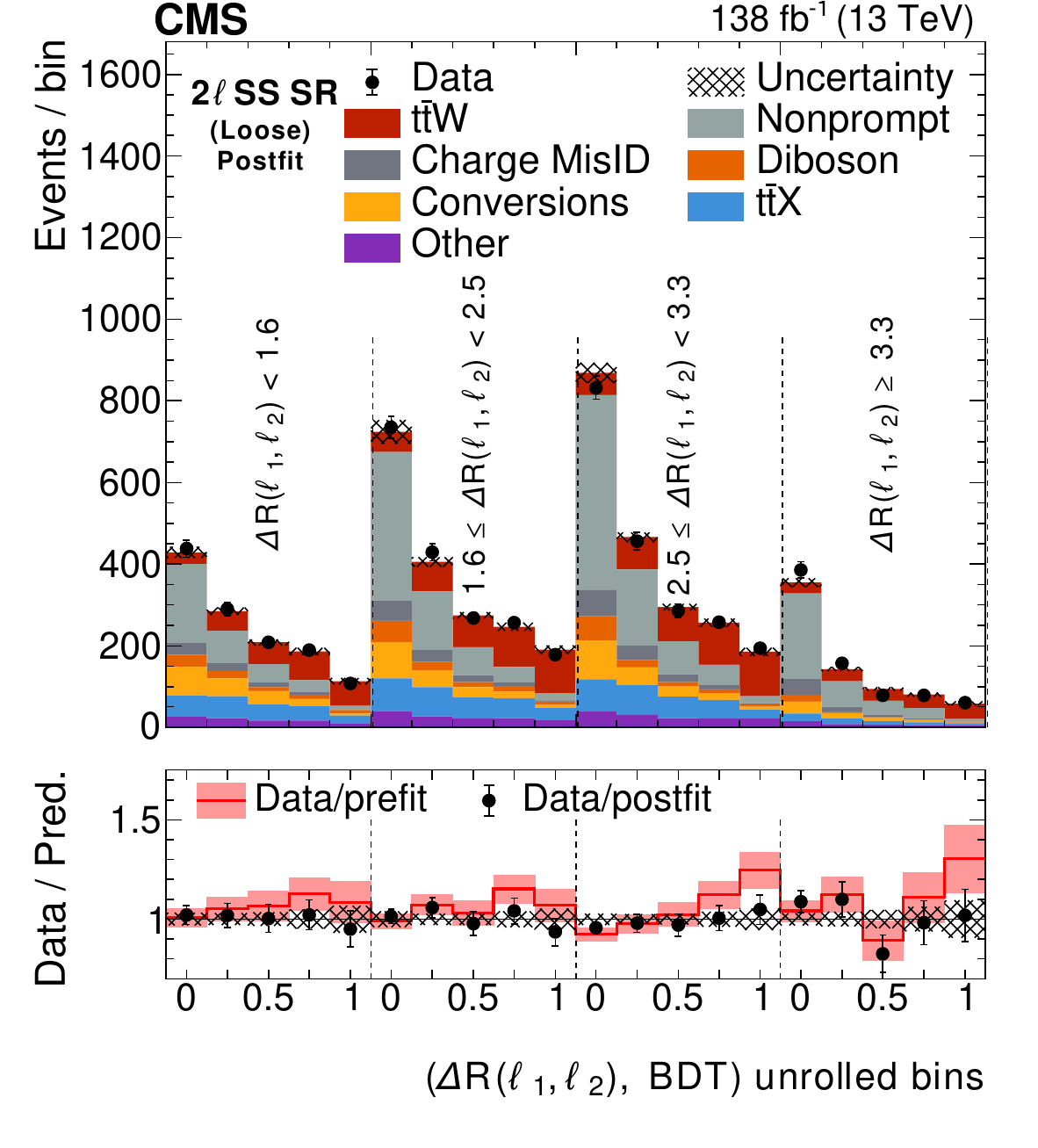}
\caption{%
    Distributions defined in a two-dimensional grid between the MVA score and the variable of interest in the two-lepton signal selection for data (points) and predictions (filled histograms) after the fit to the data in the SR and CR for the loose lepton selection.
    \ratiodescription
}

\label{fig:signalregion:dilepton_3}
\end{figure}

The signal extraction is performed by fitting the event yields in reconstruction-level bins defined in a two-dimensional grid between the MVA score and the variable of interest.
We use five bins in the MVA score and a number of bins in the variable of interest equal to the number of generator level bins to be measured, as shown in Fig.~\ref{fig:signalregion:dilepton_3}.

\begin{figure}[!pt]
\centering
\includegraphics[width=0.475\textwidth]{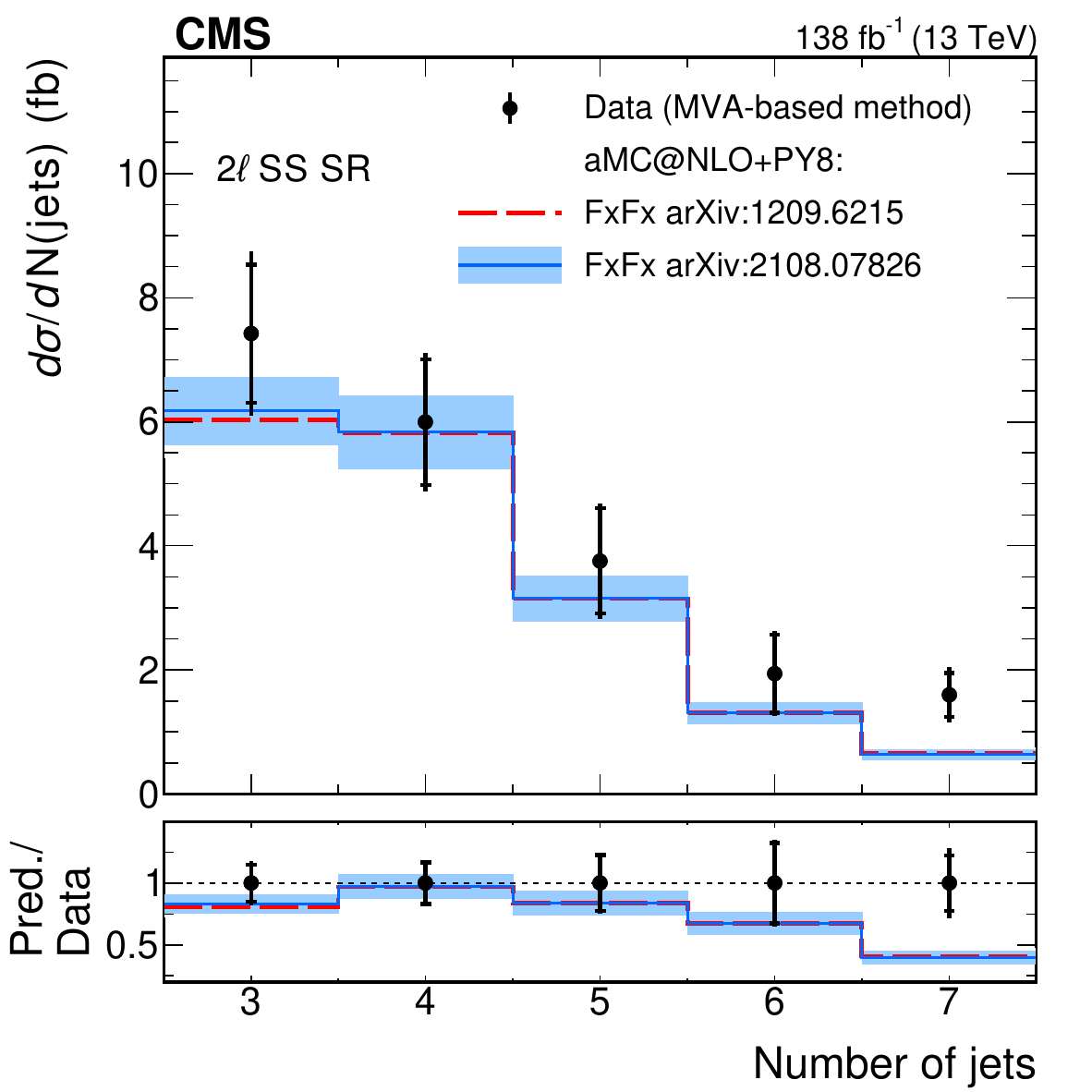}%
\hfill%
\includegraphics[width=0.475\textwidth]{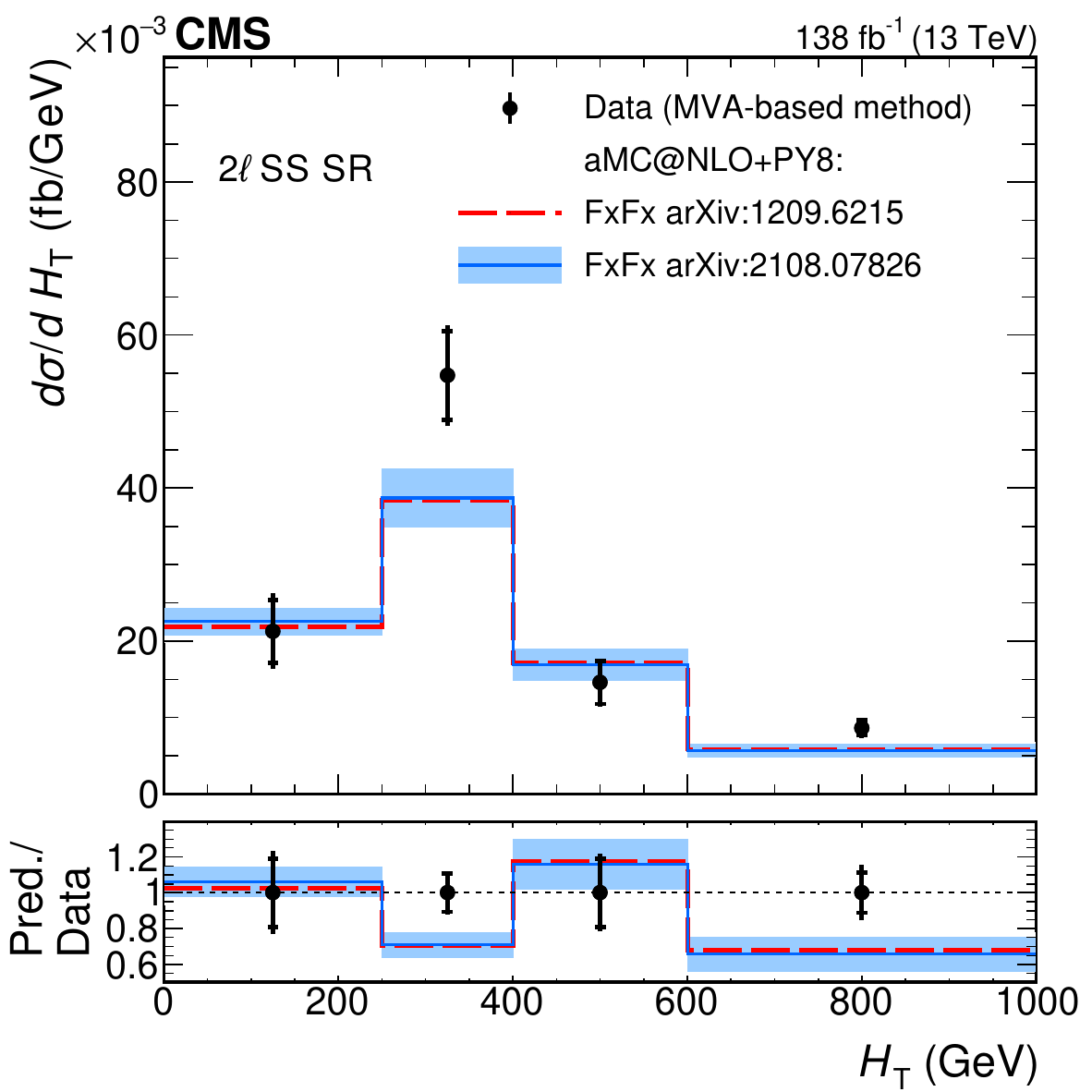} \\[\cmsTabSkip]
\includegraphics[width=0.475\textwidth]{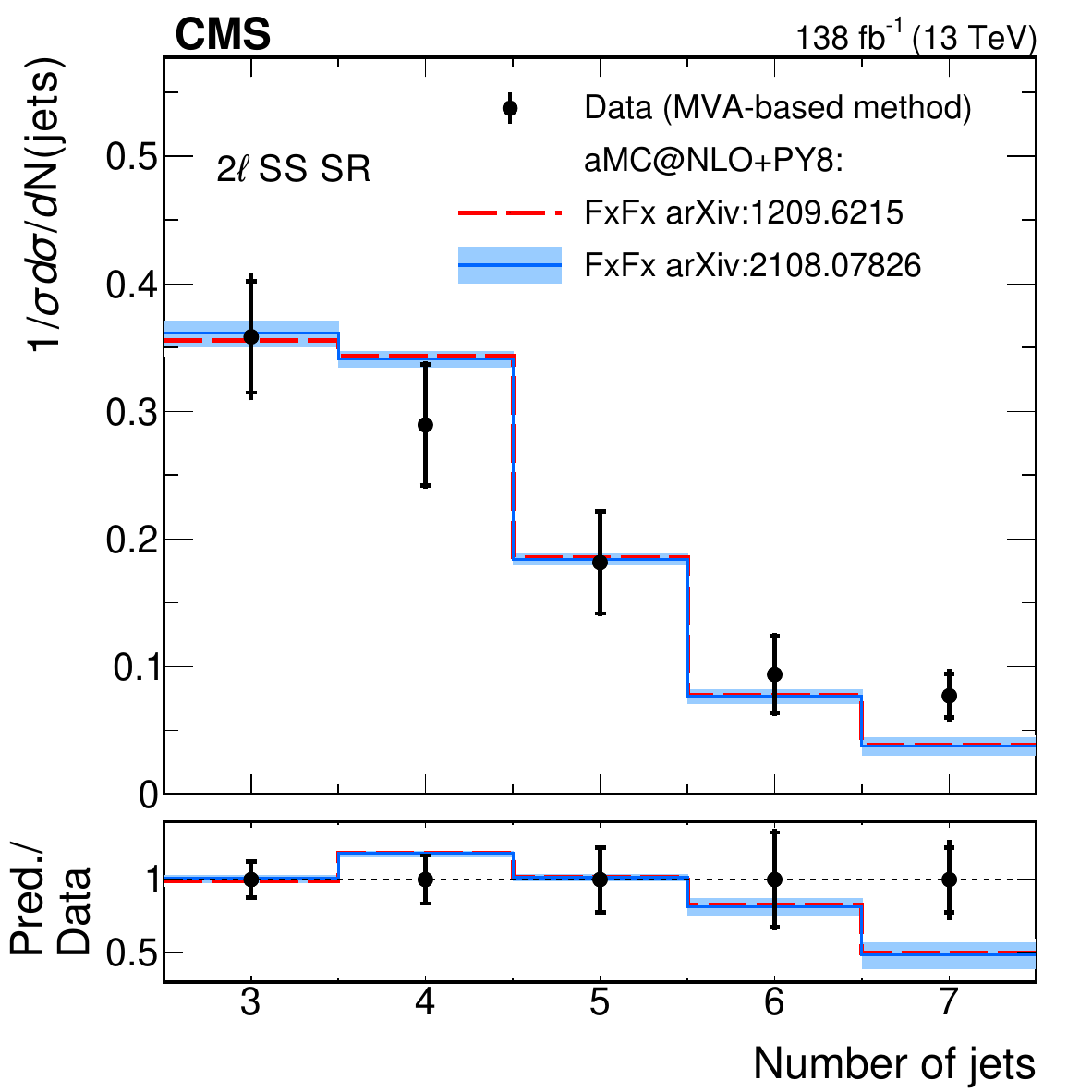}%
\hfill%
\includegraphics[width=0.475\textwidth]{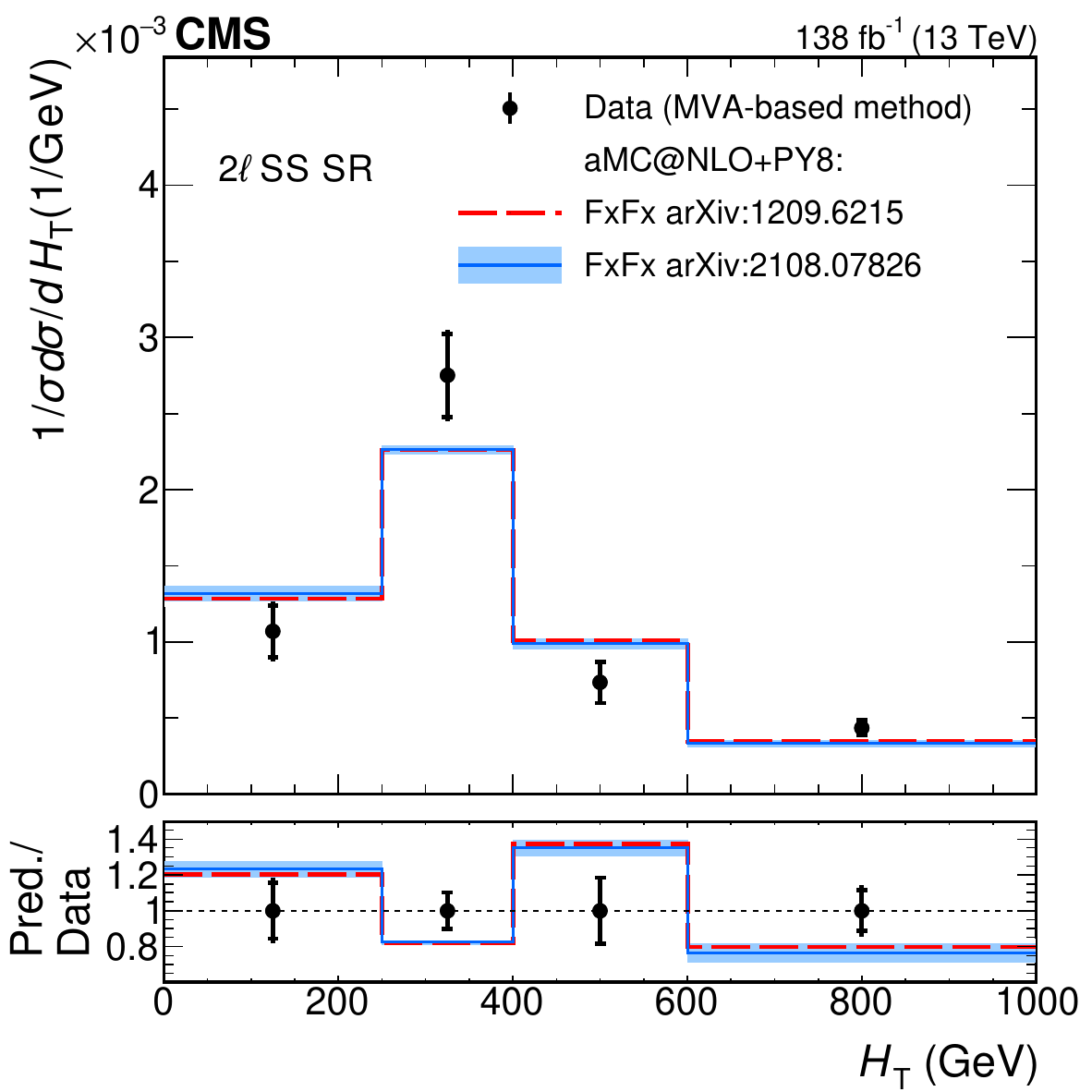}
\caption{%
    Absolute (upper row) and normalized (lower row) differential cross section measured as a function of the jet multiplicity (left) and \HT (right), using the MVA-based method.
    \diffextracaption
}
\label{fig:differential_xsec_1}
\end{figure}

\begin{figure}[!pt]
\centering
\includegraphics[width=0.475\textwidth]{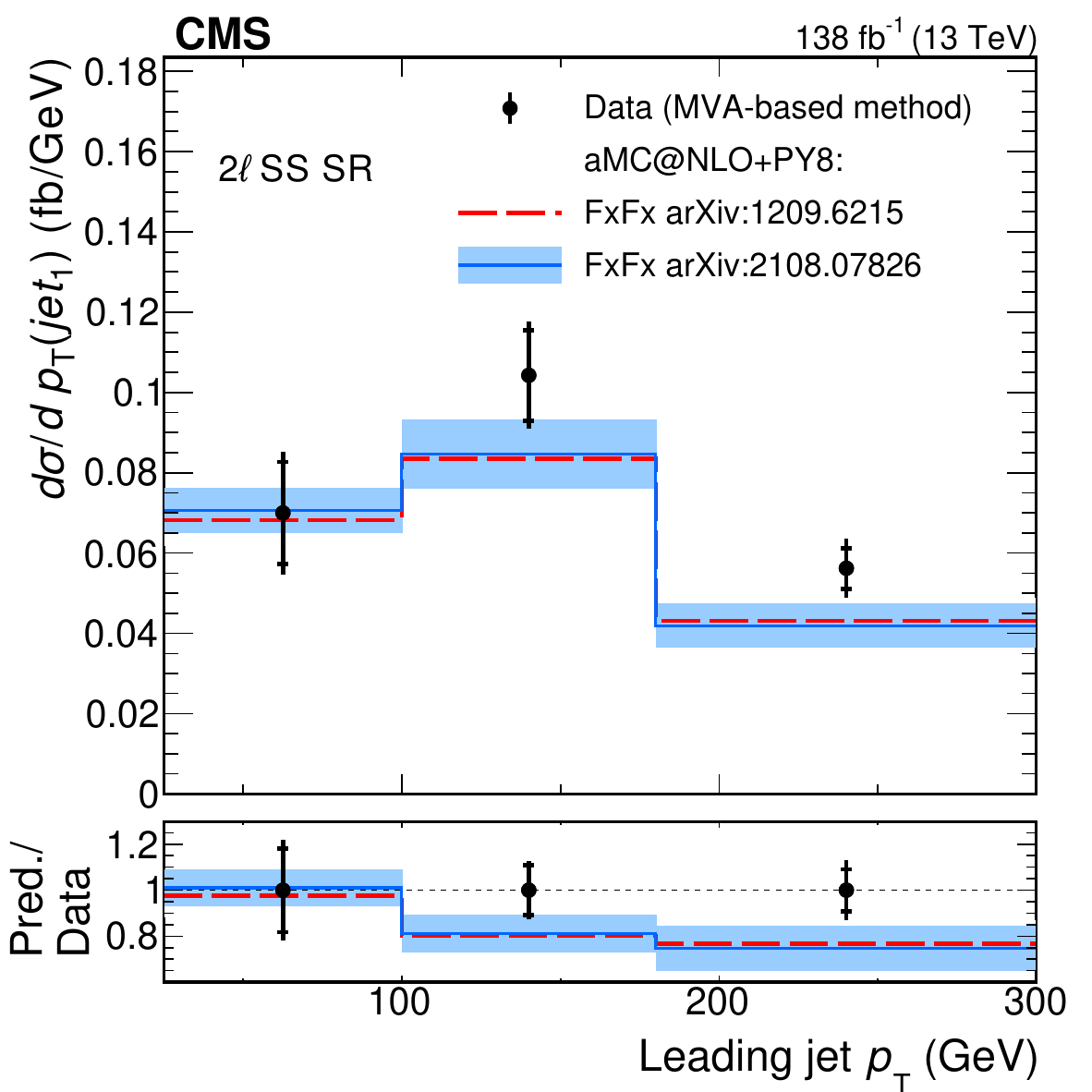}%
\hfill%
\includegraphics[width=0.475\textwidth]{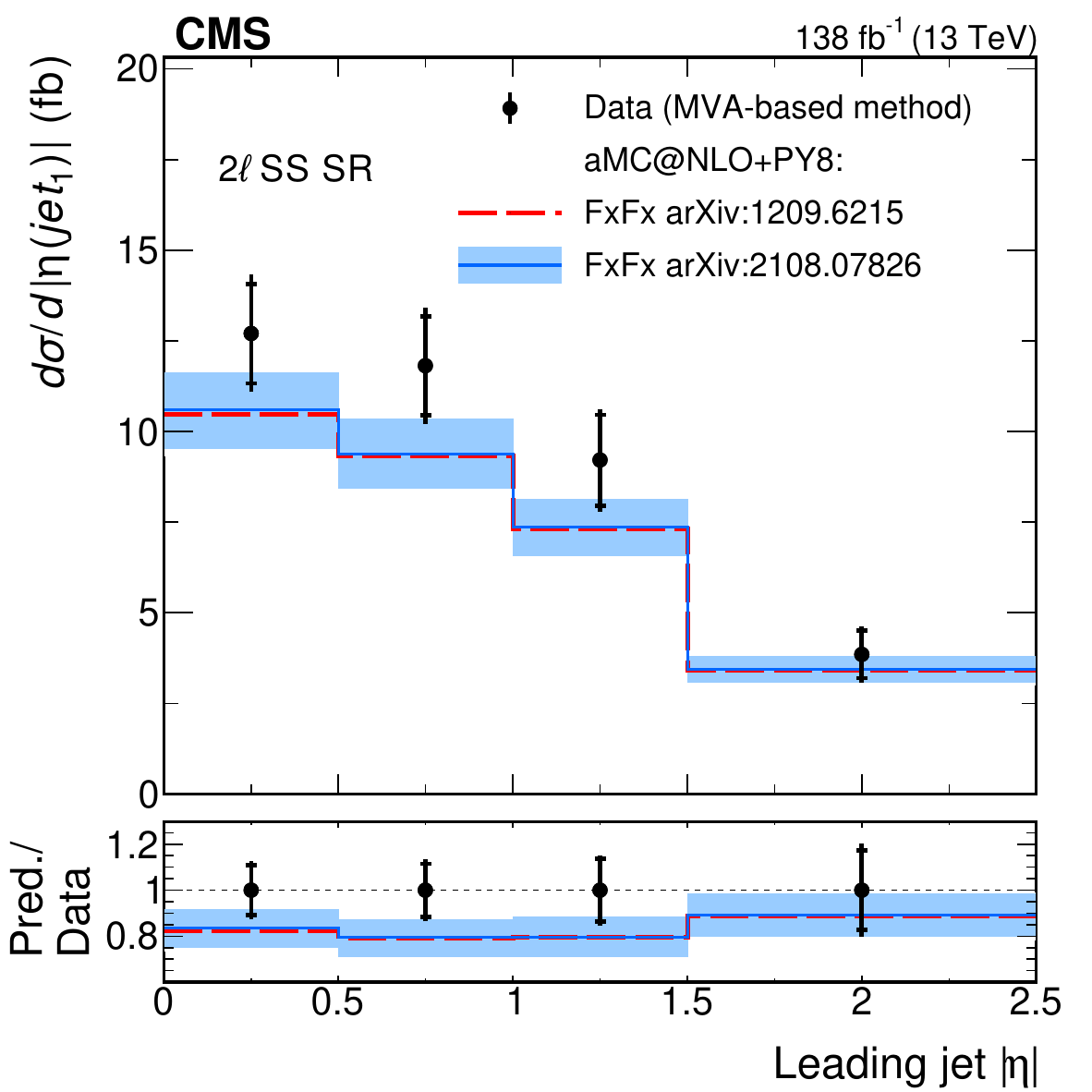} \\[\cmsTabSkip]
\includegraphics[width=0.475\textwidth]{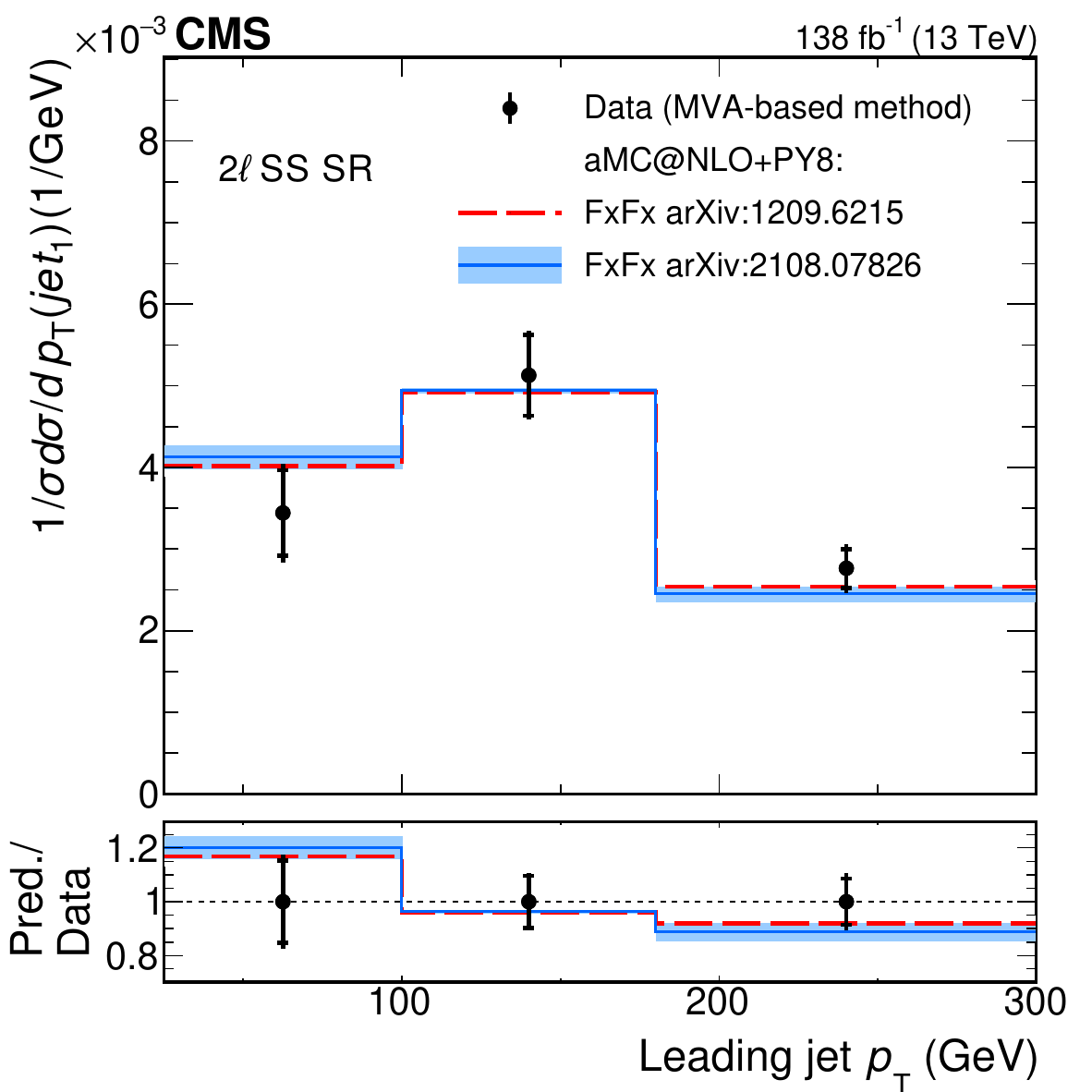}%
\hfill%
\includegraphics[width=0.475\textwidth]{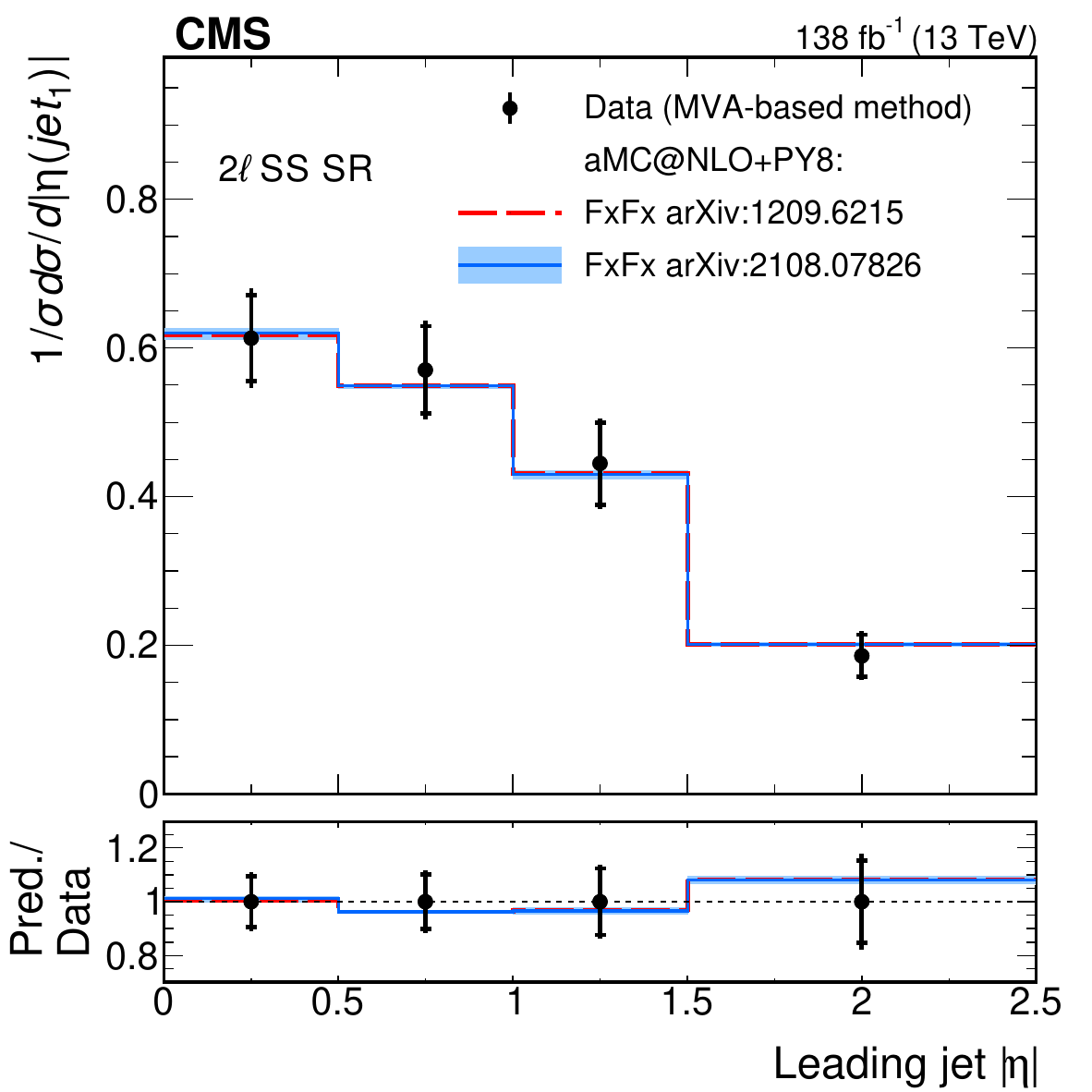}
\caption{%
    Absolute (upper row) and normalized (lower row) differential cross section measured as a function of the leading jet \pt (left) and leading jet \abseta (right), using the MVA-based method.
    \diffextracaption
}
\label{fig:differential_xsec_2}
\end{figure}

\begin{figure}[!pt]
\centering
\includegraphics[width=0.475\textwidth]{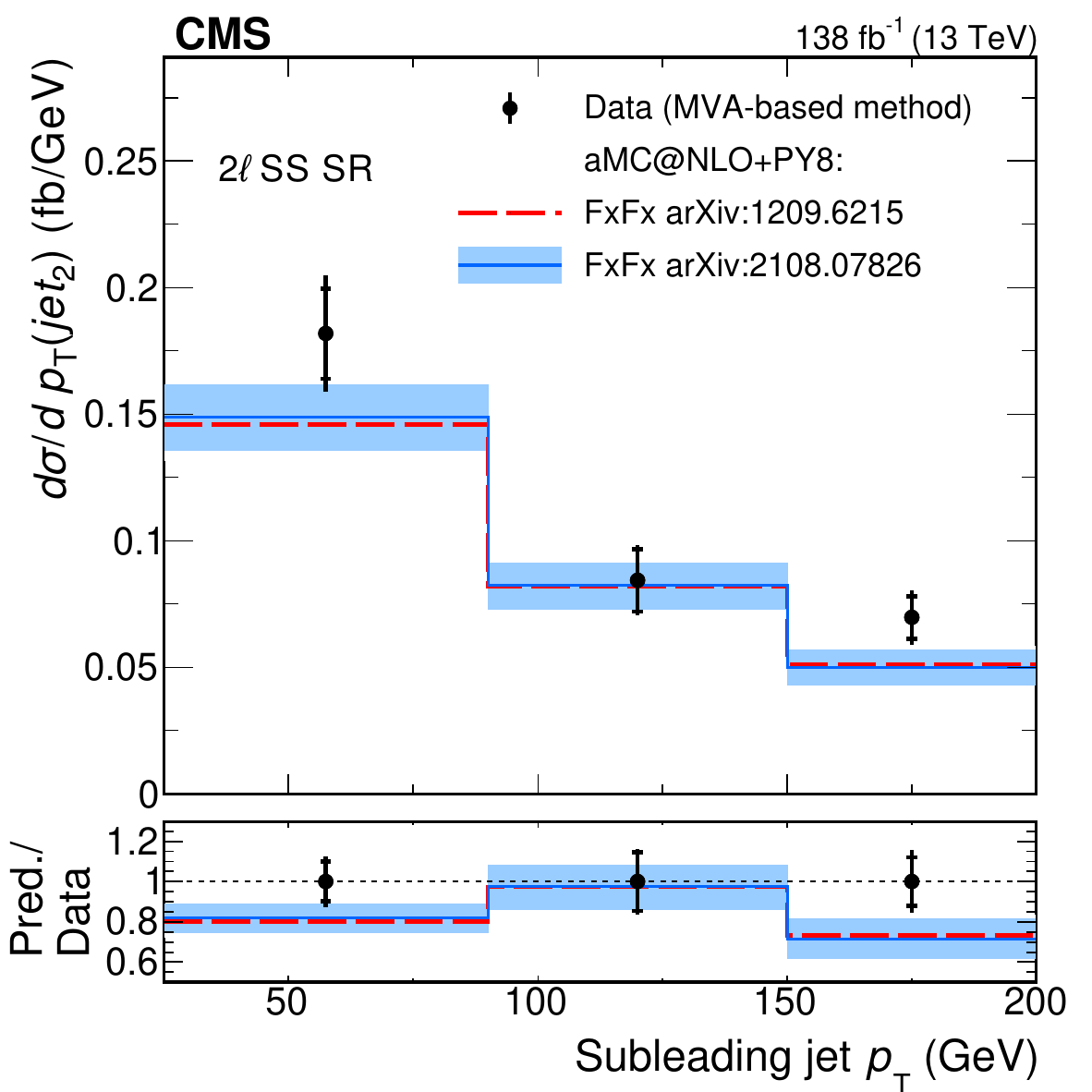}%
\hfill%
\includegraphics[width=0.475\textwidth]{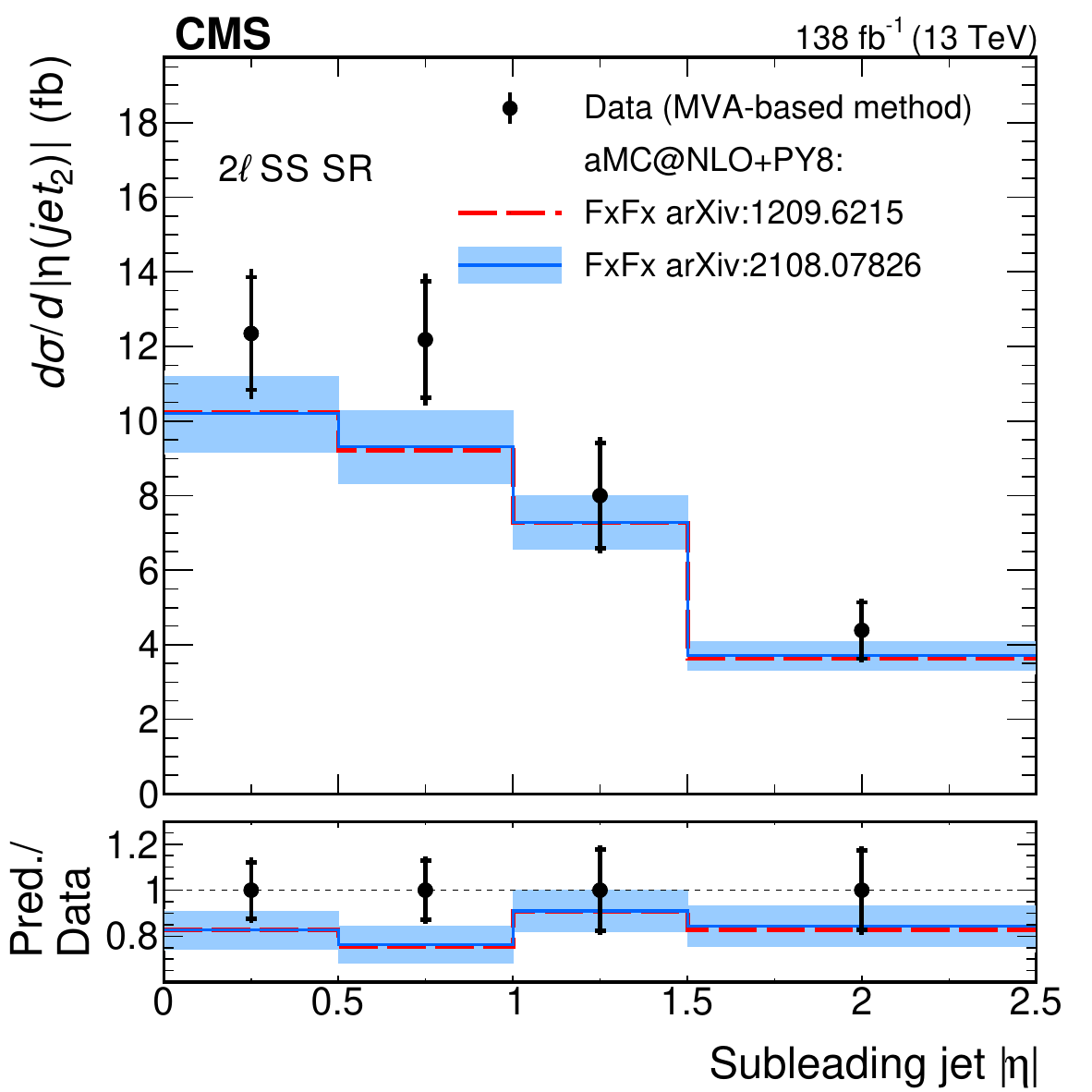} \\[\cmsTabSkip]
\includegraphics[width=0.475\textwidth]{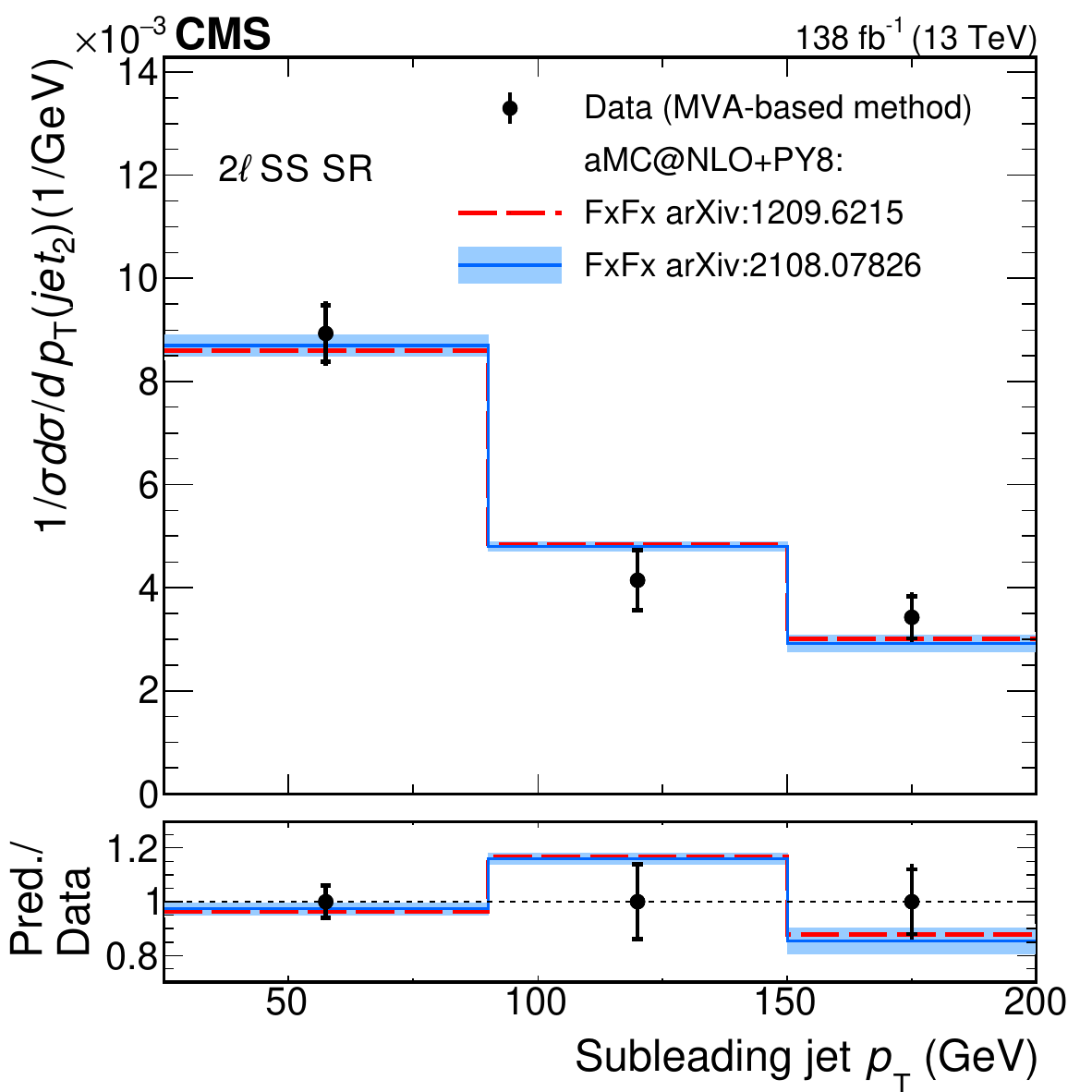}%
\hfill%
\includegraphics[width=0.475\textwidth]{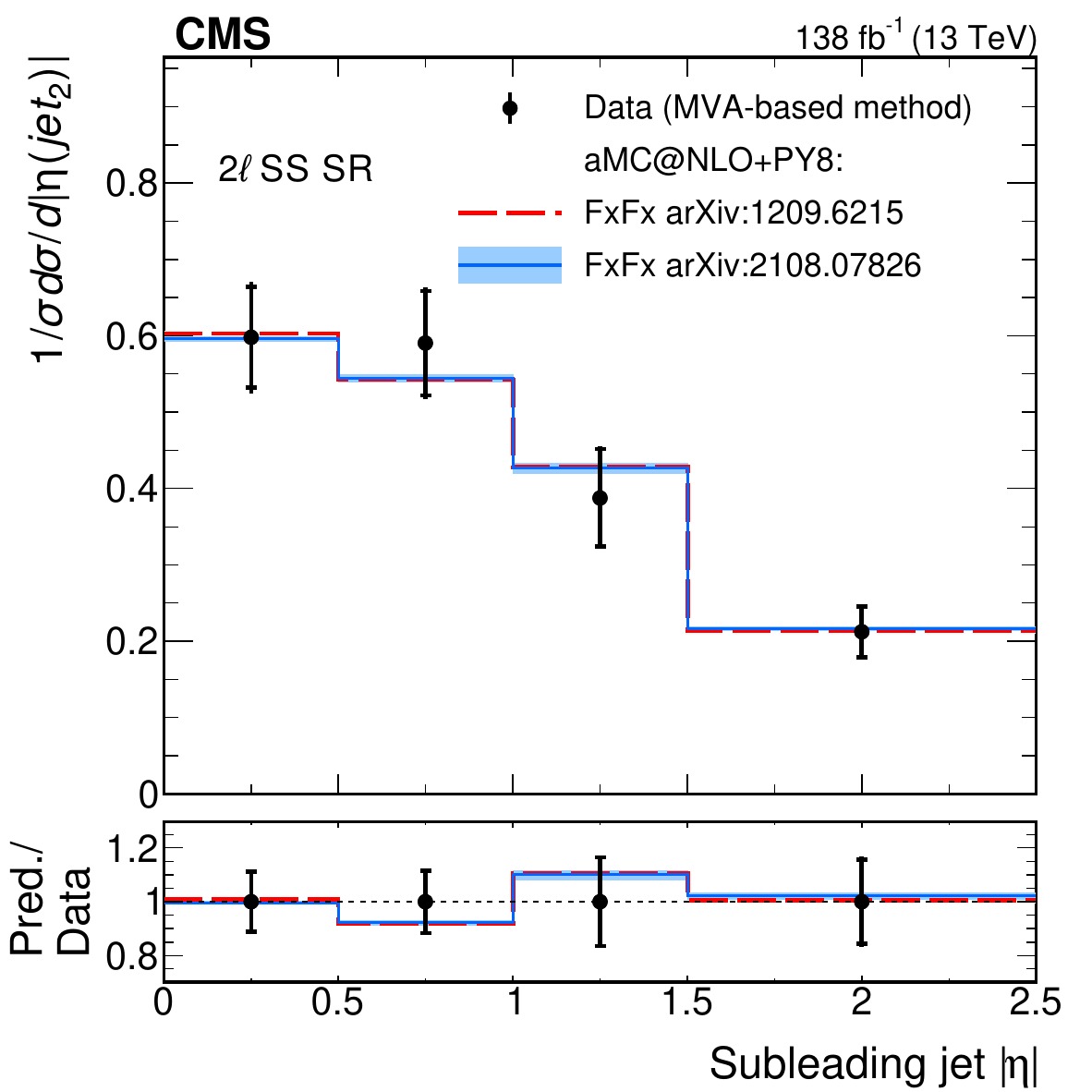}
\caption{%
    Absolute (upper row) and normalized (lower row) differential cross section measured as a function of the subleading jet \pt (left), and subleading jet \abseta (right), using the MVA-based method.
    \diffextracaption
}
\label{fig:differential_xsec_3}
\end{figure}

\begin{figure}[!pt]
\centering
\includegraphics[width=0.475\textwidth]{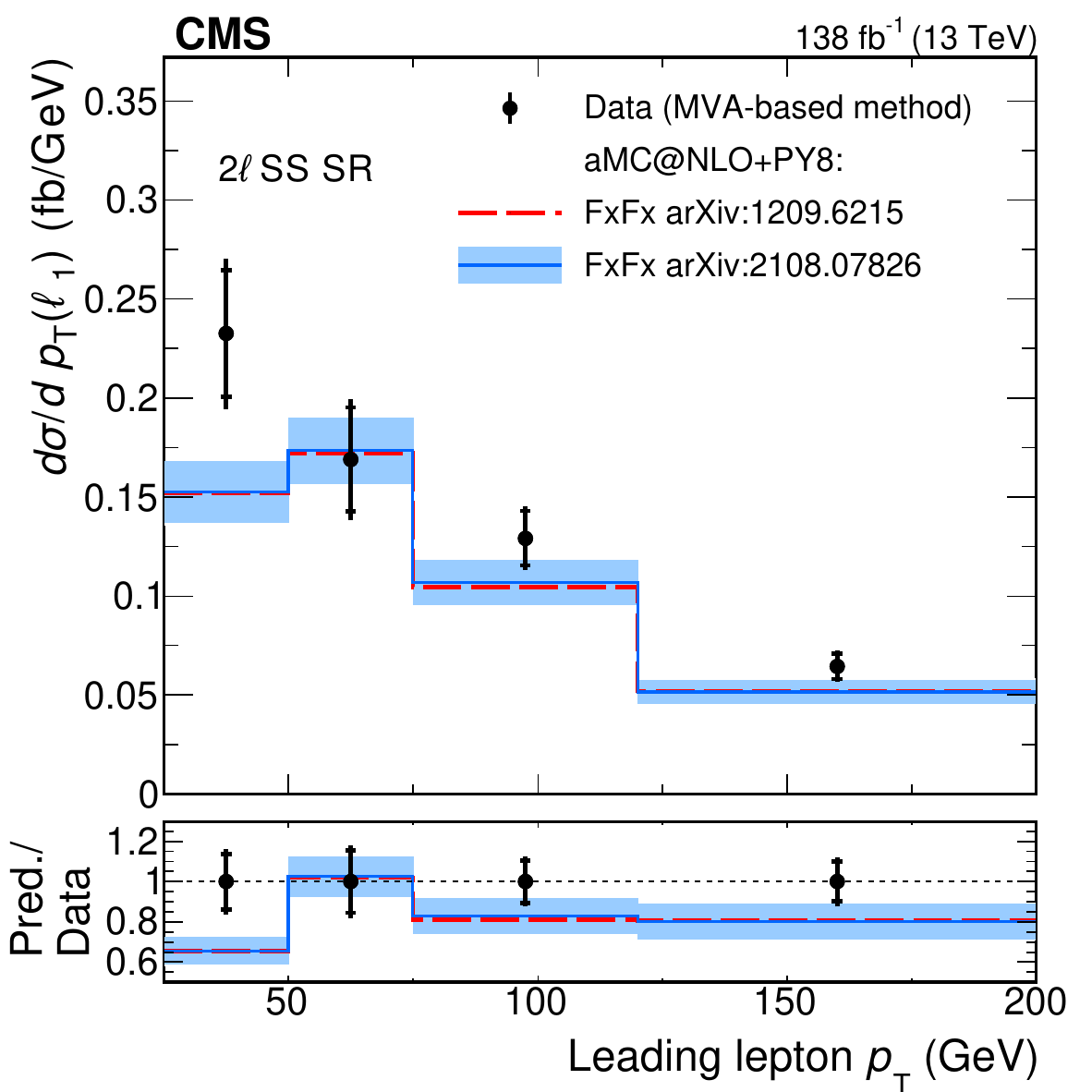}%
\hfill%
\includegraphics[width=0.475\textwidth]{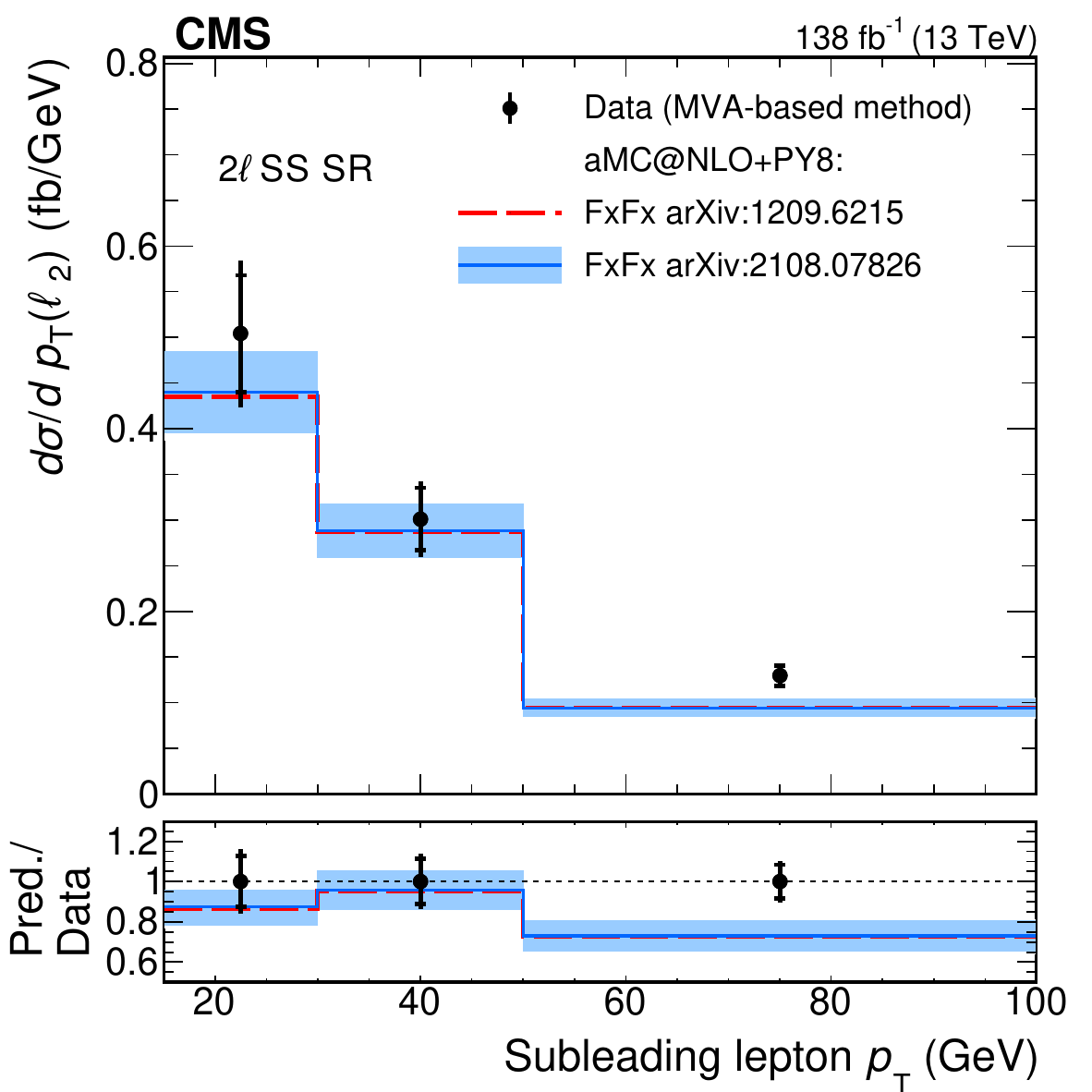} \\[\cmsTabSkip]
\includegraphics[width=0.475\textwidth]{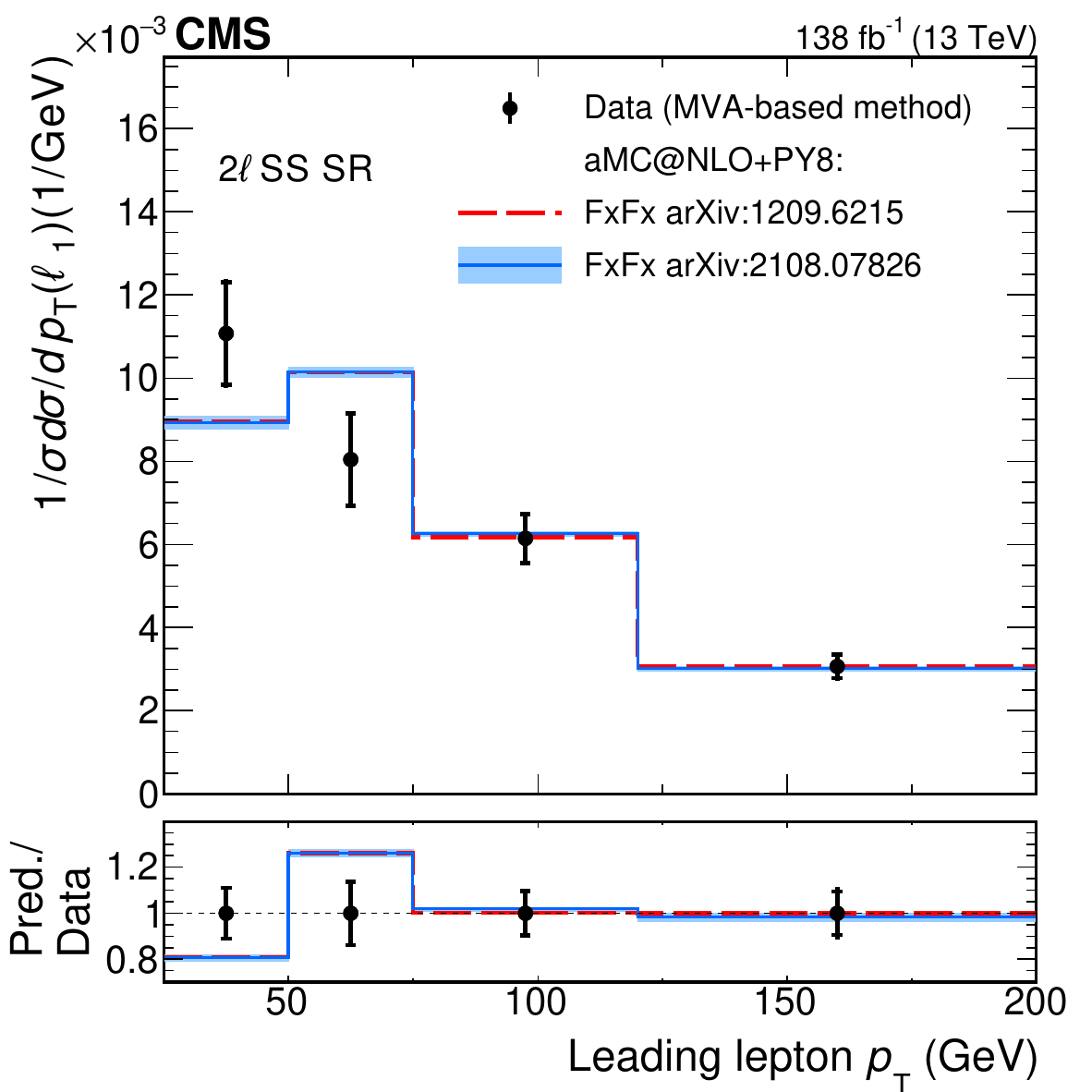}%
\hfill%
\includegraphics[width=0.475\textwidth]{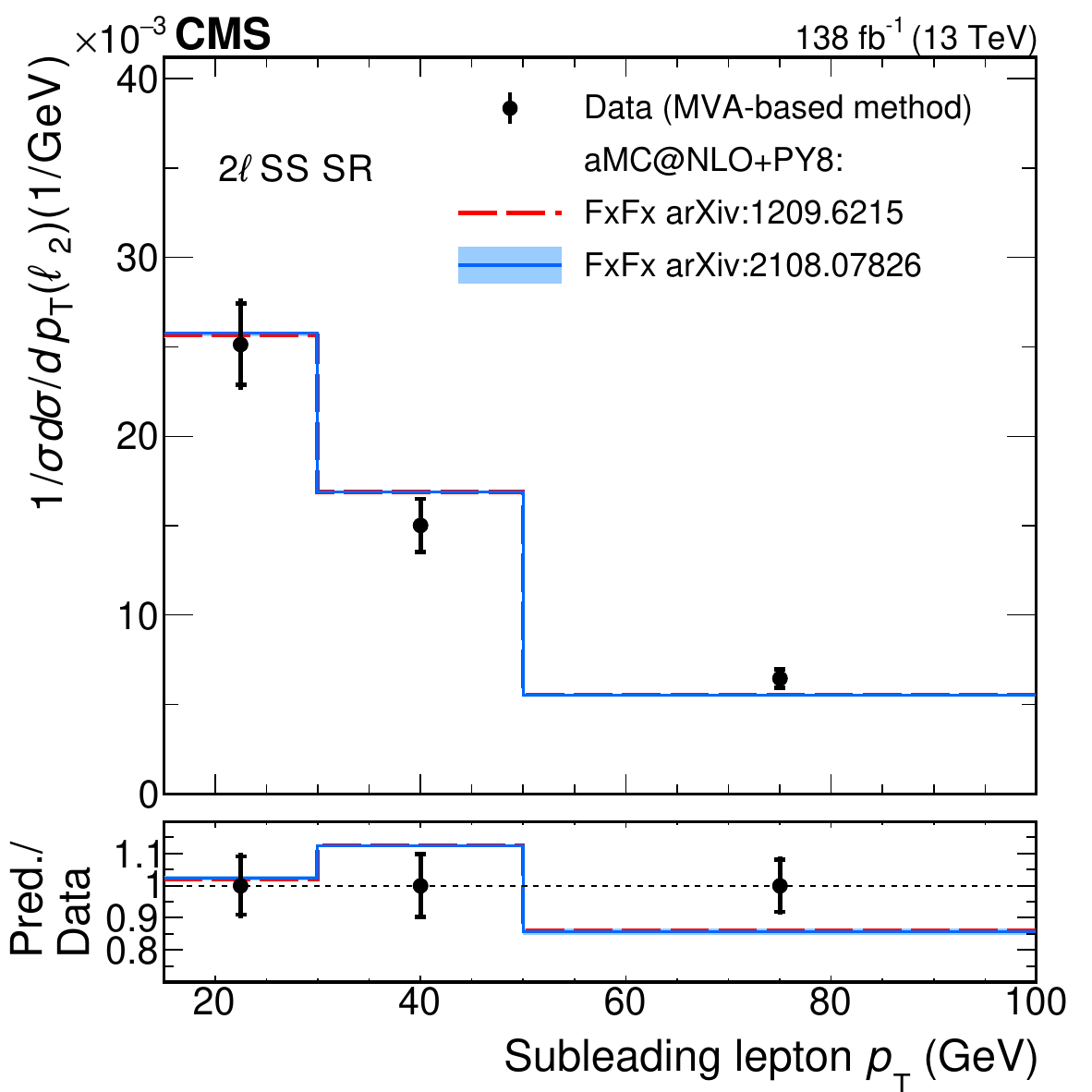}
\caption{%
    Absolute (upper row) and normalized (lower row) differential cross section measured as a function of the leading (left) and subleading (right) lepton \pt, using the MVA-based method.
    \diffextracaption
}
\label{fig:differential_xsec_4}
\end{figure}

\begin{figure}[!pt]
\centering
\includegraphics[width=0.475\textwidth]{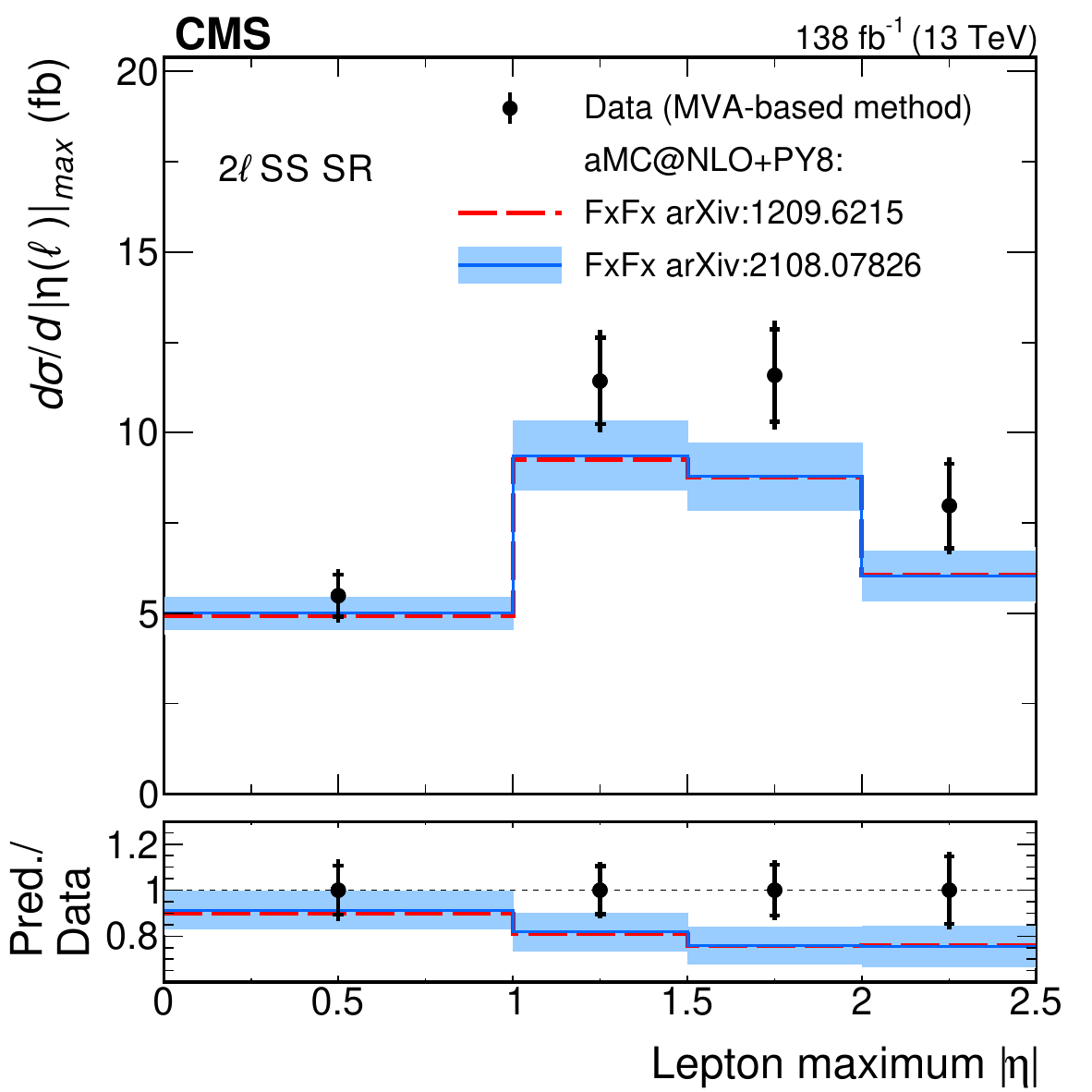}%
\hfill%
\includegraphics[width=0.475\textwidth]{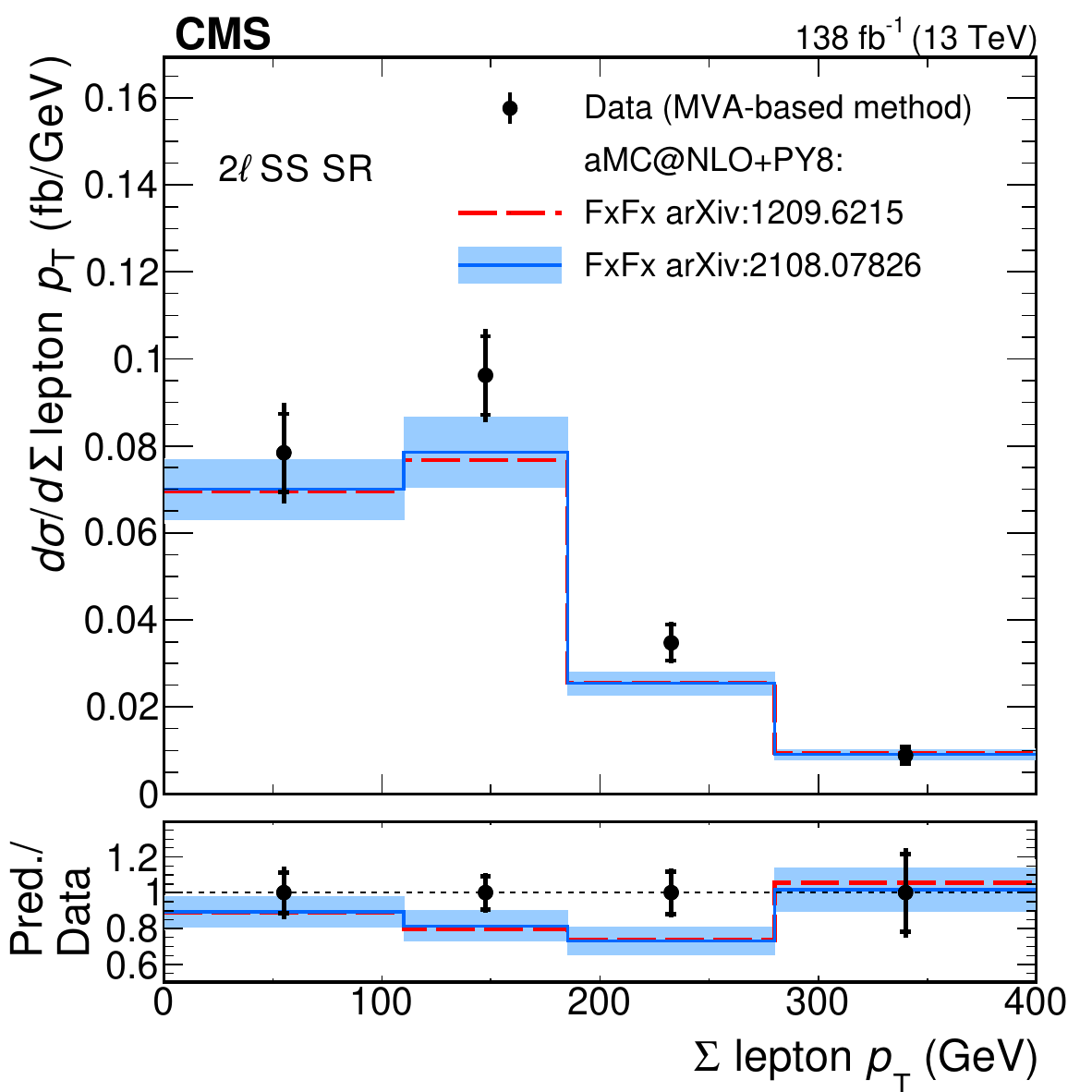} \\[\cmsTabSkip]
\includegraphics[width=0.475\textwidth]{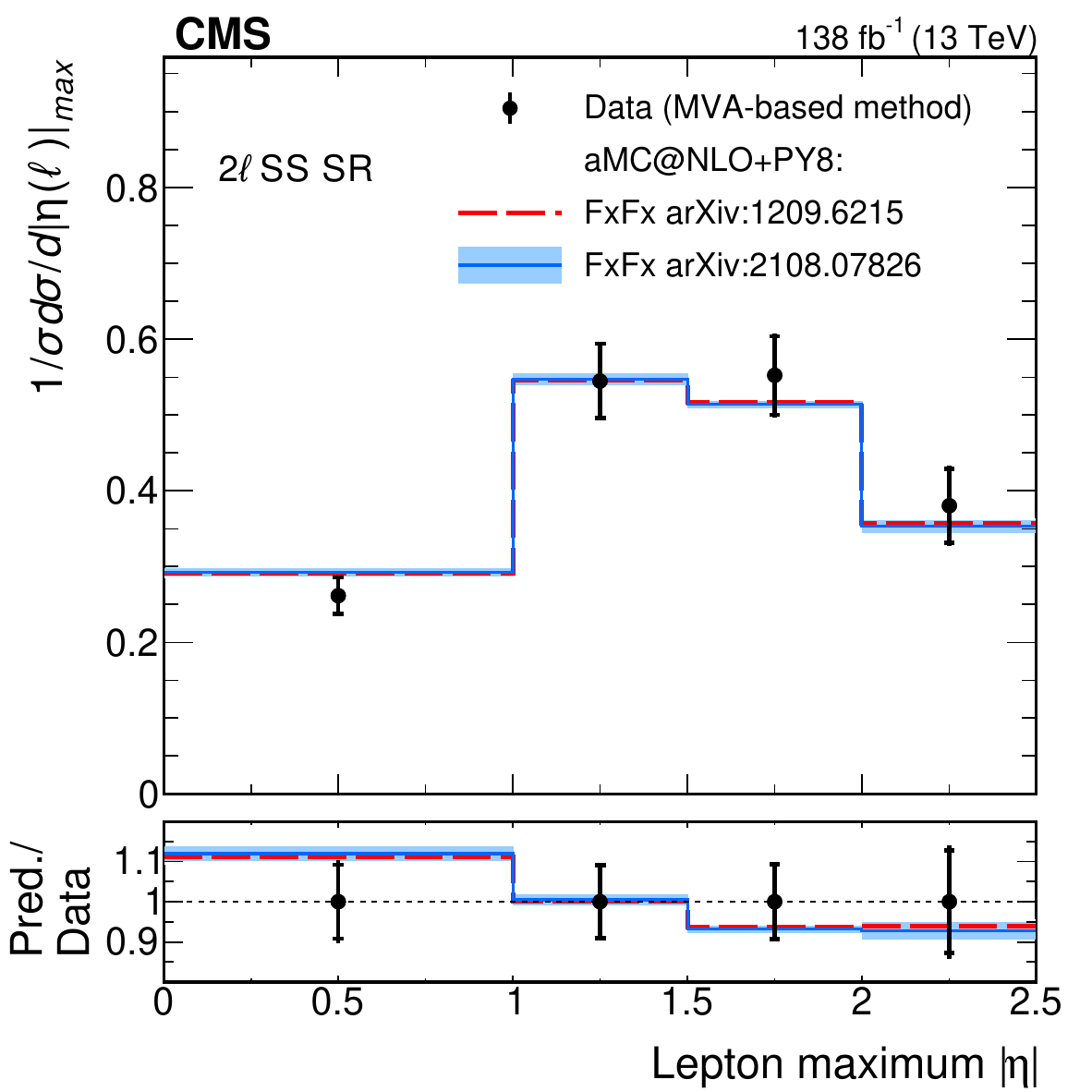}%
\hfill%
\includegraphics[width=0.475\textwidth]{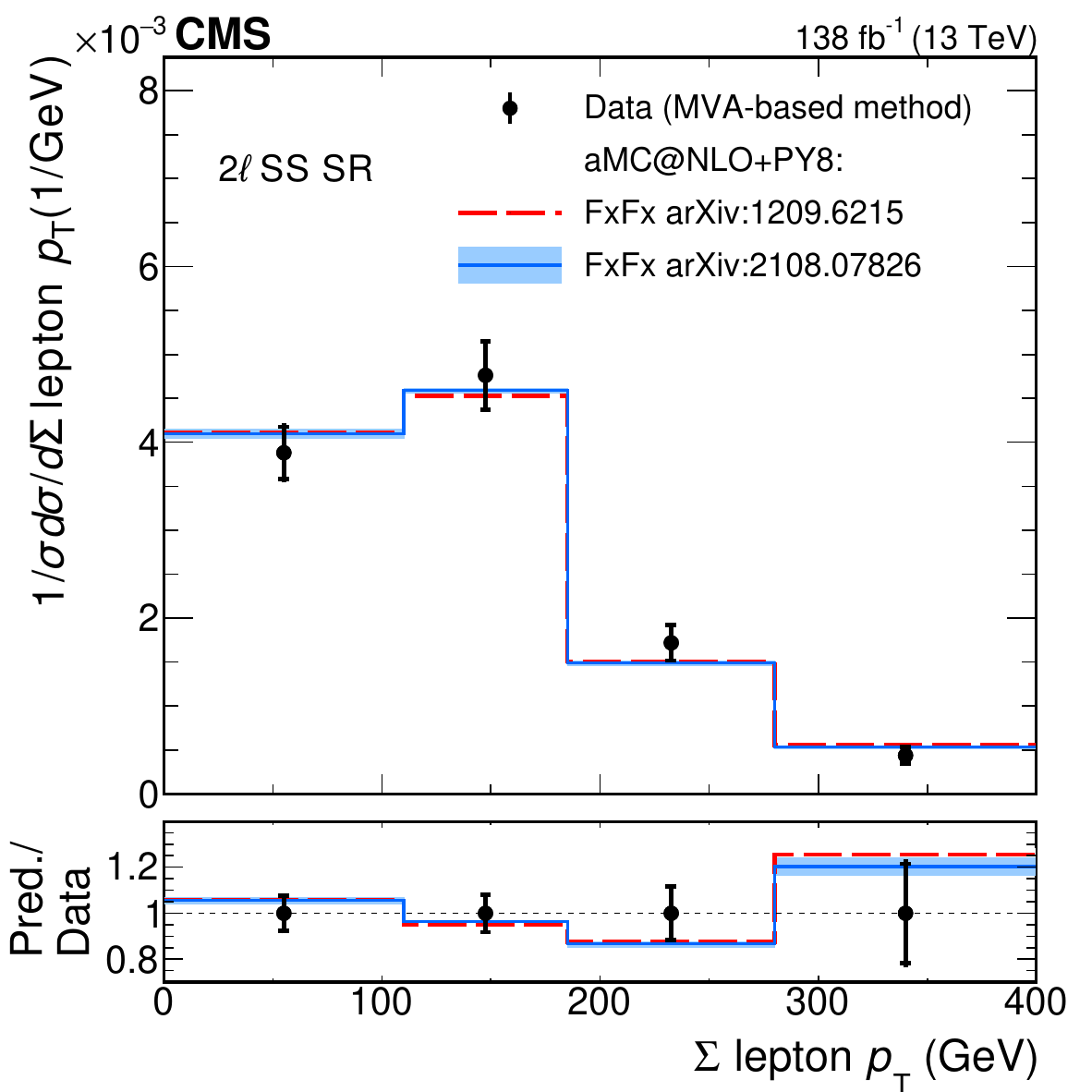}
\caption{%
    Absolute (upper row) and normalized (lower row) differential cross section measured as a function of the maximum \abseta of the selected leptons (left) and the sum of their \pt, using the MVA-based method.
    \diffextracaption
}
\label{fig:differential_xsec_5}
\end{figure}

\begin{figure}[!pt]
\centering
\includegraphics[width=0.475\textwidth]{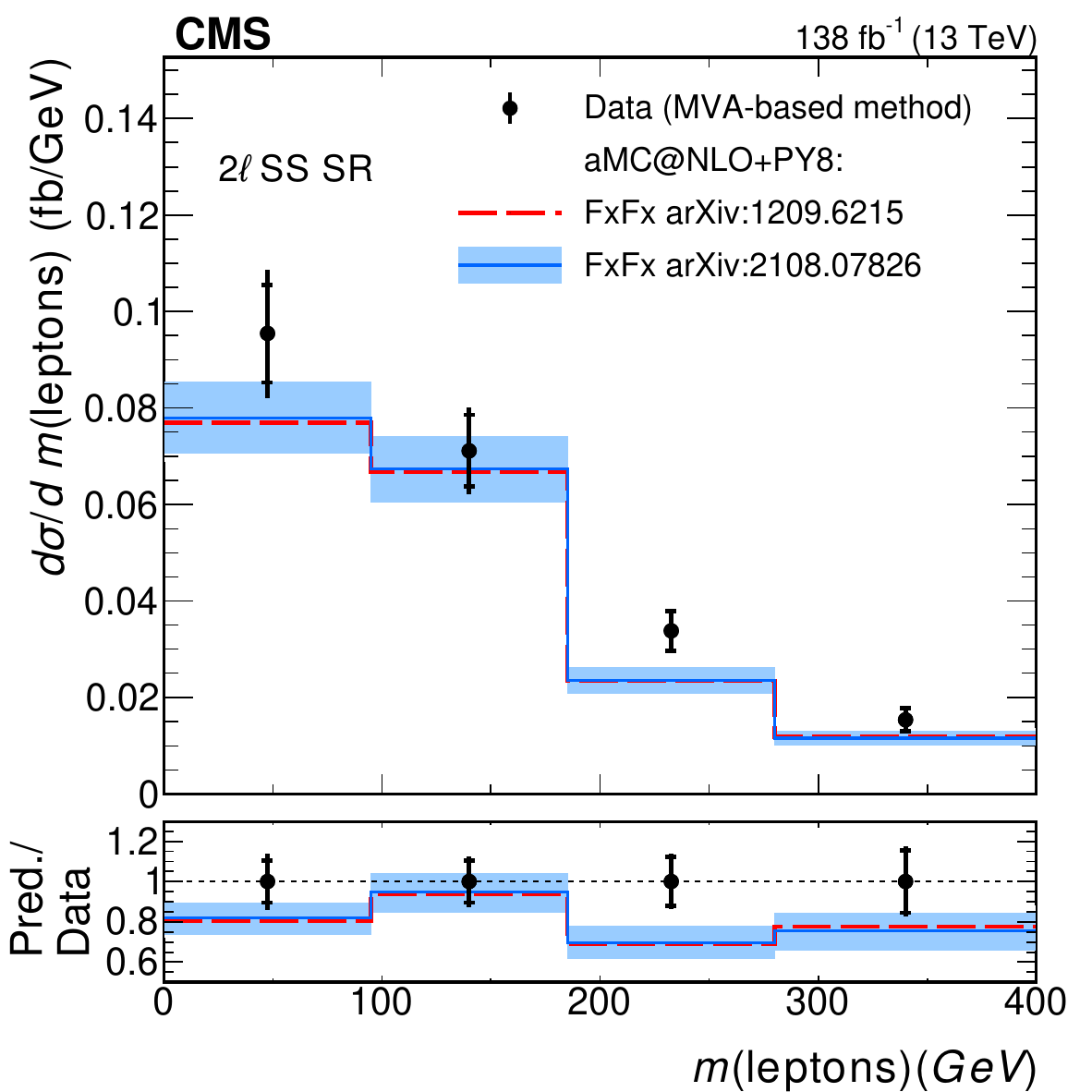}%
\hfill%
\includegraphics[width=0.475\textwidth]{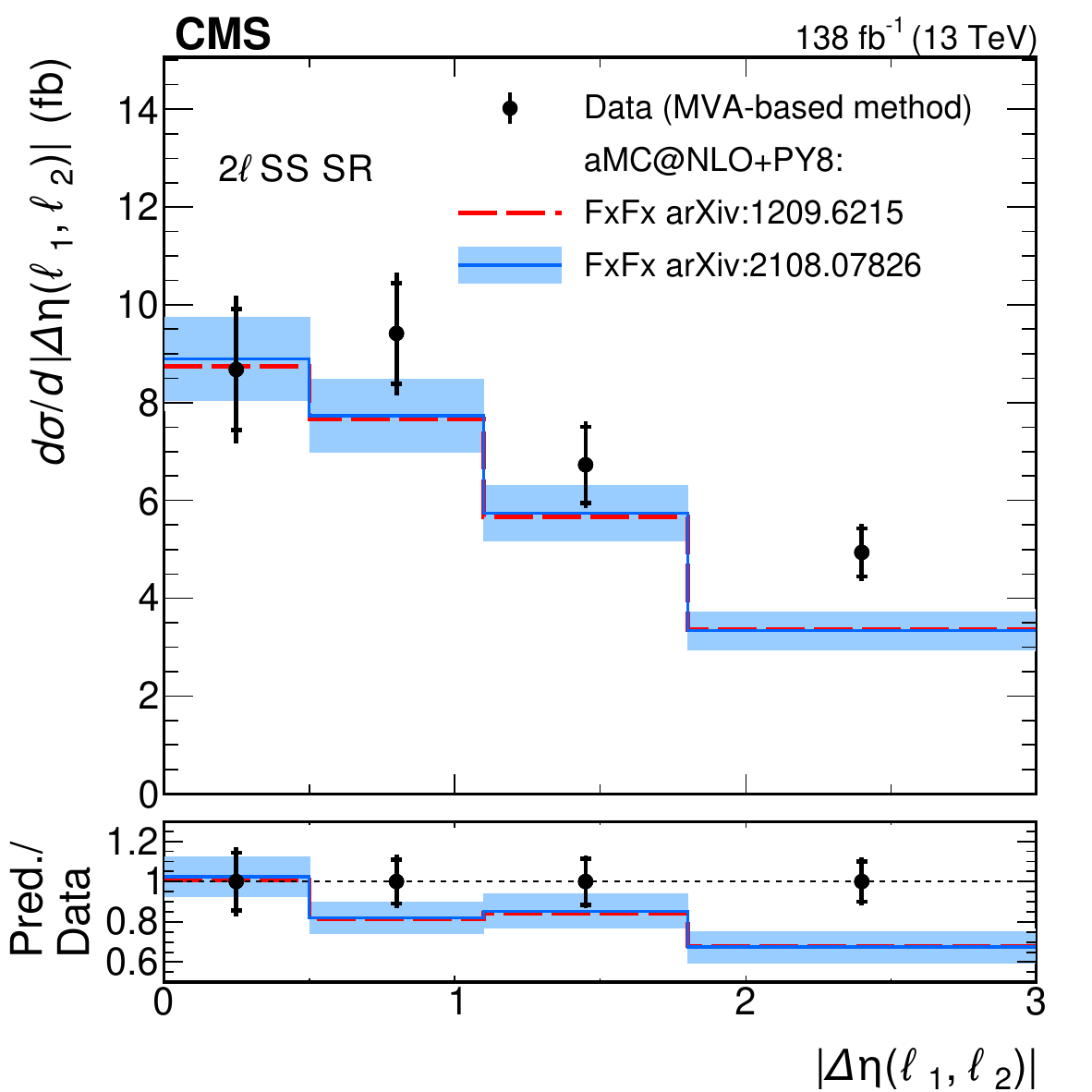} \\[\cmsTabSkip]
\includegraphics[width=0.475\textwidth]{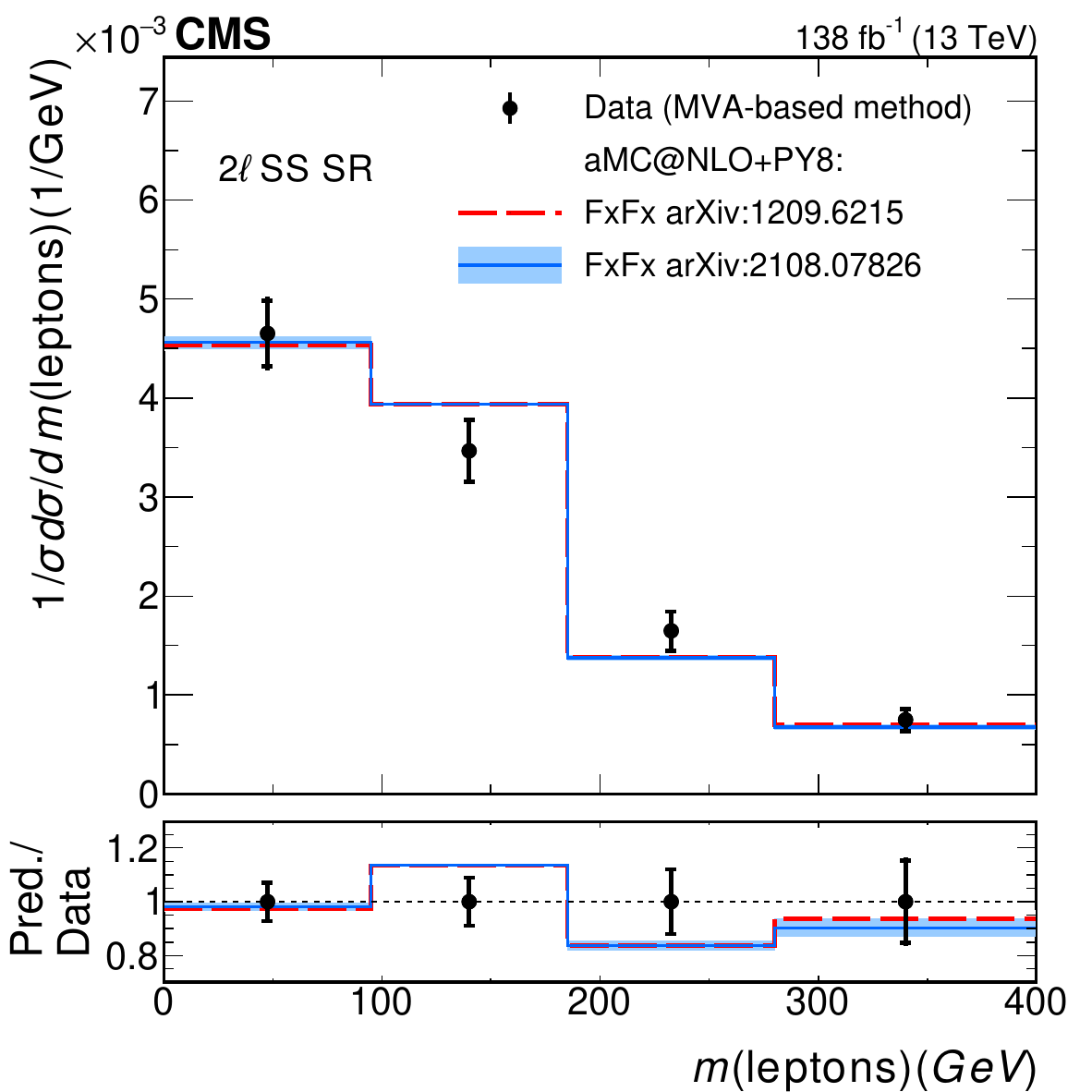}%
\hfill%
\includegraphics[width=0.475\textwidth]{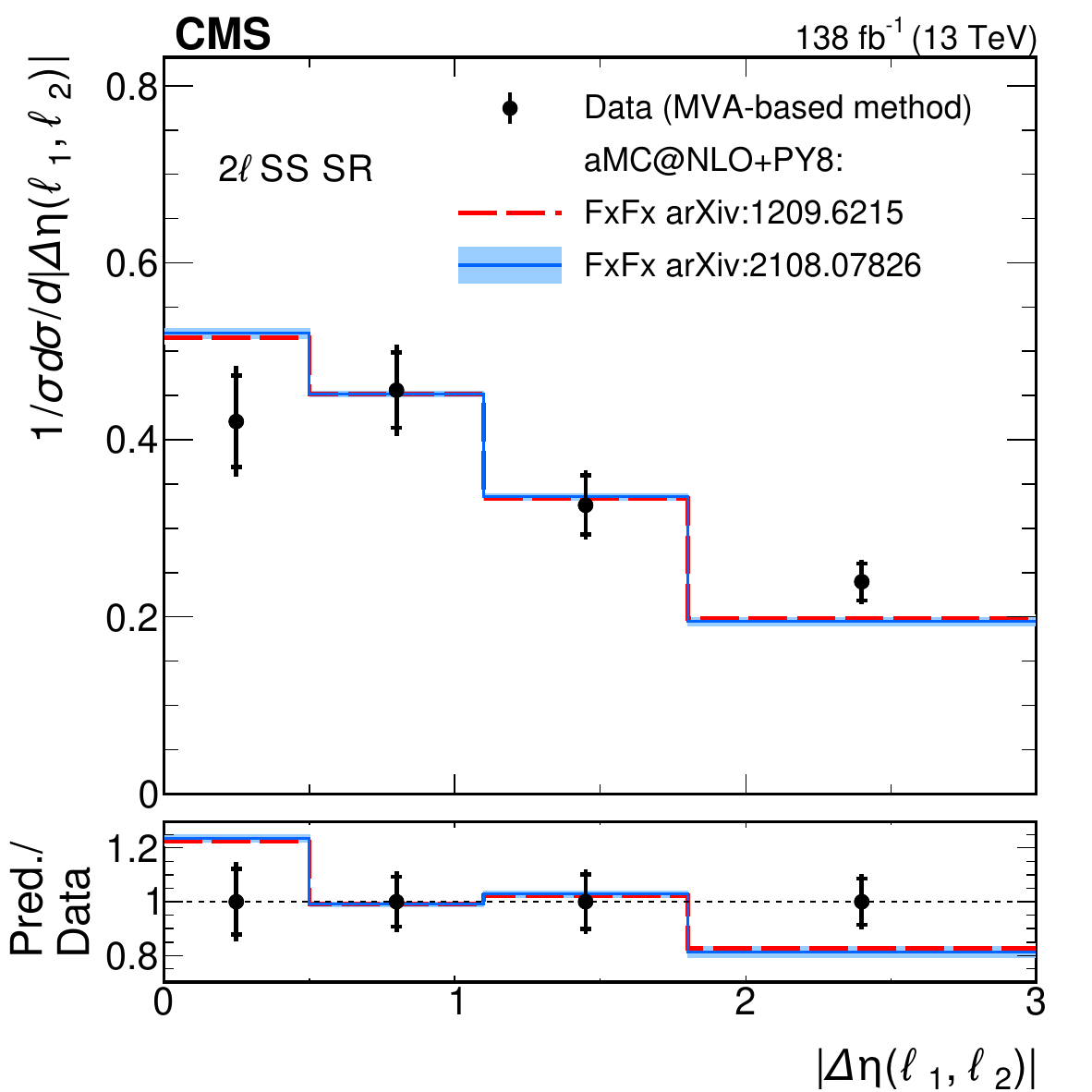}
\caption{%
    Absolute (upper row) and normalized (lower row) differential cross section measured as a function of the invariant mass of the leptons (left) and the \Detaloneltwo, using the MVA-based method.
    \diffextracaption
}
\label{fig:differential_xsec_6}
\end{figure}

\begin{figure}[!th]
\centering
\includegraphics[width=0.475\textwidth]{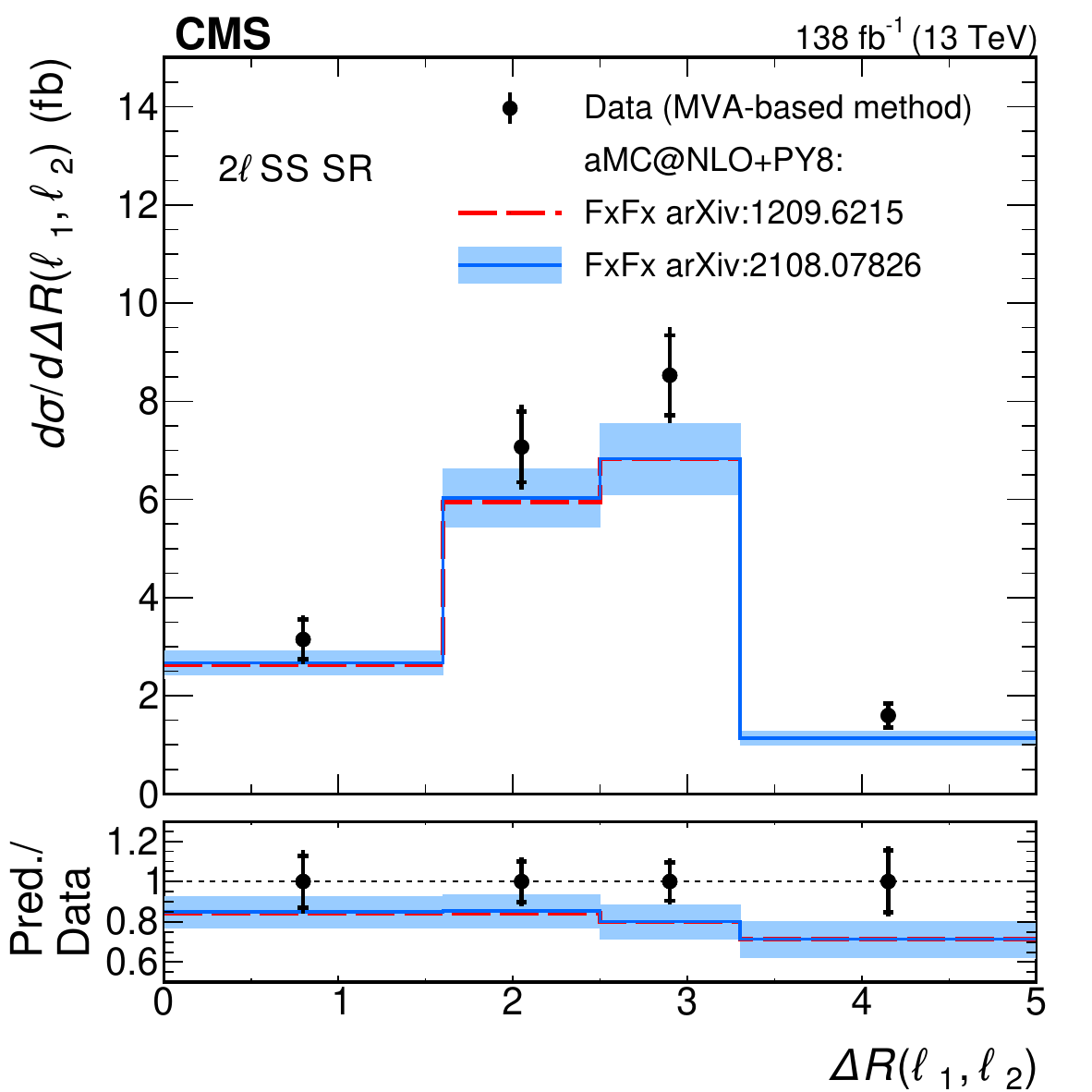}%
\hfill%
\includegraphics[width=0.475\textwidth]{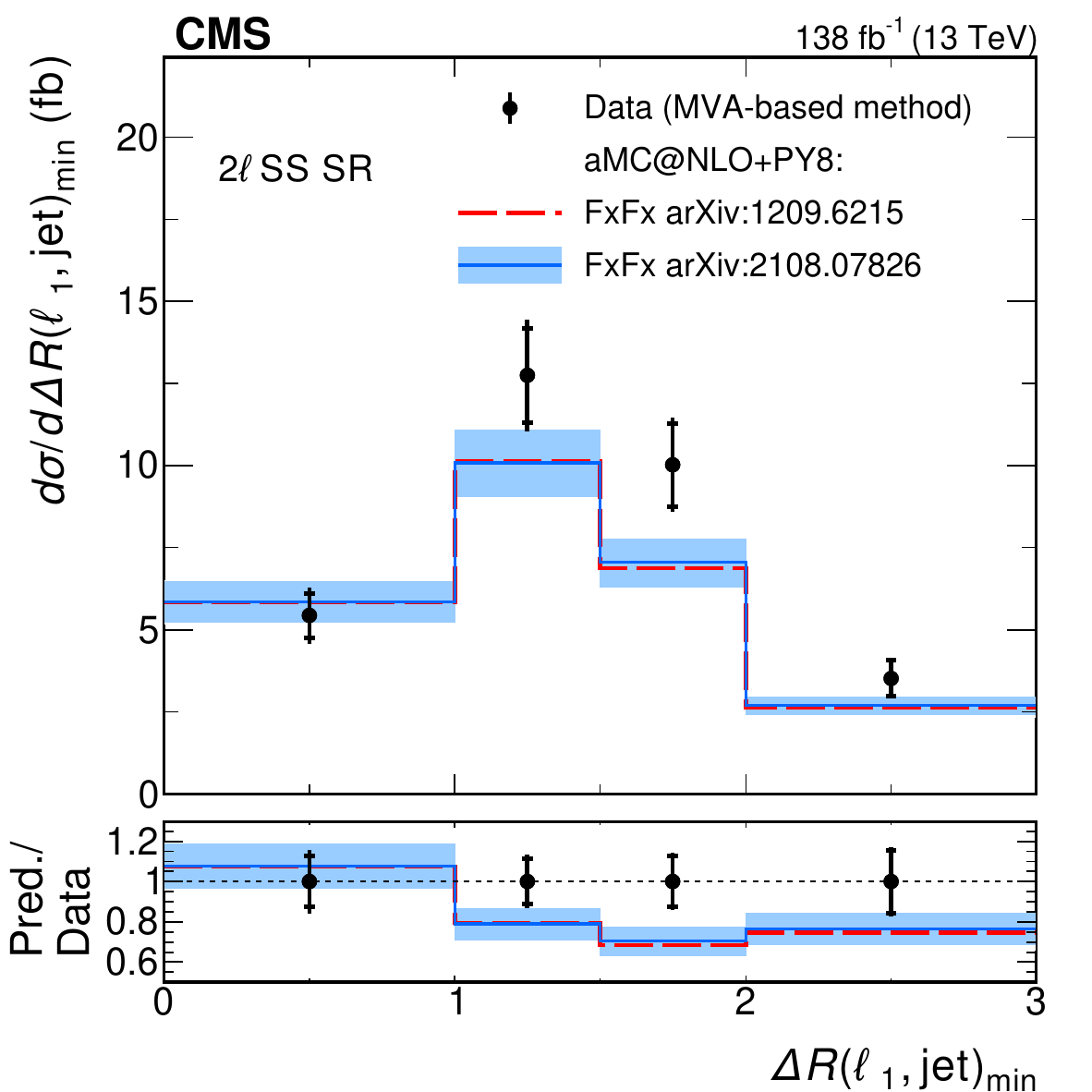} \\[\cmsTabSkip]
\includegraphics[width=0.475\textwidth]{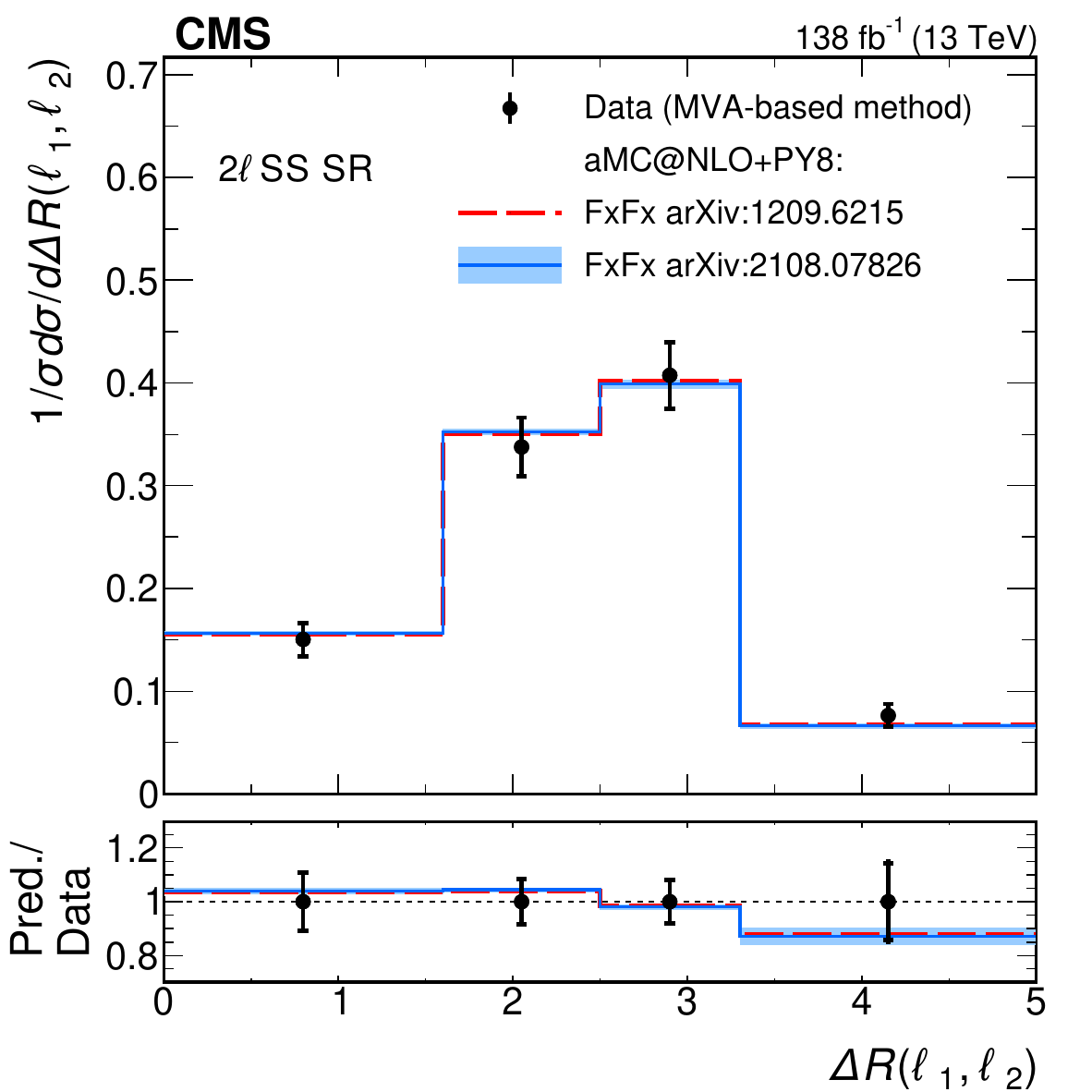}%
\hfill%
\includegraphics[width=0.475\textwidth]{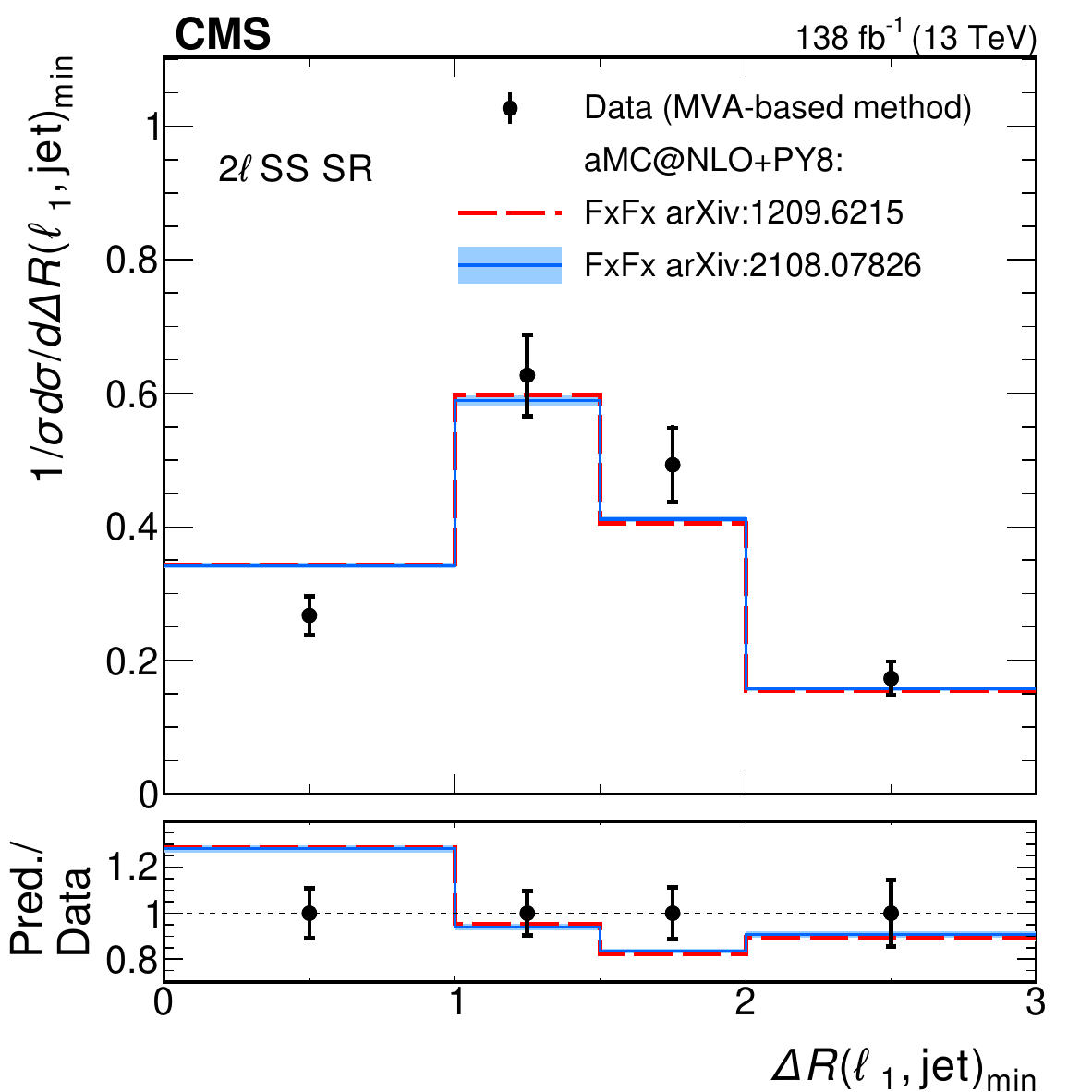}
\caption{%
    Absolute (upper row) and normalized (lower row) differential cross section differential cross section measured as a function of the \DRloneltwo and the \DRlonejetmin, using the MVA-based method.
    \diffextracaption
}
\label{fig:differential_xsec_7}
\end{figure}

To constrain some of the background processes, the number of observed events in the three- and four-lepton CRs are included in the fit as a function of the jet and \PQb-tagged jet multiplicity in the three-lepton region, and the \PQb-tagged jet multiplicity and number of \PZ boson candidates for the four-lepton region.
In addition, the number of events in the CR NP and the CR ChargeMisId are also included, to constrain the corresponding processes.
The differential cross sections measured in the \twoLeptonss signal region as functions of the variables of interest are shown separately for the absolute and normalized differential cross sections in Figs.~\ref{fig:differential_xsec_1}--\ref{fig:differential_xsec_7}.

In addition to the differential measurements, we also measured the inclusive cross section using the inclusive BDT output distribution in the \twoLeptonss SR, shown in Fig.~\ref{fig:signalregion:dilepton_2}, the inclusive BDT output distribution in the \threeLepton SR, and the same CRs as in the differential measurement.
We observe a cross section of $\sigma_{\ttW}^{\text{observed}}=938\pm45\stat\pm52\syst\unit{fb}$.
This result is well in agreement with the previous measurements by both ATLAS~\cite{ATLAS:2024moy} and CMS~\cite{CMS:2020mpn,CMS:2022tkv}, finding a cross section that is in slight tension with the prediction of $745\pm52\unit{fb}$ given by the NNLO calculations~\cite{Buonocore:2023ljm}.
This discrepancy in the normalization is also observed in the \twoLeptonss unfolded results.
The normalized measured distributions in the \twoLeptonss region are generally however in agreement with respect to any of the reference models described in Section~\ref{sec:samples}, which also agree very well with each other for the merging scale chosen.

Across all measured observables and bins, statistical uncertainties are the dominant source of uncertainty, typically ranging from 10 to 25\%.
Among the systematic uncertainties, the largest contributions arise from background estimation, particularly from the nonprompt sources, which contributes between 3 and 10\% in most bins---reaching its highest impact in regions where nonprompt backgrounds are most significant.
In bins with higher signal purity, the dominant systematic uncertainties arise from the modeling of key processes such as \ttH, \ttZ, and the signal itself, with effects ranging from 0.8 to 1.2\%.
Among experimental sources, the uncertainty in the integrated luminosity contributes most significantly, at the 2--3\% level.
Additional subleading experimental uncertainties, including jet energy calibration, lepton identification, and \PQb tagging, contribute at the 1--3\% level, depending on the sensitivity of the observable to the properties of the corresponding reconstructed objects.

\subsection{Consistency checks}

In order to better understand the robustness of our measurement, extensive stress tests have been performed to evaluate our method in terms of its dependence on the signal model.
Firstly, we employ the counting method used in the \threeLepton region, described in Section~\ref{subsec:3l}, also in the \twoLeptonss region.
This allows us to provide a complementary measurement.
This strategy profits from the high signal purity achieved by the tighter lepton selection, allowing us not to rely on the BDT discrimination while remaining competitive in terms of sensitivity.
This results in a better control over the measured fiducial phase space and, therefore, in a more model-independent measurement.
In this region, events are categorized at reconstruction level according to the sum of the lepton charges and their flavor, resulting in six categories.
In addition, we consider two reconstruction-level bins per generator-level bin of the variable of interest.
The differential cross sections measured by both strategies in the \twoLeptonss signal region as functions of the variables of interest are shown separately for the absolute and normalized differential cross sections in the appendix in Figs.~\ref{fig:differential_xsectogether_1}--\ref{fig:differential_xsectogether_7}.
We observe an overall good agreement between both methods.
In addition, we have computed the inclusive fiducial cross sections by integrating the different measurements independently for each measurement strategy, showing consistent results within uncertainties between the two methods and among the different variables measured.
The normalization of backgrounds and behavior of other uncertainties are furthermore found to be consistent within uncertainties between the two methods and among the different variables measured.
The agreement between the two different differential measurements and the reference models is quantified with a $\chi^2$ goodness-of-fit tests done for each variable using the full covariance matrix of the result.
These are shown in Table~\ref{tab:gofs}.
The lower values of the normalized $\chi^2$ $p$-values for the measured unfolded jet multiplicity, subleading lepton \pt, \HT, and \DRlonejetmin show slight indications of mismodelling. In addition, the counting method also reports slightly lower values for the leading jet \pt distribution.
These trends are however not yet conclusive.

\begin{table}[!ht]
\centering
\topcaption{%
    The $p$-values from the $\chi^2$ goodness-of-fit tests comparing the absolute (column 1 and 2) and normalized (column 3 and 4) differential cross sections predicted by the new FxFx model~\cite{Frederix:2021agh} with the two measurements in the \twoLeptonss region.
}
\label{tab:gofs}
\renewcommand{\arraystretch}{1.1}
\begin{tabular}{lllll}
    & \multicolumn{4}{c}{$p$-value from $\chi^2$ goodness-of-fit} \\
    & \multicolumn{2}{c}{Absolute} & \multicolumn{2}{c}{Normalized}  \\
    Variable               & MVA-based & Counting & MVA-based & Counting \\  \hline
    \Detaloneltwo          & 0.10      & 0.38     & 0.24      & 0.81     \\
    \DRloneltwo            & 0.40      & 0.60     & 0.84      & 0.92     \\
    \DRlonejetmin          & 0.08      & 0.08     & 0.11      & 0.20     \\
    \HT                    & 0.05      & $<$0.01  & 0.05      & $<$0.01  \\
    Leading jet \abseta    & 0.40      & 0.37     & 0.92      & 0.32     \\
    Leading jet \pt        & 0.24      & $<$0.01  & 0.49      & $<$0.01  \\
    Subleading jet \abseta & 0.47      & 0.46     & 0.91      & 0.31     \\
    Subleading jet \pt     & 0.28      & $<$0.01  & 0.47      & 0.01     \\
    Leading lepton \pt     & 0.15      & 0.19     & 0.31      & 0.49     \\
    Subleading lepton \pt  & 0.11      & 0.29     & 0.12      & 0.89     \\
    Lepton maximum \abseta & 0.35      & 0.35     & 0.78      & 0.46     \\
    $\sum$ lepton \pt      & 0.28      & 0.06     & 0.39      & 0.16     \\
    \mleptons              & 0.20      & 0.19     & 0.31      & 0.20     \\
    Number of jets         & 0.18      & 0.46     & 0.31      & 0.64     \\
\end{tabular}
\end{table}

As an additional stress test, pseudodata samples were generated by reweighting signal events based on a linear function of each differential variable we want to measure.
Each reweighting function was chosen to induce a linear shape difference of 30--40\% between the first and last bins of the reweighting variable distribution.
The stress test then consisted of trying to capture the reweighted distributions of all other variables based on this linear reweighting function, estimating the sensitivity of the fitting procedure to a mismodelling of this variable.
We also perform another signal injection test in which pseudodata are constructed from signal samples incorporating the effect of a number of EFT operators from the \texttt{dim6top} model.
We considered the effect of 22 different EFT operators, setting values to their associated Wilson coefficients so that the modification they introduce in the particle-level kinematic distributions is comparable to or higher than the analysis sensitivity.
This set-up allows us to study the effect of modifications in multidimensional kinematic distributions of our signal.
The results of these studies show that both methods are very robust to such alternate signal models, with the first test showing no bias in both measurement strategies.
In the EFT-based injection test, most modified distributions could also be recovered without bias.
It is however noted that the counting method is more robust to extreme modifications of more than 100\% in the final bin of jet-energy related variables, such as \HT, but is only biased for models that largely modify the charge asymmetry of \ttW.
Overall, the counting method is identified as the most resilient and should hence be the preferred result for most BSM interpretations that typically involve large deviations from the SM predictions.
As the MVA-based method also shows good robustness and achieves slightly better precision, it is the preferred result only for use-cases where the kinematic distributions of the considered models is close to that of our reference models, such as alternative SM predictions or some BSM scenarios.

\subsection{Signal extraction in the \texorpdfstring{\threeLepton}{3l} region}
\label{subsec:3l}

\begin{figure}[!htp]
\centering
\includegraphics[width=0.4\textwidth]{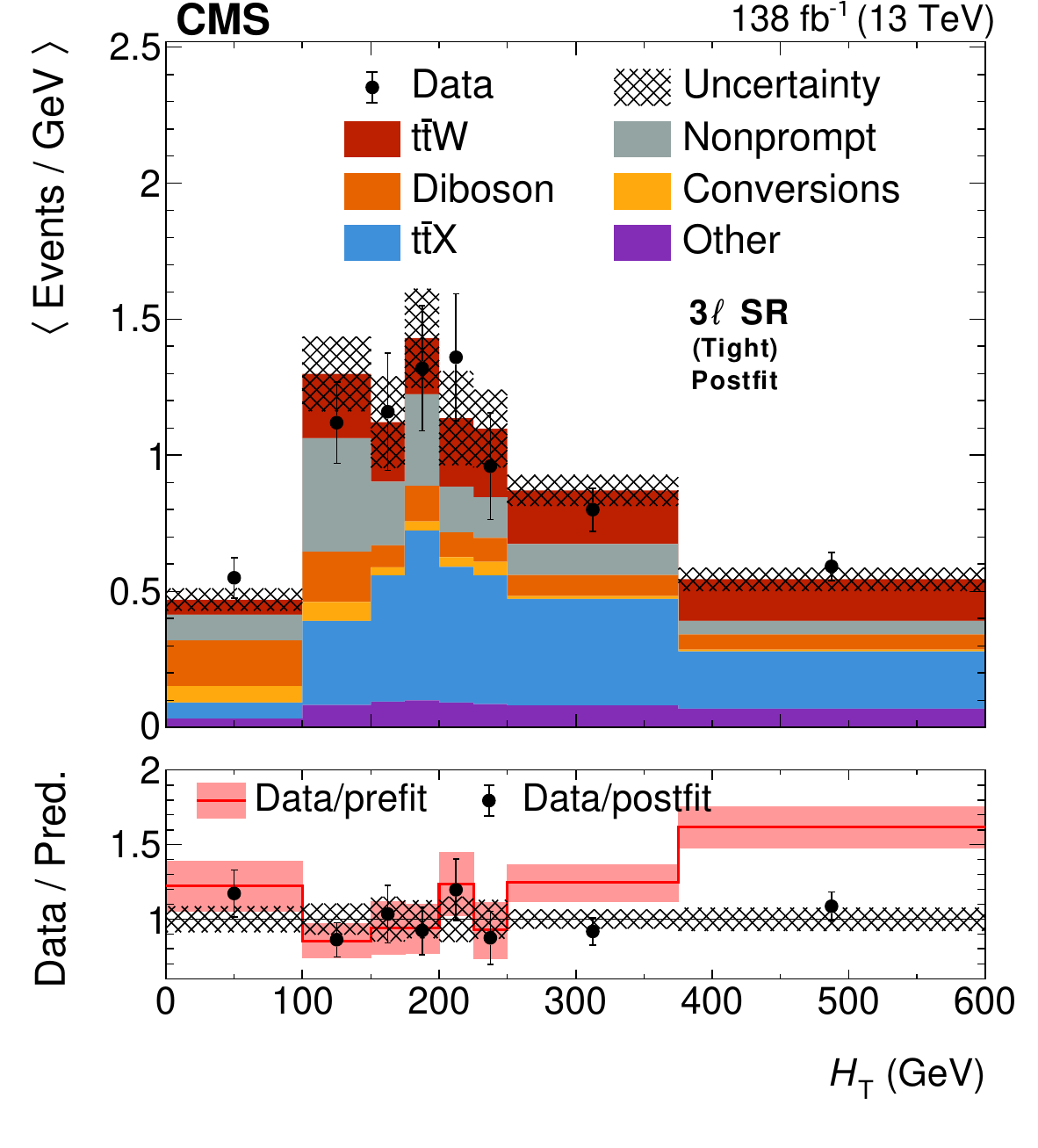}%
\hspace*{0.05\textwidth}%
\includegraphics[width=0.4\textwidth]{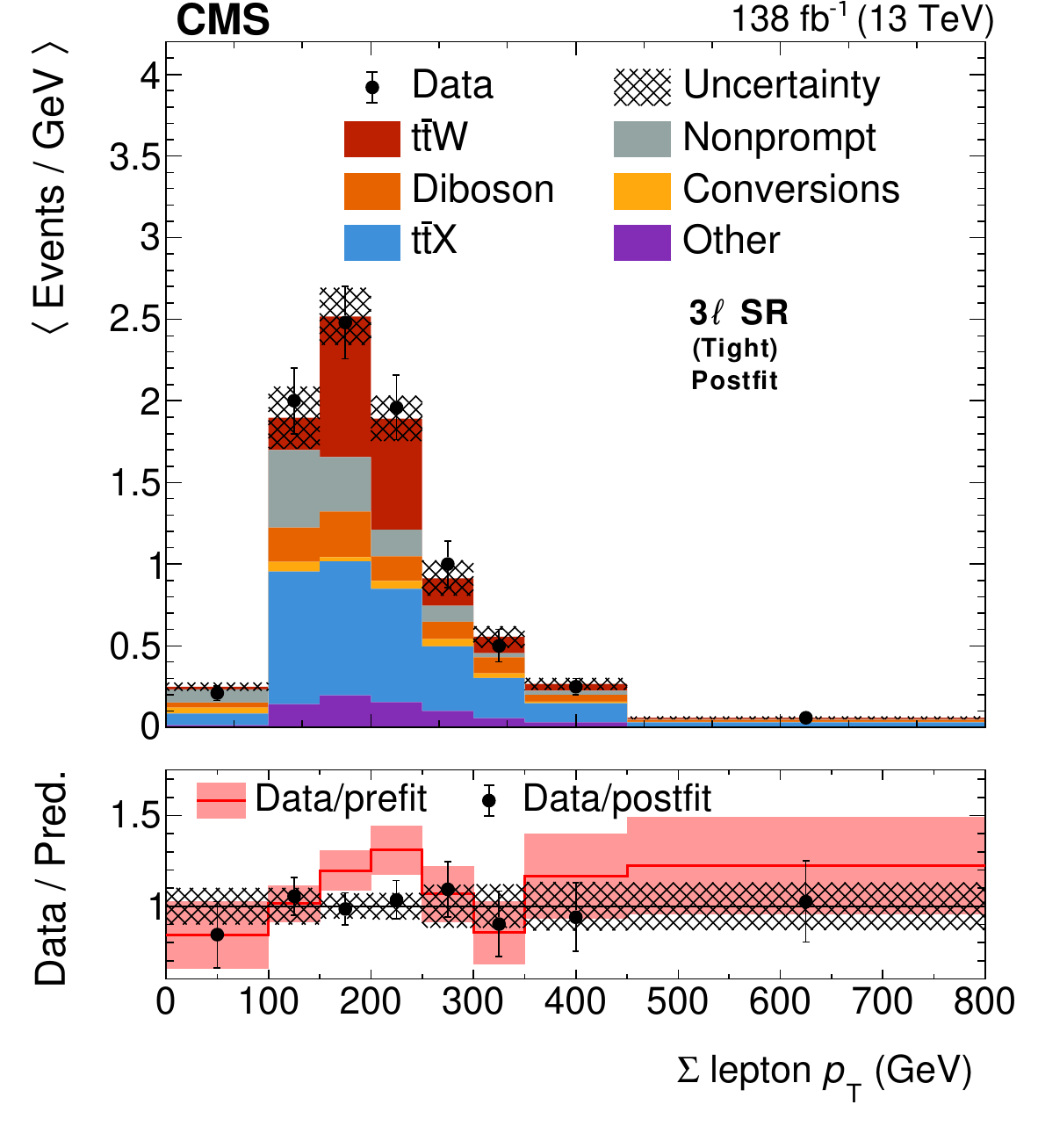}
\caption{%
    Scalar \pt sum of the jets in the event (left) and scalar \pt sum of the leptons in the event (right); in the trilepton signal selection for data (points) and predictions (filled histograms) after the fit to the data for the tight lepton selection.
    \ratiodescription
    \overflow
    \binwidth
}
\label{fig:signalregion:trilepton_tight}
\end{figure}

In the \threeLepton region we do not use the event-level MVA as the ML algorithm does not provide a better separability in this region while the signal purity diminishes significantly.
Instead, we select events in which the leptons pass the tight selection criteria described in Section~\ref{sec:objects_events} to ensure a high-purity selection while relying less on the discrimination between signal and background based on event-level kinematic variables.
The \HT and $\sum$ lepton \pt distributions in the \threeLepton region are shown in Fig.~\ref{fig:signalregion:trilepton_tight}.
To further increase the analysis sensitivity, we categorize events at reconstruction level in four categories according to the flavor of the leptons.
To constrain some of the background processes, the number of observed events in the three- and four-lepton CRs are included in the fit as a function of the same variables as in the \twoLeptonss region fitting strategy.
The VR NP is not included in the fit to guarantee the nonprompt background will not be calibrated in the measurement phase space.
The ChargeMisId VR is also not included as the charge-misidentification background is not relevant in the \threeLepton region.

\begin{figure}[!htp]
\centering
\includegraphics[width=0.475\textwidth]{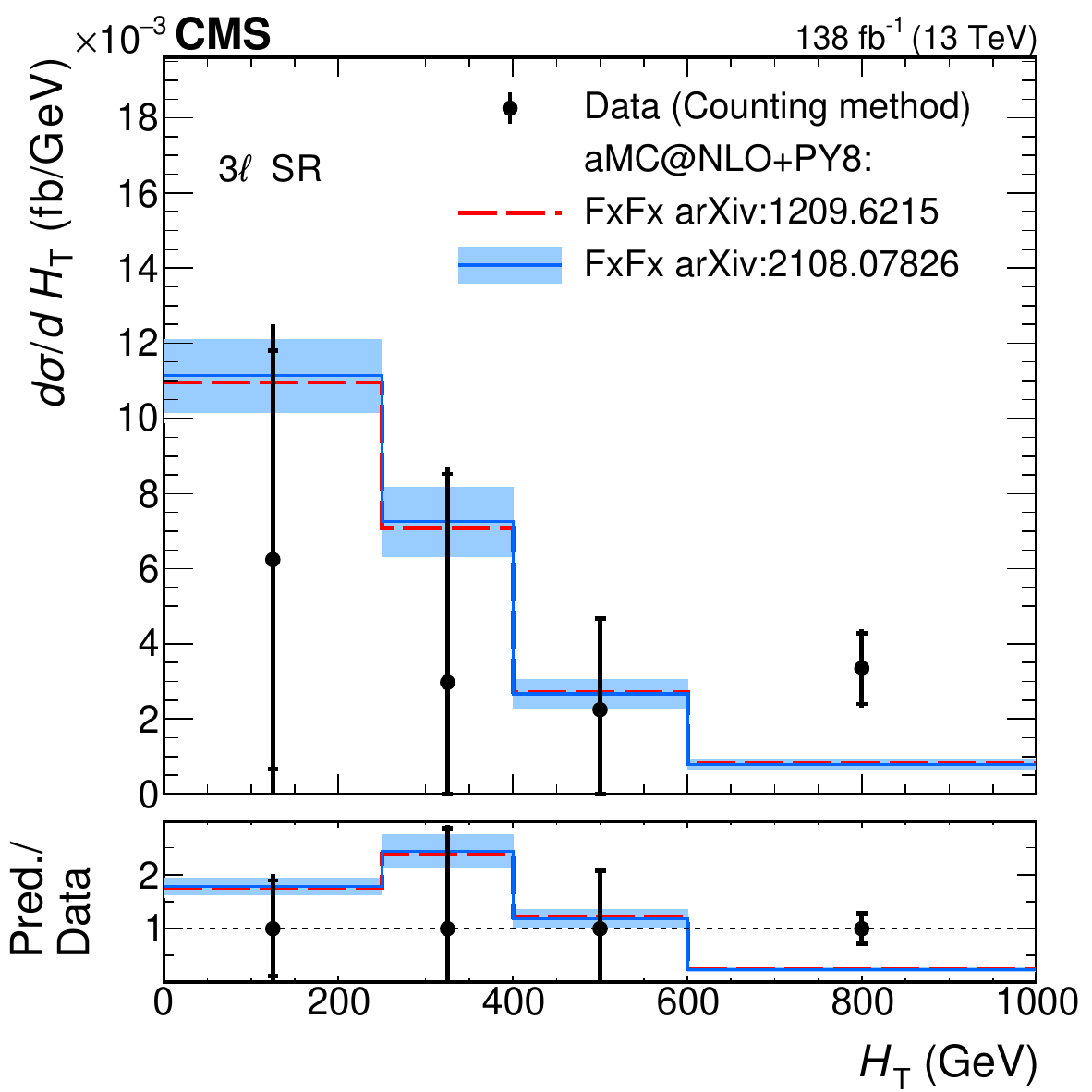}%
\hfill%
\includegraphics[width=0.475\textwidth]{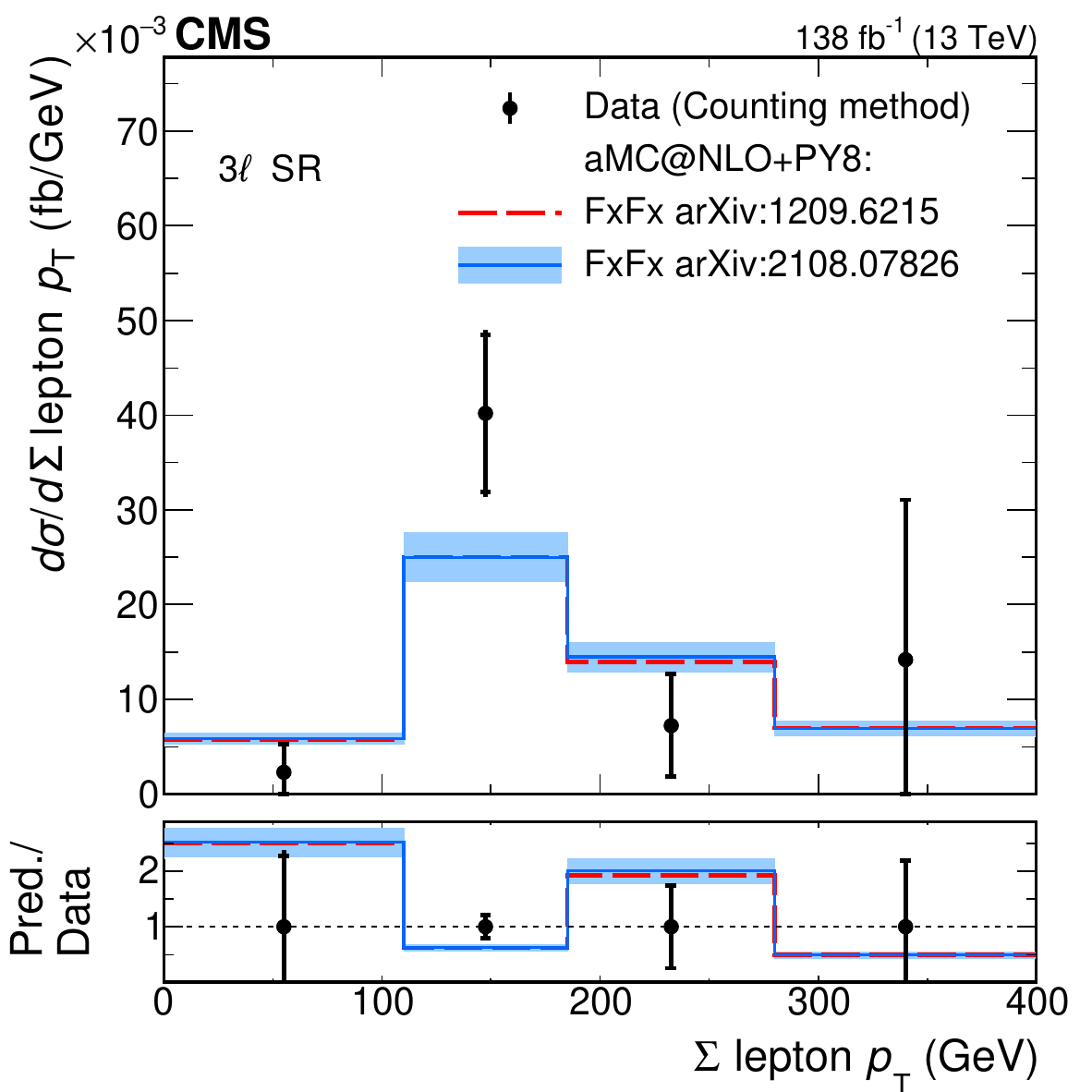}\\[\cmsTabSkip]
\includegraphics[width=0.475\textwidth]{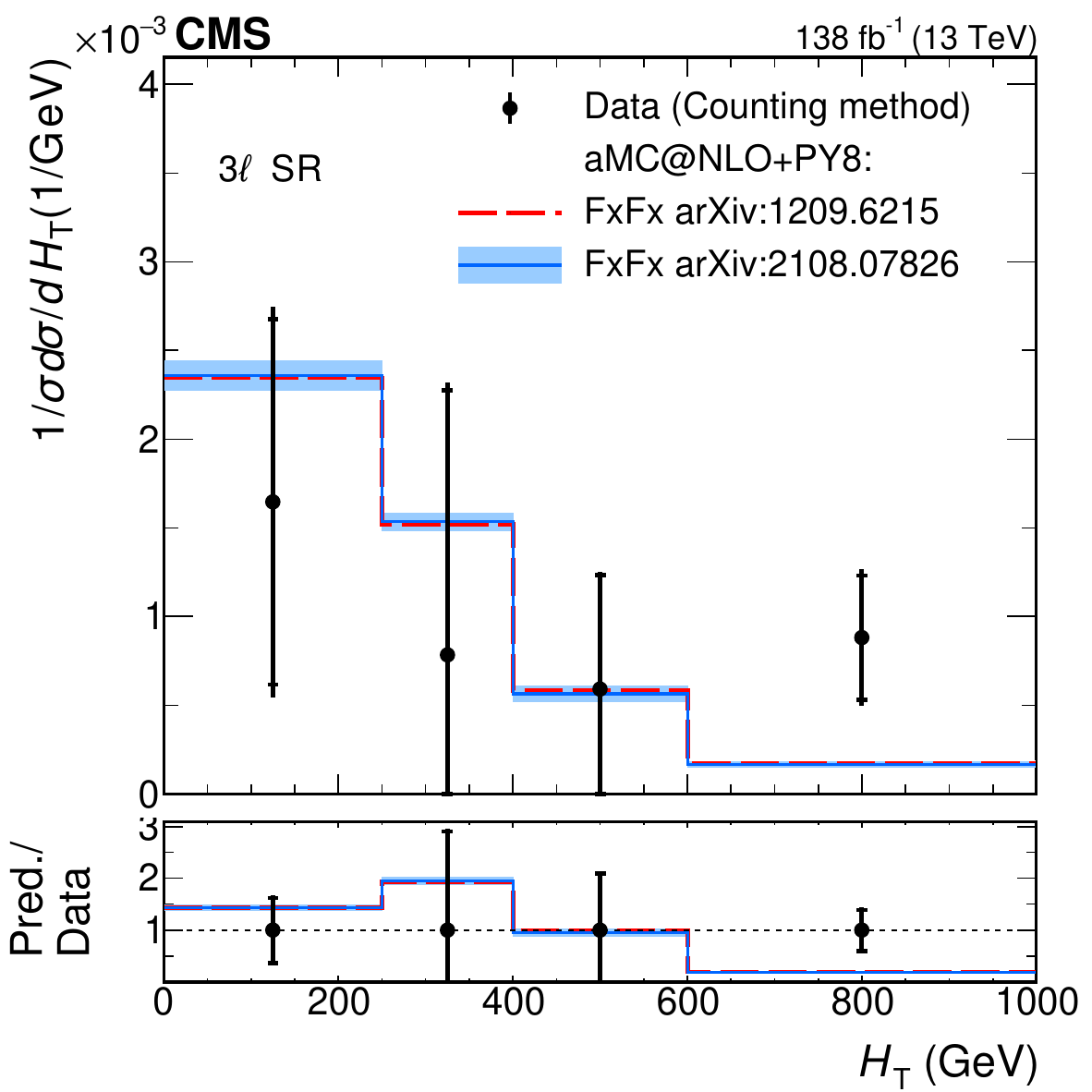}%
\hfill%
\includegraphics[width=0.475\textwidth]{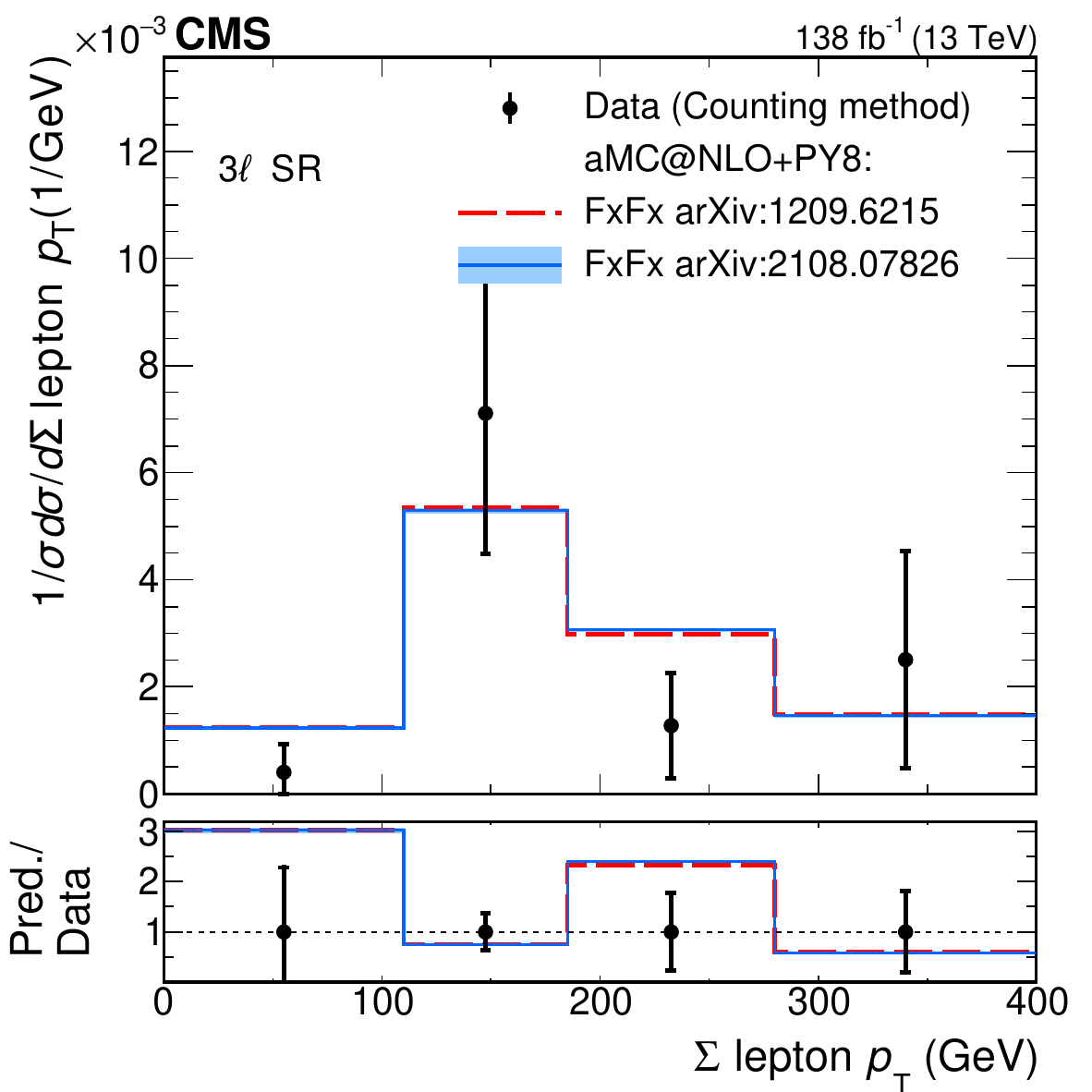}
\caption{%
    Absolute (upper row) and normalized (lower row) differential cross section measured as a function of the \HT (left) and the scalar sum of the lepton \pt's, using the counting method in the \threeLepton SR.
    \diffextracaption
}
\label{fig:differential_xsec_3l}
\end{figure}

Figure~\ref{fig:differential_xsec_3l} shows the absolute and normalized differential cross section measured in the \threeLepton SR as a function of \HT and $\sum$ lepton \pt.
These two variables were selected as the information contained by the two variables is different from that in the \twoLeptonss SR, as \HT in the \threeLepton SR does not contain the two jets in the hadronically decaying \PW boson present in the \twoLeptonss.
The $\sum$ lepton \pt variable in this region includes three leptons, in contrast to the two that are considered in the \twoLeptonss SR.

\section{Measurement of the charge asymmetry}
\label{sec:charge_asym}

We measure the leptonic charge asymmetry, \Acl, in signal events using events with three leptons.
The asymmetry \Acl is defined as
\begin{equation}
    \Acl=\frac{N(\Dyl>0)-N(\Dyl<0)}{N(\Dyl>0)+N(\Dyl<0)}.
\end{equation}
\Dyl denotes $\abs{y_{\Pell^+}}-\abs{y_{\Pell^-}}$, where $y_{\Pell^+}$ ($y_{\Pell^-}$) are the rapidity of the lepton produced in the top quark (antiquark) decays.
This quantity is only well-defined in the \threeLepton region, since for signal events in this region, both top quarks must have produced leptons in their decay.

To reconstruct \Dyl we must determine the two leptons that have been produced in the top quark decays out of the three in the event.
In a three-lepton \ttWp (\ttWm) event, there are two leptons with positive (negative) charge and a third one with negative (positive) charge.
Since the two top quarks have the opposite electric charge, the negatively (positively) charged lepton must come from a top antiquark (quark) decay.
To determine which one of the remaining two is produced in the other top quark decay and which one in the spectator \PW boson decay, we employ a neural network that assigns to each lepton the probability of belonging to each of the categories.
The neural network is trained using \textsc{PyTorch}~\cite{paszke2017torch} considering a sample of positive (negative) leptons from \ttWp (\ttWm) production, passing the selection of the fiducial region, described in Section~\ref{sec:differential}.
We define as signal those leptons produced in a top quark decay and as background those produced in the \PW boson decay.
The network, which yields as output a per-lepton quantity, takes as an input a set of variables associated to the given lepton.
In addition to the lepton, we consider two jets that are proxies to the \PQb jets produced in the top quark decays, which correspond to the \PQb-tagged jets with the highest \pt.
In case there is only one \PQb-tagged jet in the event, we consider that one and the jet closest in \DR to the lepton.
The input variables to the discriminator are
\begin{itemize}
\item the lepton \pt, $\eta$, and $\phi$;
\item the invariant mass of the system formed by the lepton and each of the two proxy jets, and
\item the \DR distance between the lepton and each of the two proxy jets.
\end{itemize}
Out of the two SS leptons, we take the one with the largest neural network score.
This selection gives the correct pairing in 71\% of the cases.
We then compute \Dylreco using the selected lepton and the one with the opposite sign.

\begin{figure}[!ht]
\centering
\includegraphics[width=0.8\textwidth]{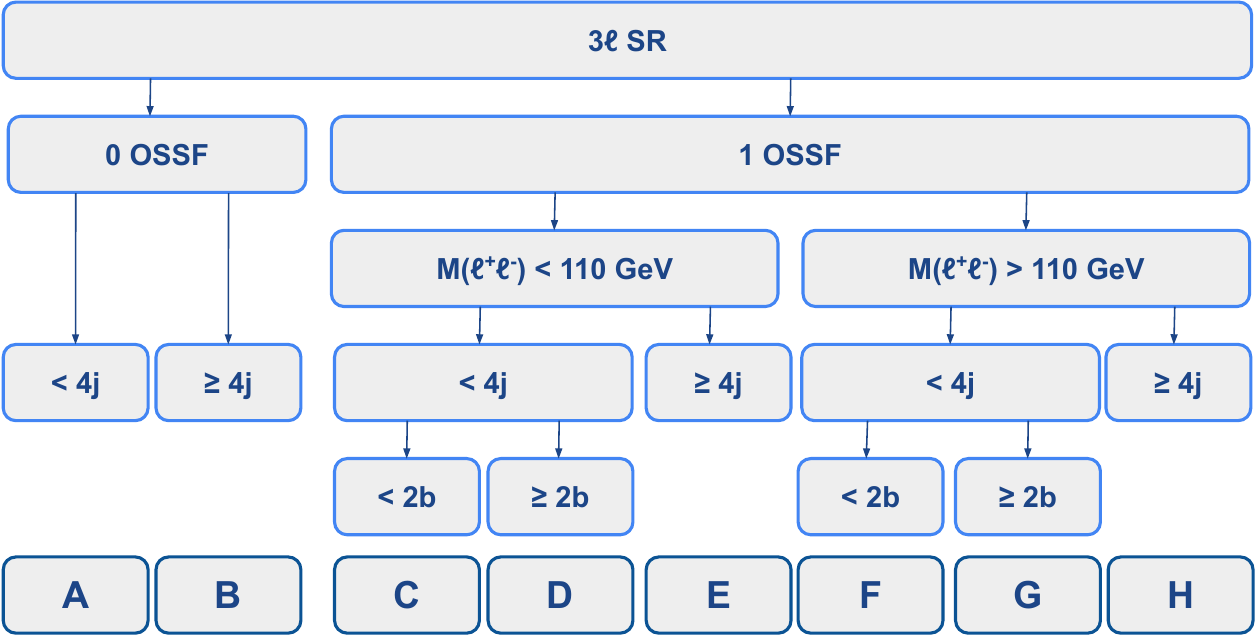}
\caption{%
    Categorization of three-lepton events for the measurement of the leptonic charge asymmetry.
}
\label{fig:charge_categorization}
\end{figure}

In order to measure \Acl, we consider events in the \threeLepton region, with the exception that we replace the requirement that events have two loose \PQb-tagged jets with a requirement of at least one medium \PQb-tagged jet is present in the event.
In addition, we require events with an opposite-sign same-flavor lepton pair, less than 4 jets, and less than 2 \PQb-tagged jets to pass a selection of $\ptmiss>50\GeV$ to reject \DYjets events with nonprompt leptons and photon conversions from $\PZ\PGg$ events.
Events are then classified based on their kinematic properties with the twofold goal of building regions pure in \ttW signal and separating events with  $\Dyl>0$ and $\Dyl<0$.
A set of 8 categories is built based on the presence of a pair of opposite-sign same-flavor lepton pair in the event, the jet and \PQb-tagged jet multiplicities, and the invariant mass of the opposite-sign same-flavor lepton pair closest to the \PZ boson mass.
The sequential categorization is shown in Fig.~\ref{fig:charge_categorization}.
In addition, we classify events in four bins of \Dylreco with edges $(-\infty,-0.5,0,0.5,\infty)$, resulting in a total of 32 analysis bins.

\begin{figure}[!ht]
\centering
\includegraphics[width=0.475\textwidth]{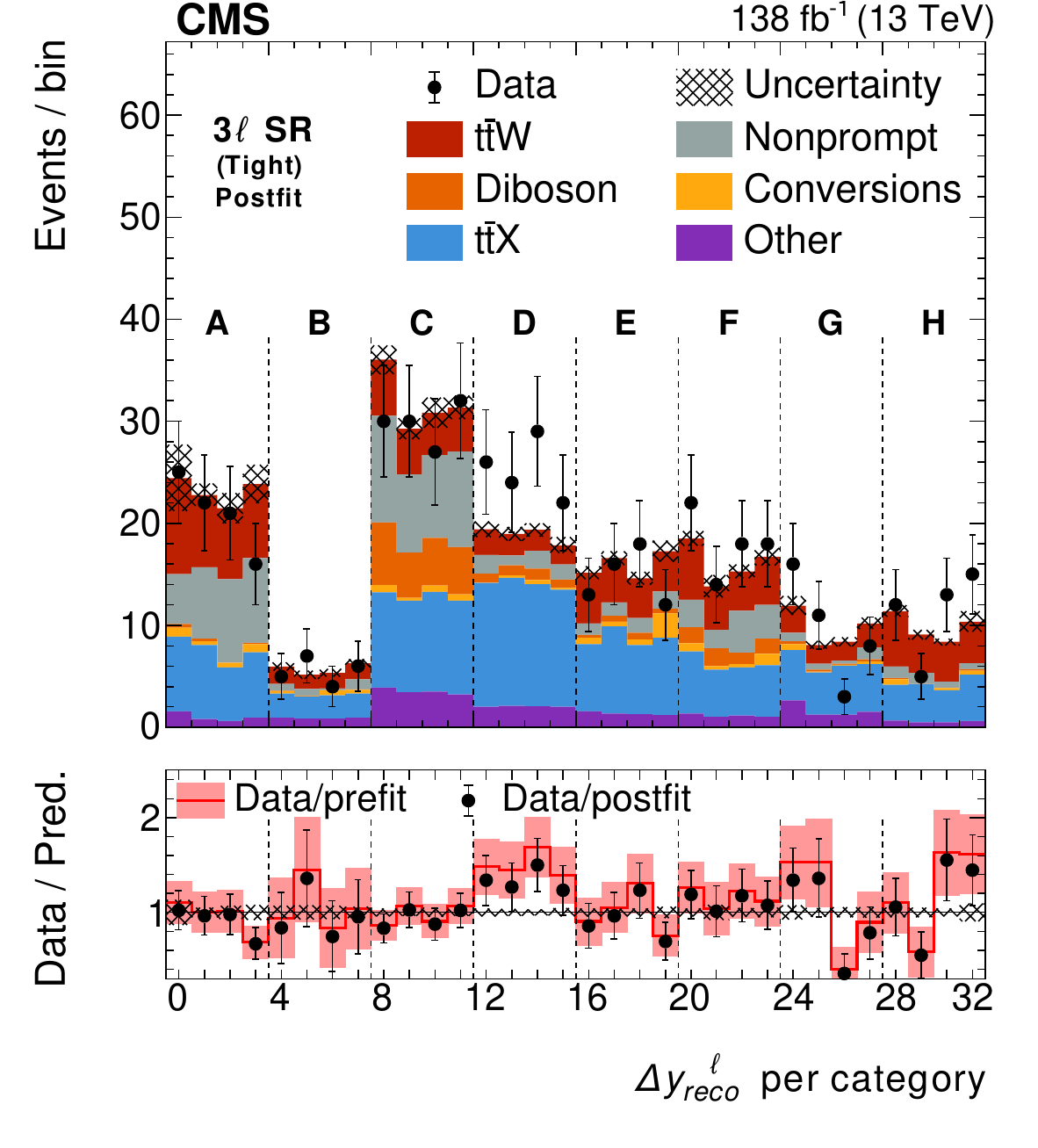}%
\hfill%
\raisebox{2.14cm}{\includegraphics[width=0.425\textwidth]{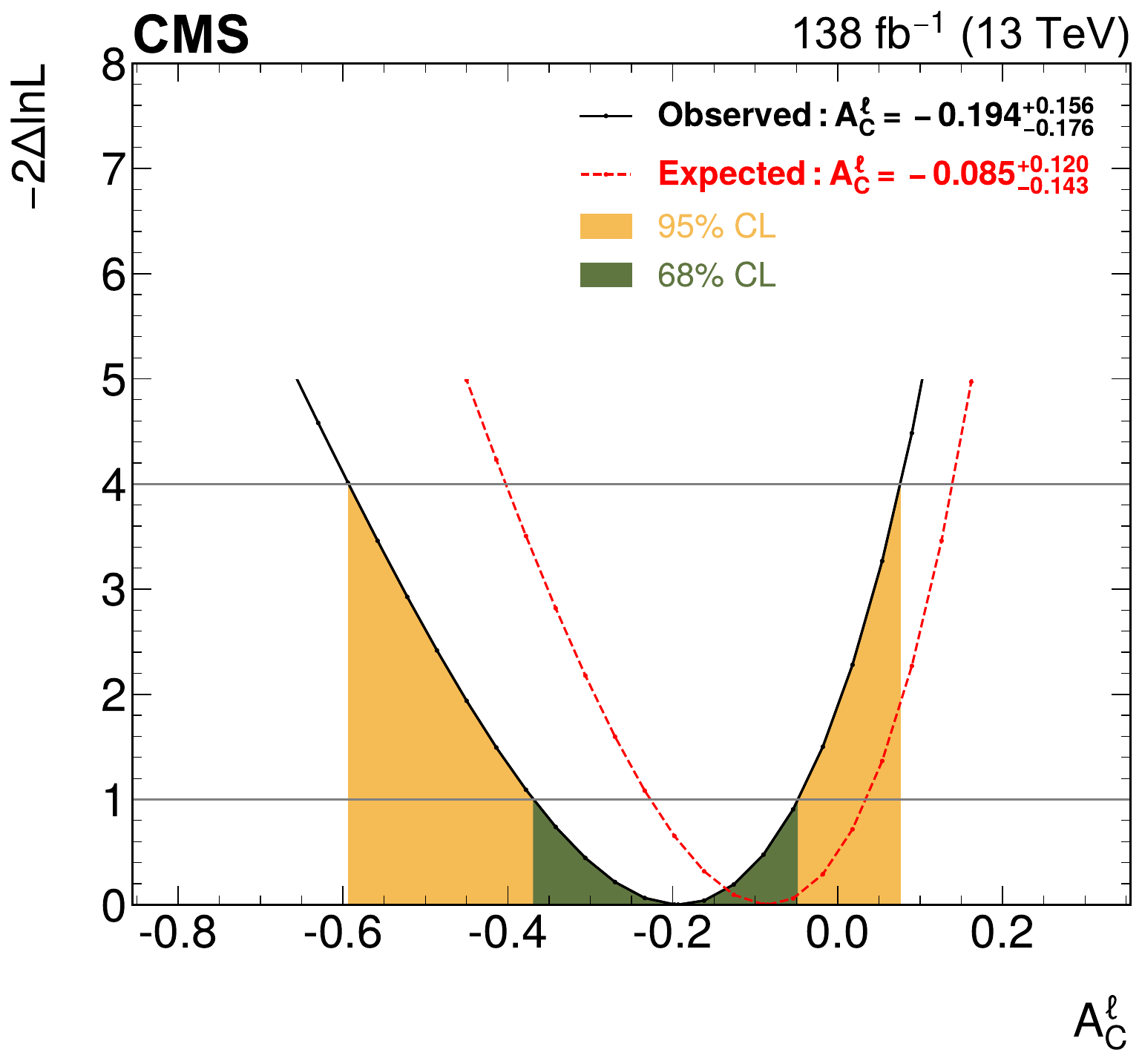}}
\caption{%
    Left: Observed yields in the 32 bins used in the analysis for data (points) and predictions (filled histograms) after the fit to the data for the tight lepton selection.
    \ratiodescription
    Right: the observed (black) and expected (red) likelihood scan as a function of \Acl.
    The shaded areas correspond to the 68 and 95\% confidence level intervals around the best-fit value, respectively.
}
\label{fig:charge_yields}
\end{figure}

To measure \Acl, we use a likelihood model similar to the one shown in Eq.~\eqref{eq:likelihood}, in which the parameters of interest are \Acl and the signal normalization.
The expected signal yield dependence on \Acl is introduced by means of the MC simulation templates with $\Dyl>0$ and $\Dyl<0$.
Figure~\ref{fig:charge_yields} (left) shows the observed number of events in each of these bins, together with the signal and background prediction obtained setting all fit parameters to the values that maximize the likelihood.
Figure~\ref{fig:charge_yields} (right) shows the likelihood as a function of \Acl, profiling over the systematic uncertainties and the signal normalization.
This procedure results in an observed (expected) value of $\Acl=-0.19\,^{+0.05}_{-0.06}\syst\,^{+0.14}_{-0.16}\stat$ ($-0.09\,^{+0.04}_{-0.05}\syst\,^{+0.11}_{-0.11}\stat$), consistent with the theoretical expectations from next-to-leading order simulations of $-0.085\pm0.006$, and achieving a better sensitivity than the previous measurement by ATLAS~\cite{ATLAS:2023xay}.
Additionally, we performed a goodness-of-fit test using the saturated model, obtaining a $p$-value of 0.83, showing no indication of significant tensions between data and the fitted model. 

\section{Summary}
\label{sec:summary}

Measurements of properties of top quark-antiquark pair production in association with a \PW boson (\ttW) in proton-proton collisions at a center-of-mass energy of 13\TeV are performed using a data sample corresponding to an integrated luminosity of 138\fbinv, recorded by the CMS experiment.
Events with either two leptons with the same electric charge or three leptons, and multiple jets and \PQb-tagged jets are used.
The differential cross sections as a function of kinematic variables sensitive to the modeling of the \ttW process are measured, using a multivariate discriminator for events in the two-lepton region and a counting method for events in the three-lepton region.
Overall, the normalized differential cross section measurements are generally consistent with the standard model expectations, while the absolute cross sections are above the theoretical predictions by approximately one standard deviation, consistent with previous inclusive cross section measurements.
In addition to the differential cross sections, the leptonic charge asymmetry of this process is measured to be $\Acl=-0.19\,^{+0.16}_{-0.18}$, consistent with the expectation of $-0.085\pm0.006$ predicted by next-to-leading order simulations.

\begin{acknowledgments}
We congratulate our colleagues in the CERN accelerator departments for the excellent performance of the LHC and thank the technical and administrative staffs at CERN and at other CMS institutes for their contributions to the success of the CMS effort. In addition, we gratefully acknowledge the computing centers and personnel of the Worldwide LHC Computing Grid and other centers for delivering so effectively the computing infrastructure essential to our analyses. Finally, we acknowledge the enduring support for the construction and operation of the LHC, the CMS detector, and the supporting computing infrastructure provided by the following funding agencies: SC (Armenia), BMBWF and FWF (Austria); FNRS and FWO (Belgium); CNPq, CAPES, FAPERJ, FAPERGS, and FAPESP (Brazil); MES and BNSF (Bulgaria); CERN; CAS, MoST, and NSFC (China); MINCIENCIAS (Colombia); MSES and CSF (Croatia); RIF (Cyprus); SENESCYT (Ecuador); ERC PRG, TARISTU24-TK10 and MoER TK202 (Estonia); Academy of Finland, MEC, and HIP (Finland); CEA and CNRS/IN2P3 (France); SRNSF (Georgia); BMFTR, DFG, and HGF (Germany); GSRI (Greece); NKFIH (Hungary); DAE and DST (India); IPM (Iran); SFI (Ireland); INFN (Italy); MSIT and NRF (Republic of Korea); MES (Latvia); LMTLT (Lithuania); MOE and UM (Malaysia); BUAP, CINVESTAV, CONACYT, LNS, SEP, and UASLP-FAI (Mexico); MOS (Montenegro); MBIE (New Zealand); PAEC (Pakistan); MES, NSC, and NAWA (Poland); FCT (Portugal); MESTD (Serbia); MICIU/AEI and PCTI (Spain); MOSTR (Sri Lanka); Swiss Funding Agencies (Switzerland); MST (Taipei); MHESI and NSTDA (Thailand); TUBITAK and TENMAK (T\"{u}rkiye); NASU (Ukraine); STFC (United Kingdom); DOE and NSF (USA).

\hyphenation{Rachada-pisek} Individuals have received support from the Marie-Curie program and the European Research Council and Horizon 2020 Grant, contract Nos.\ 675440, 724704, 752730, 758316, 765710, 824093, 101115353, 101002207, 101001205, and COST Action CA16108 (European Union); the Leventis Foundation; the Alfred P.\ Sloan Foundation; the Alexander von Humboldt Foundation; the Science Committee, project no. 22rl-037 (Armenia); the Fonds pour la Formation \`a la Recherche dans l'Industrie et dans l'Agriculture (FRIA-Belgium); the Beijing Municipal Science \& Technology Commission, No. Z191100007219010, the Fundamental Research Funds for the Central Universities, the Ministry of Science and Technology of China under Grant No. 2023YFA1605804, and the Natural Science Foundation of China under Grant No. 12061141002 (China); the Ministry of Education, Youth and Sports (MEYS) of the Czech Republic; the Shota Rustaveli National Science Foundation, grant FR-22-985 (Georgia); the Deutsche Forschungsgemeinschaft (DFG), among others, under Germany's Excellence Strategy -- EXC 2121 ``Quantum Universe" -- 390833306, and under project number 400140256 - GRK2497; the Hellenic Foundation for Research and Innovation (HFRI), Project Number 2288 (Greece); the Hungarian Academy of Sciences, the New National Excellence Program - \'UNKP, the NKFIH research grants K 131991, K 133046, K 138136, K 143460, K 143477, K 146913, K 146914, K 147048, 2020-2.2.1-ED-2021-00181, TKP2021-NKTA-64, and 2021-4.1.2-NEMZ\_KI-2024-00036 (Hungary); the Council of Science and Industrial Research, India; ICSC -- National Research Center for High Performance Computing, Big Data and Quantum Computing, FAIR -- Future Artificial Intelligence Research, and CUP I53D23001070006 (Mission 4 Component 1), funded by the NextGenerationEU program (Italy); the Latvian Council of Science; the Ministry of Education and Science, project no. 2022/WK/14, and the National Science Center, contracts Opus 2021/41/B/ST2/01369, 2021/43/B/ST2/01552, 2023/49/B/ST2/03273, and the NAWA contract BPN/PPO/2021/1/00011 (Poland); the Funda\c{c}\~ao para a Ci\^encia e a Tecnologia, grant CEECIND/01334/2018 (Portugal); the National Priorities Research Program by Qatar National Research Fund;  MICIU/AEI/10.13039/501100011033, ERDF/EU, "European Union NextGenerationEU/PRTR", and Programa Severo Ochoa del Principado de Asturias (Spain); the Chulalongkorn Academic into Its 2nd Century Project Advancement Project, and the National Science, Research and Innovation Fund via the Program Management Unit for Human Resources \& Institutional Development, Research and Innovation, grant B39G680009 (Thailand); the Kavli Foundation; the Nvidia Corporation; the SuperMicro Corporation; the Welch Foundation, contract C-1845; and the Weston Havens Foundation (USA).
\end{acknowledgments}

\bibliography{auto_generated}

\appendix
\numberwithin{figure}{section}
\section{Results in the \texorpdfstring{\twoLeptonss}{2lSS} region using the counting method}
\label{app:countingmethod_2lss_results}

\begin{figure}[!ht]
\centering
\includegraphics[width=0.33\textwidth]{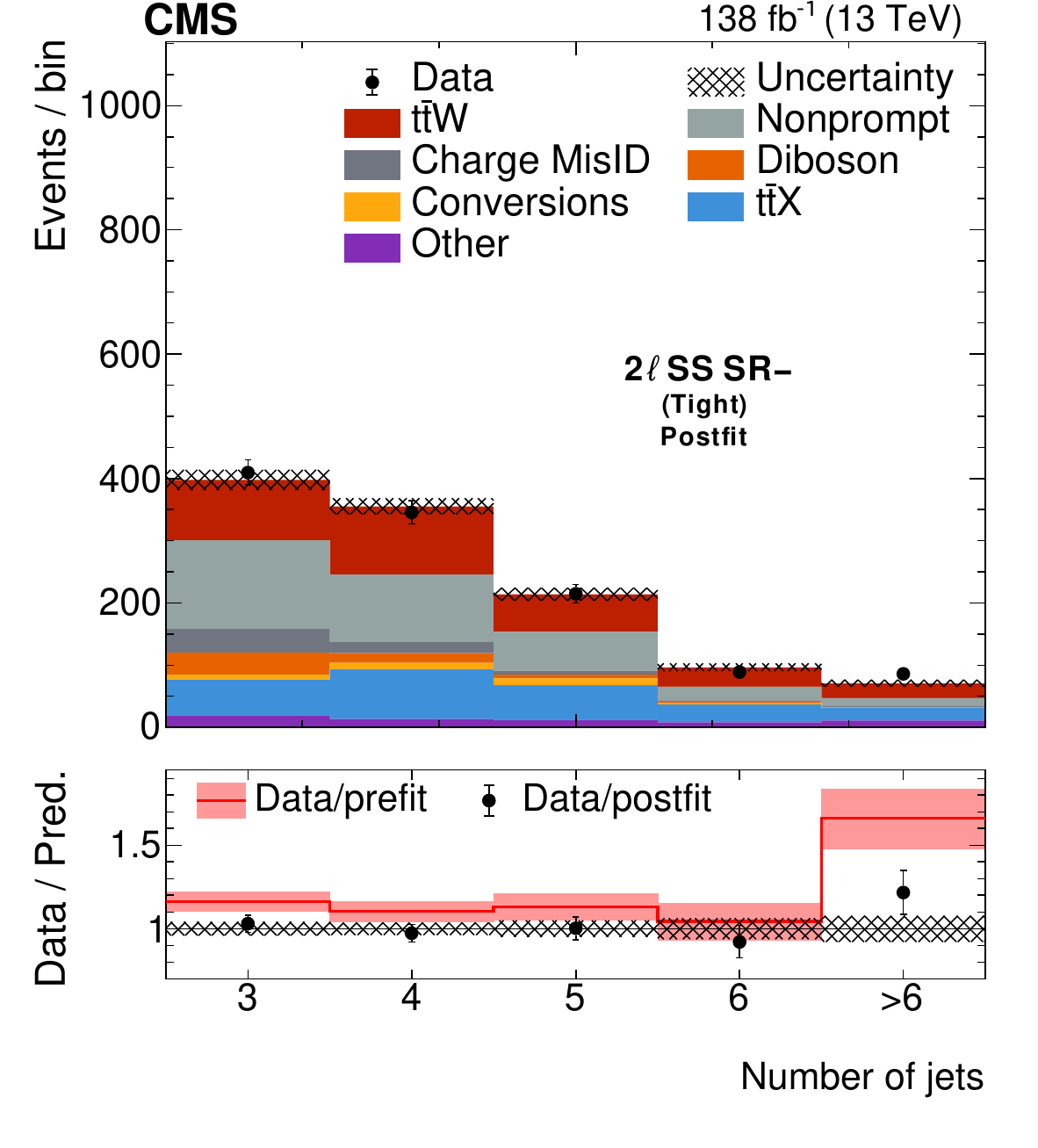}%
\hfill%
\includegraphics[width=0.33\textwidth]{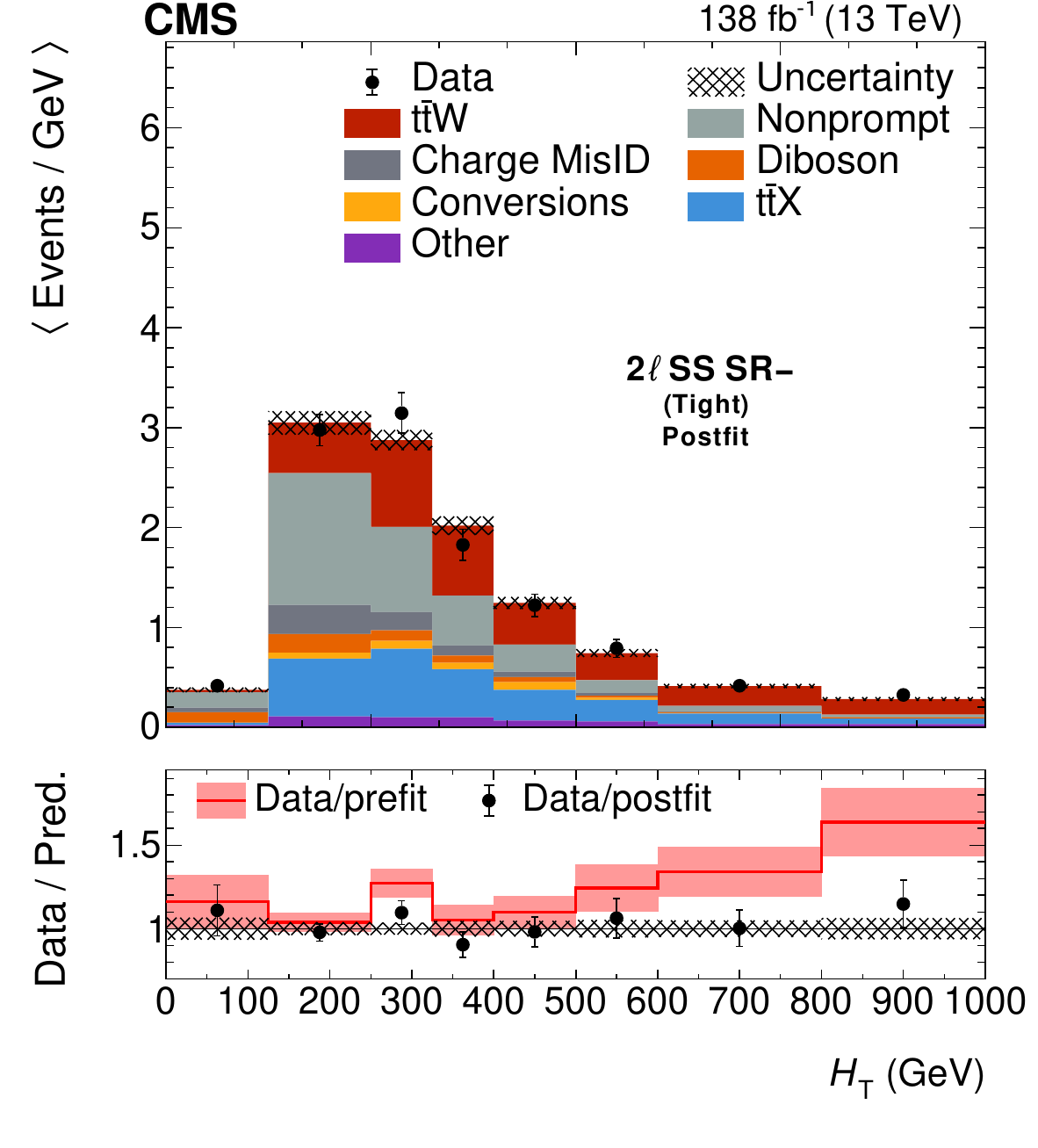}%
\hfill%
\includegraphics[width=0.33\textwidth]{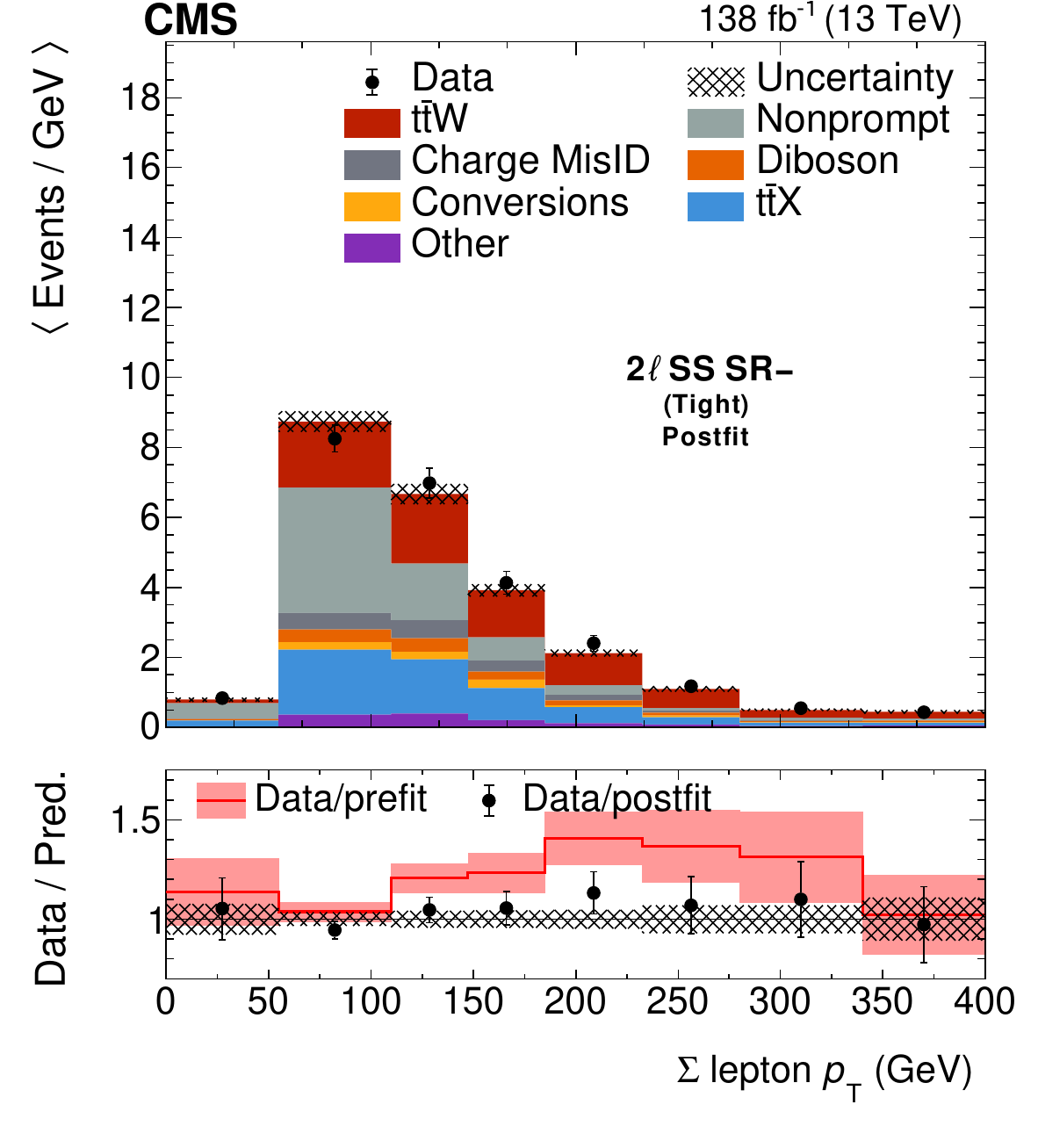}\\
\includegraphics[width=0.33\textwidth]{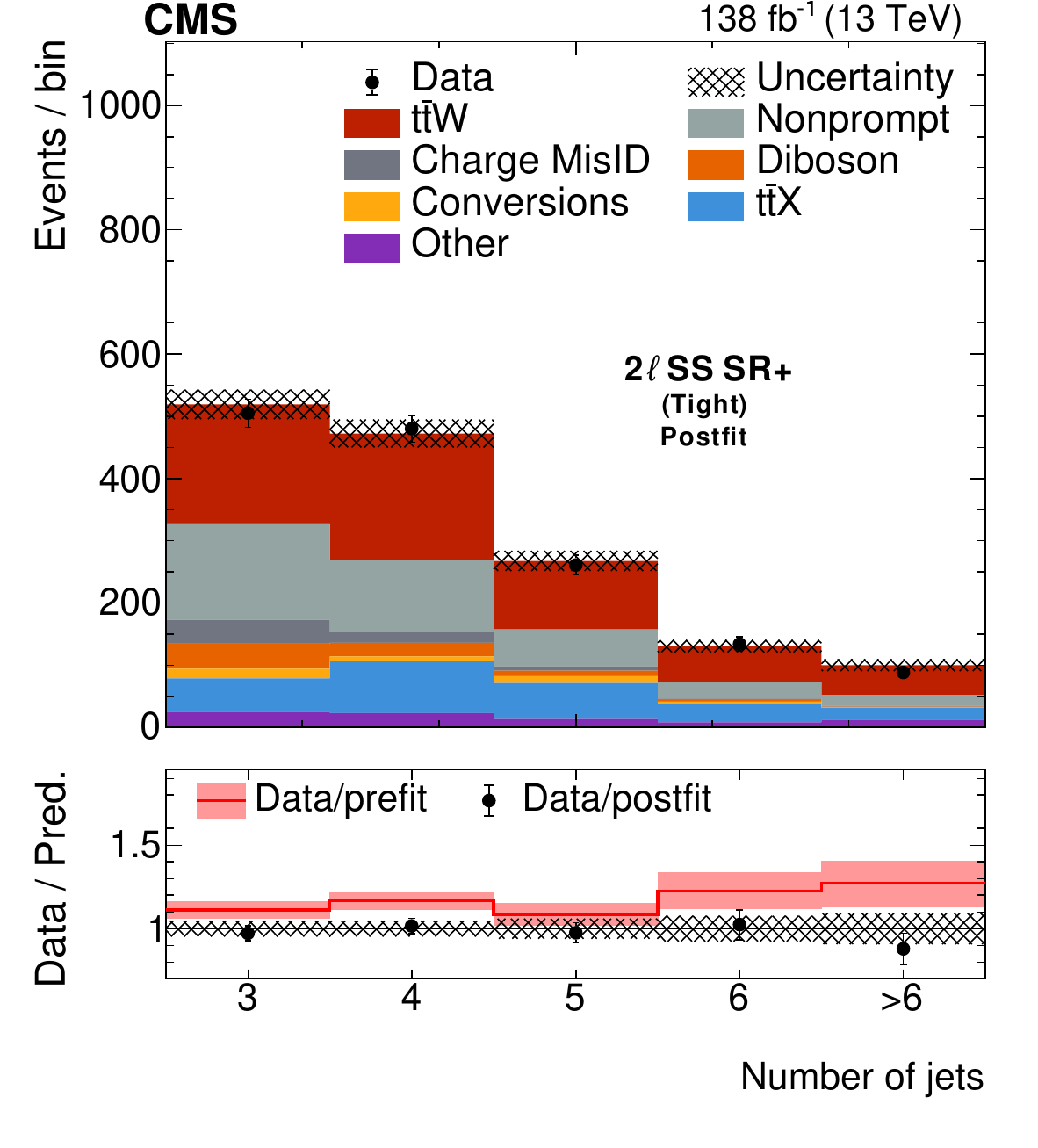}%
\hfill%
\includegraphics[width=0.33\textwidth]{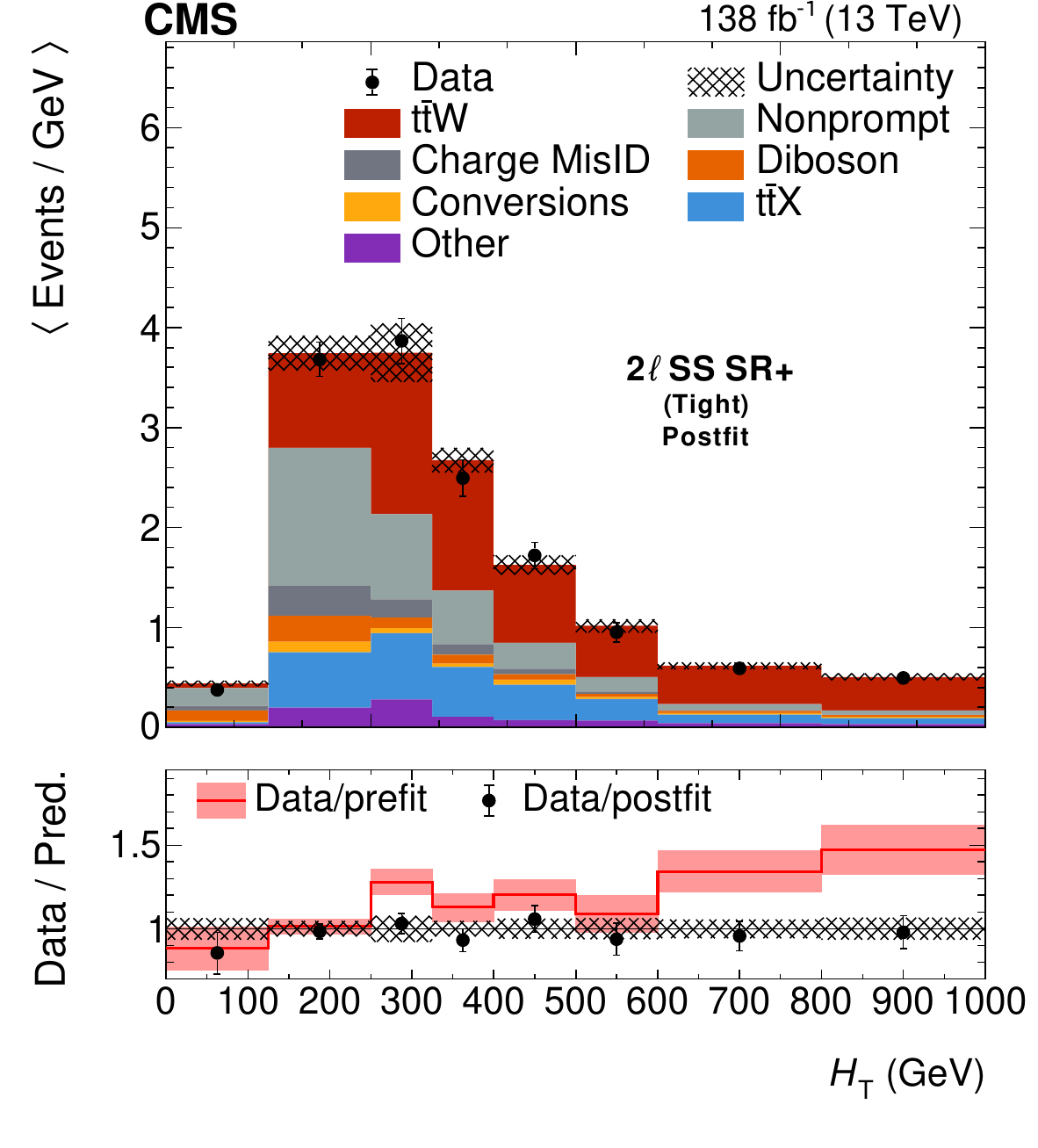}%
\hfill%
\includegraphics[width=0.33\textwidth]{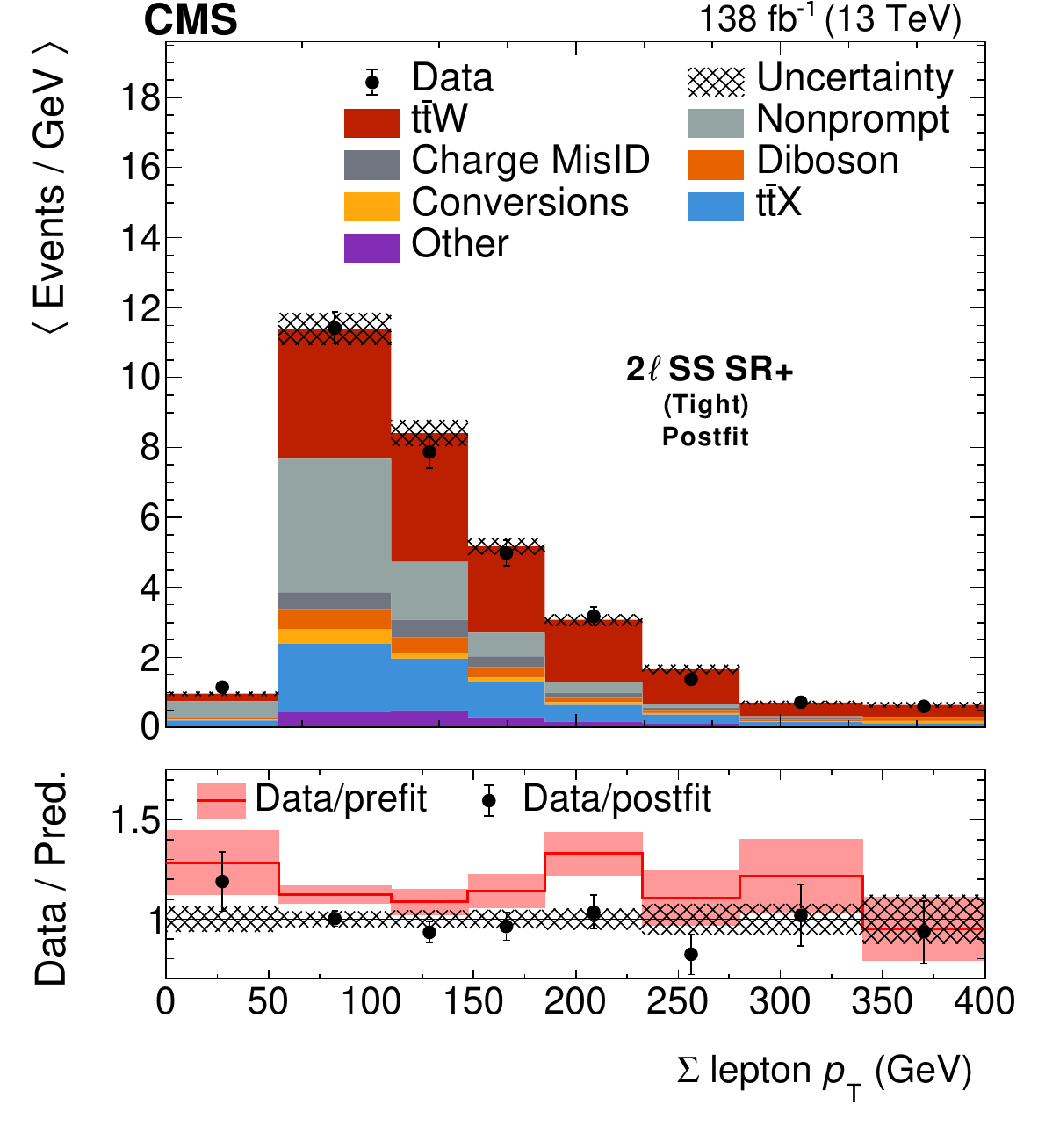}
\caption{%
    Distributions of some variables of interest in the two-lepton signal selection for data (points) and predictions (filled histograms) after the fit to the data for the tight lepton selection.
    Left: number of selected jets in the event (negative, positive signed leptons SR).
    Middle: scalar \pt sum of selected jets in the event (negative, positive signed leptons SR).
    Right: scalar \pt sum of the leptons in the event (negative, positive signed leptons SR).
    \ratiodescription
    \overflow
    The bin contents of the middle and right side plots are divided by the width of the bin.
}
\label{fig:signalregion:dilepton_tight}
\end{figure}

\begin{figure}[!ptt]
\centering
\includegraphics[width=0.475\textwidth]{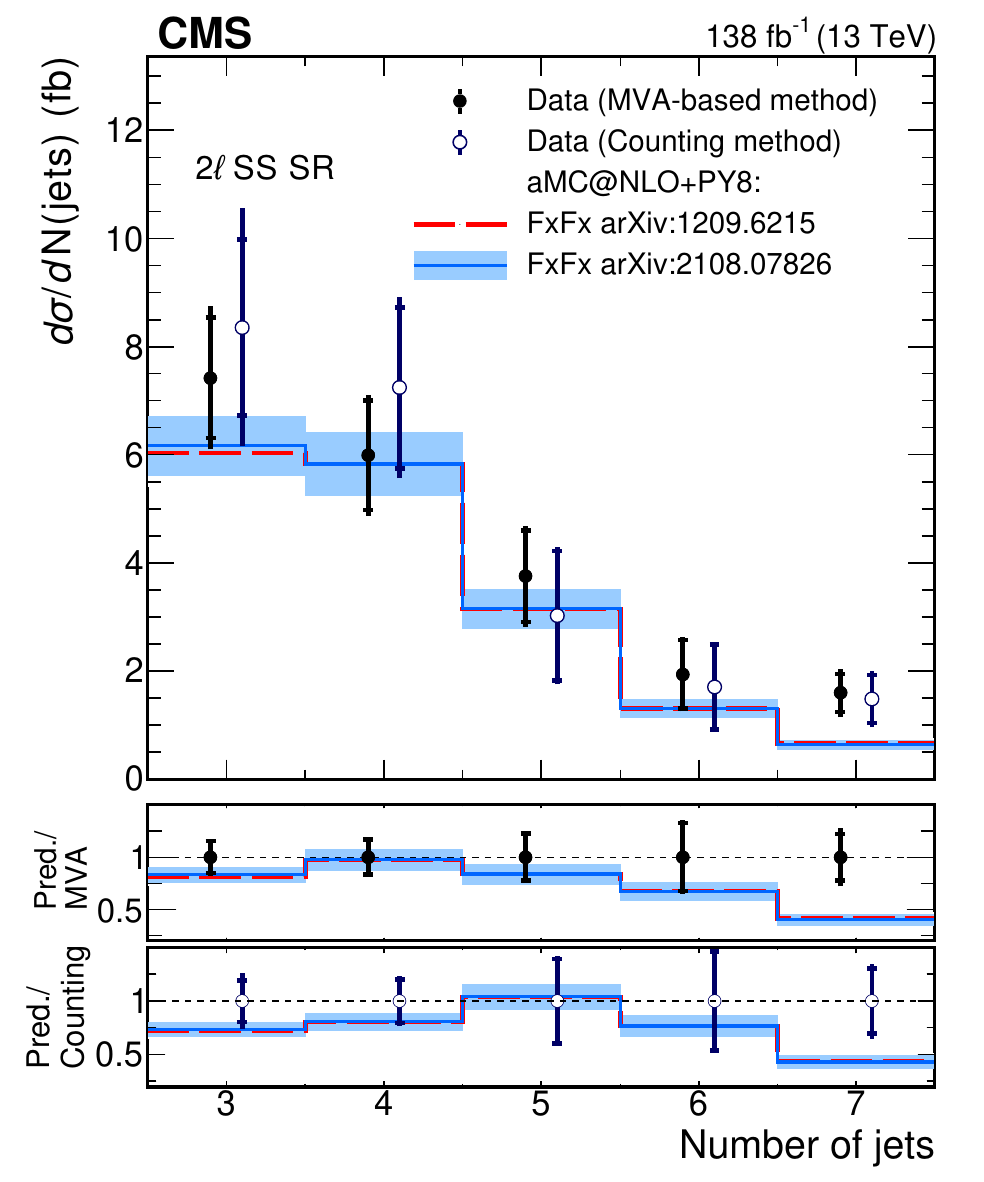}%
\hfill%
\includegraphics[width=0.475\textwidth]{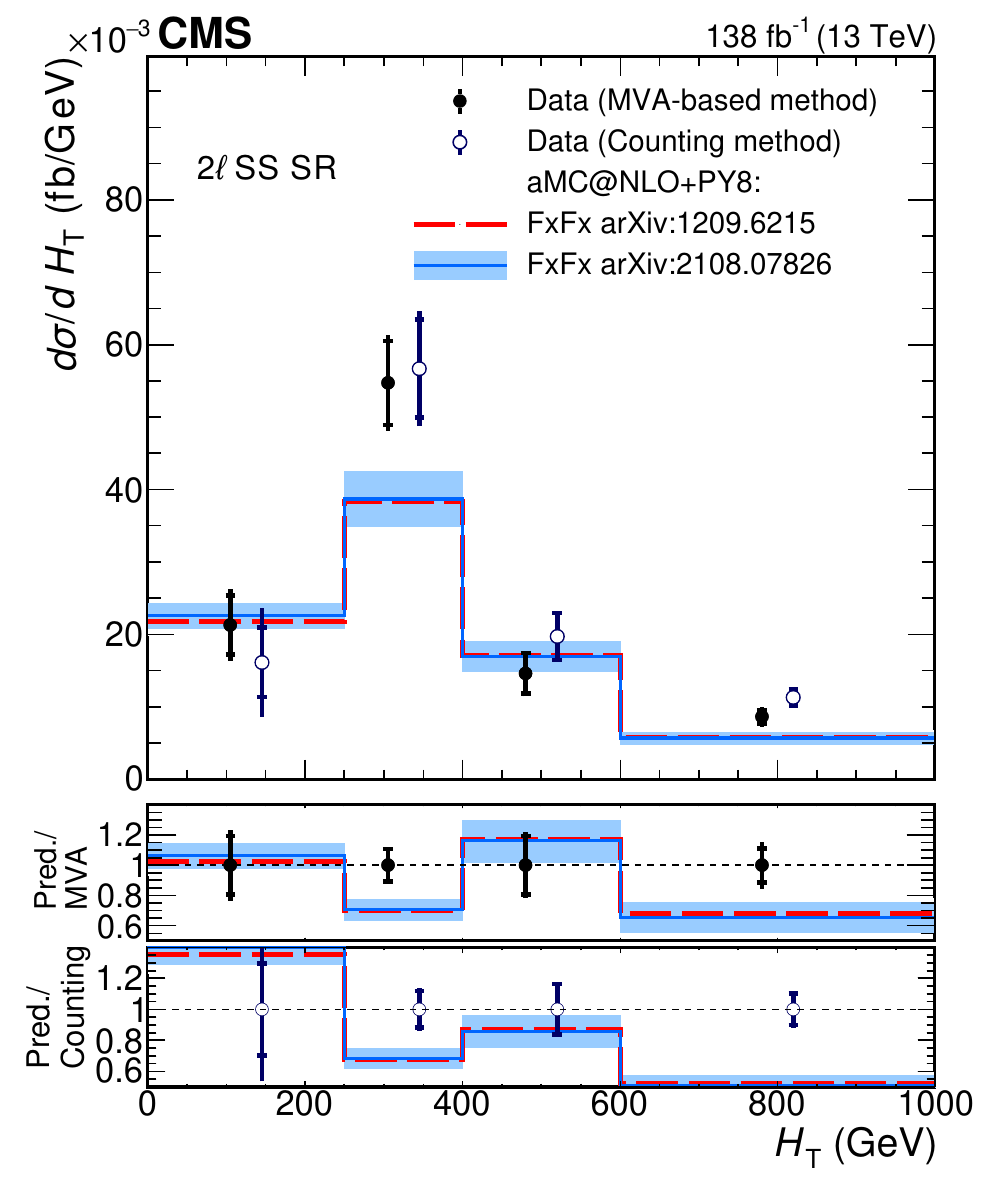} \\[\cmsTabSkip]
\includegraphics[width=0.475\textwidth]{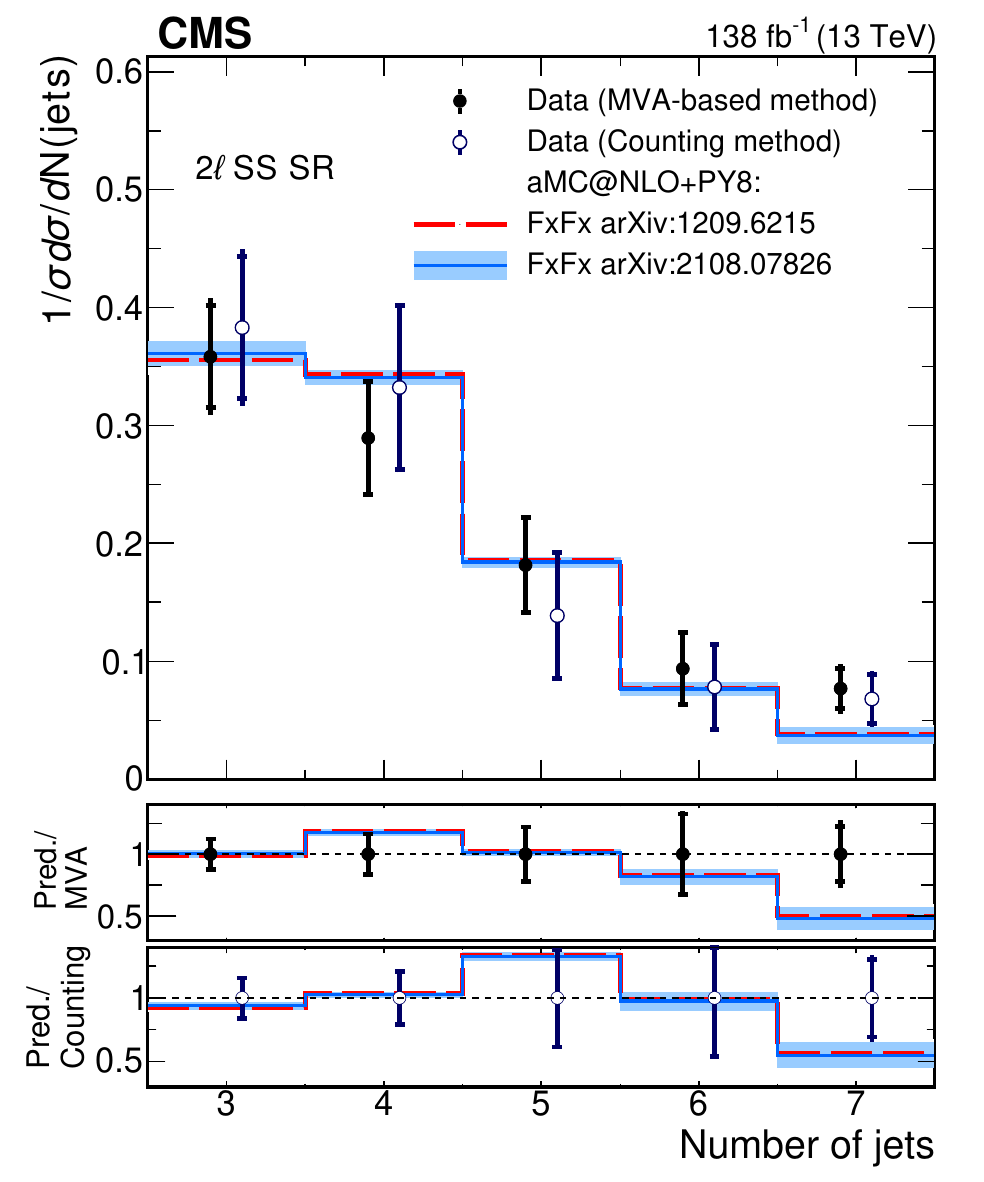}%
\hfill%
\includegraphics[width=0.475\textwidth]{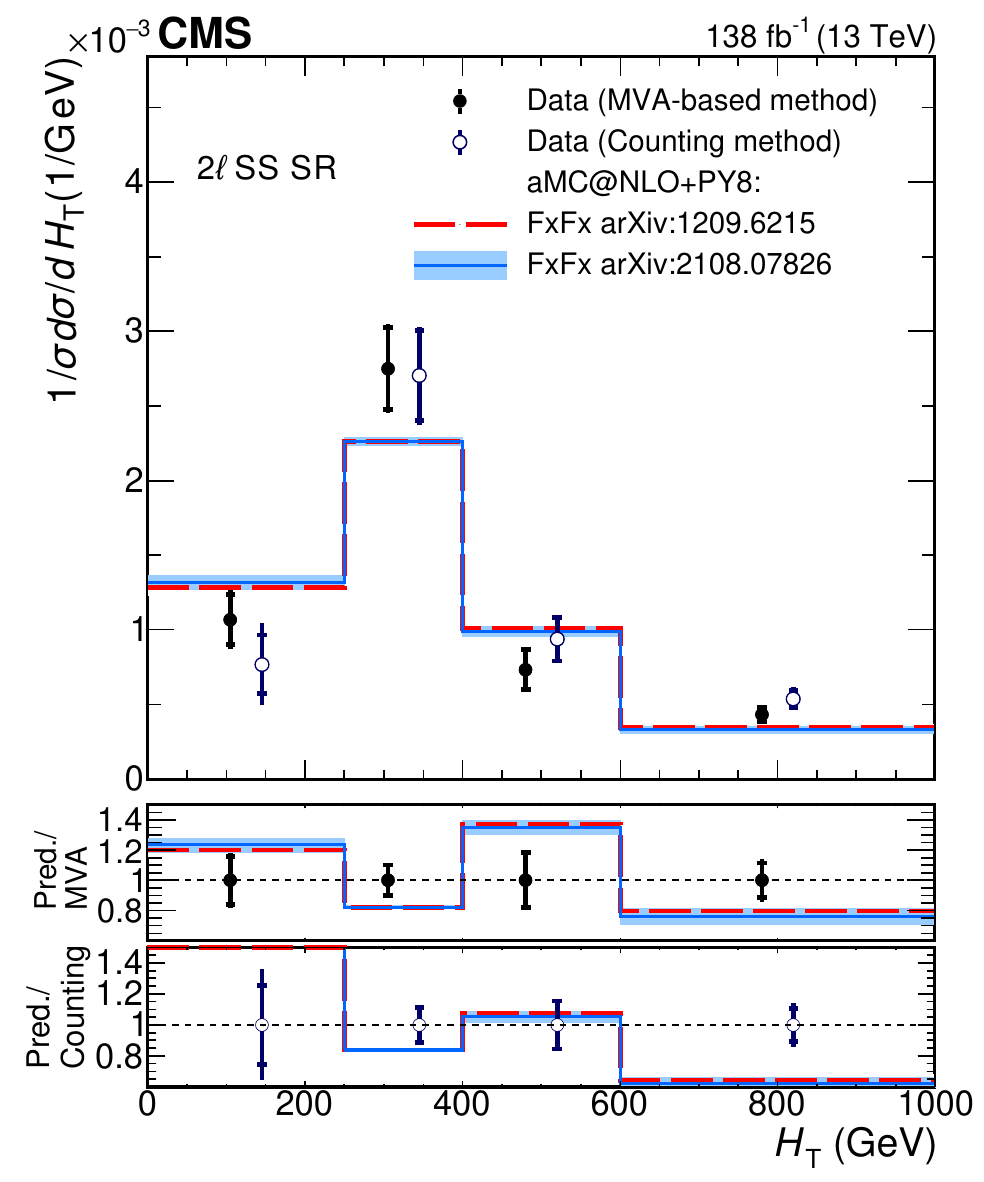}
\caption{%
    Absolute (upper row) and normalized (lower row) differential cross section measured as a function of the jet multiplicity (left) and \HT (right), using the MVA-based and the counting method.
    \diffextracaptionTogether
}
\label{fig:differential_xsectogether_1}
\end{figure}

\begin{figure}[!ptt]
\centering
\includegraphics[width=0.475\textwidth]{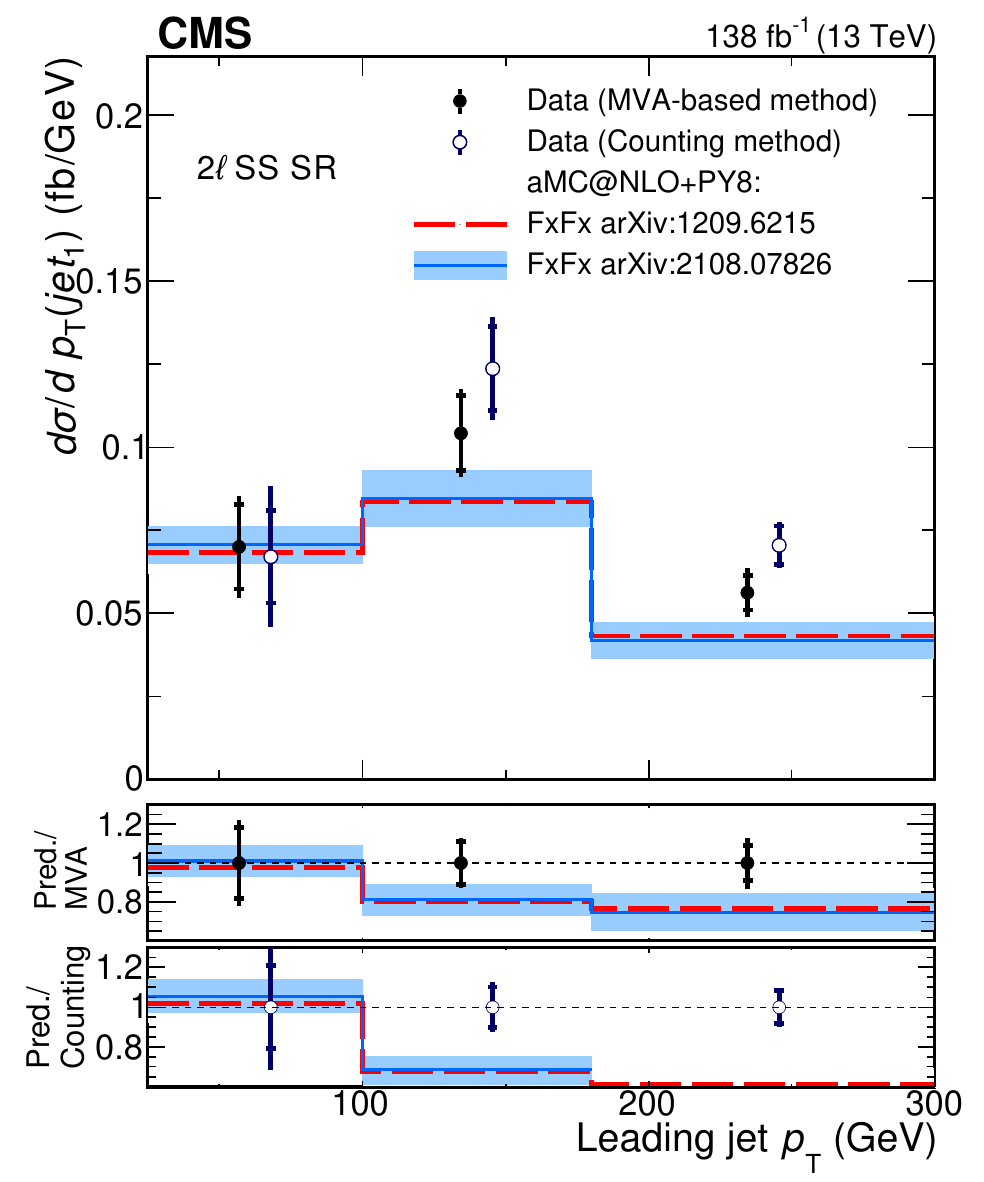}%
\hfill%
\includegraphics[width=0.475\textwidth]{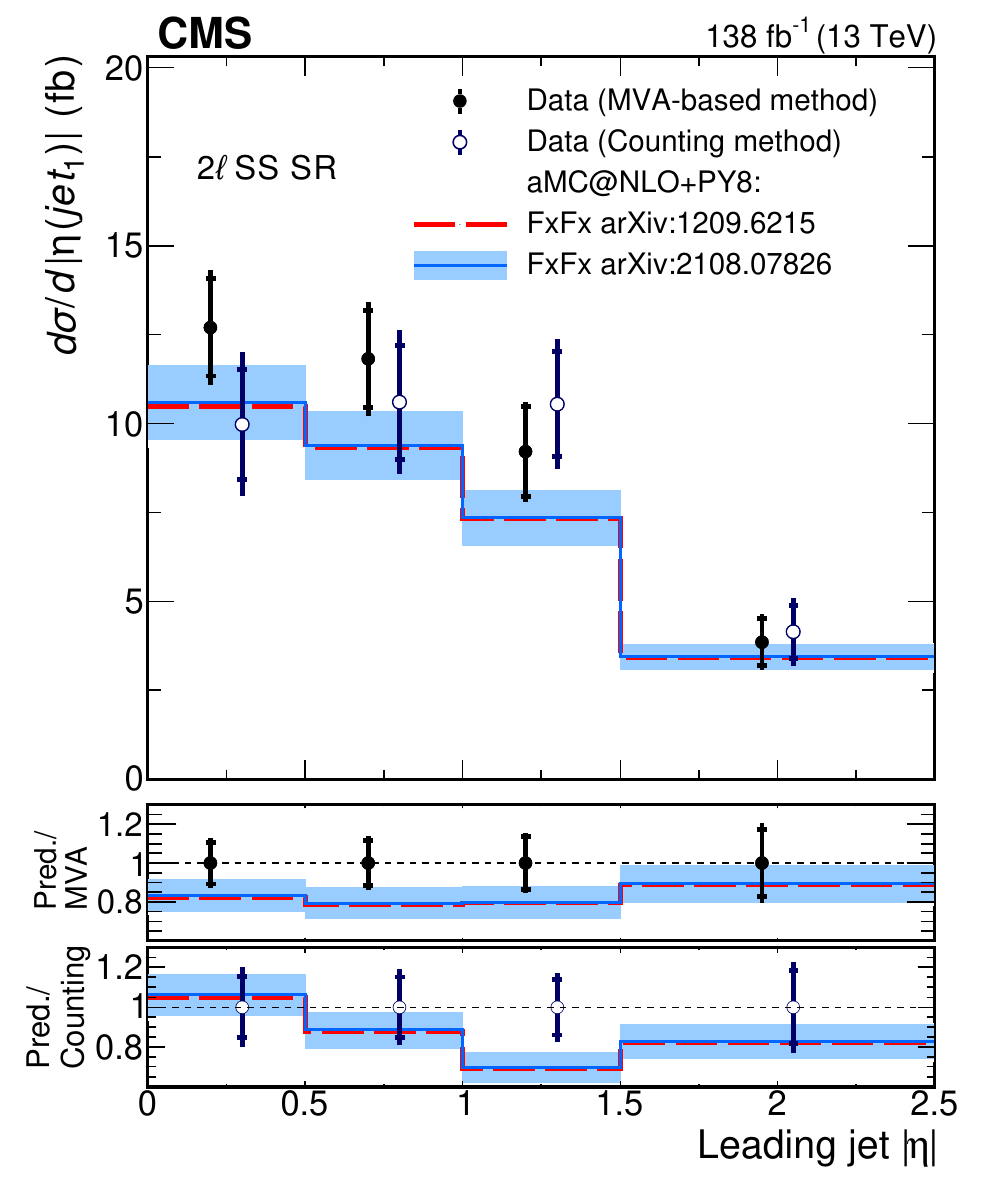} \\[\cmsTabSkip]
\includegraphics[width=0.475\textwidth]{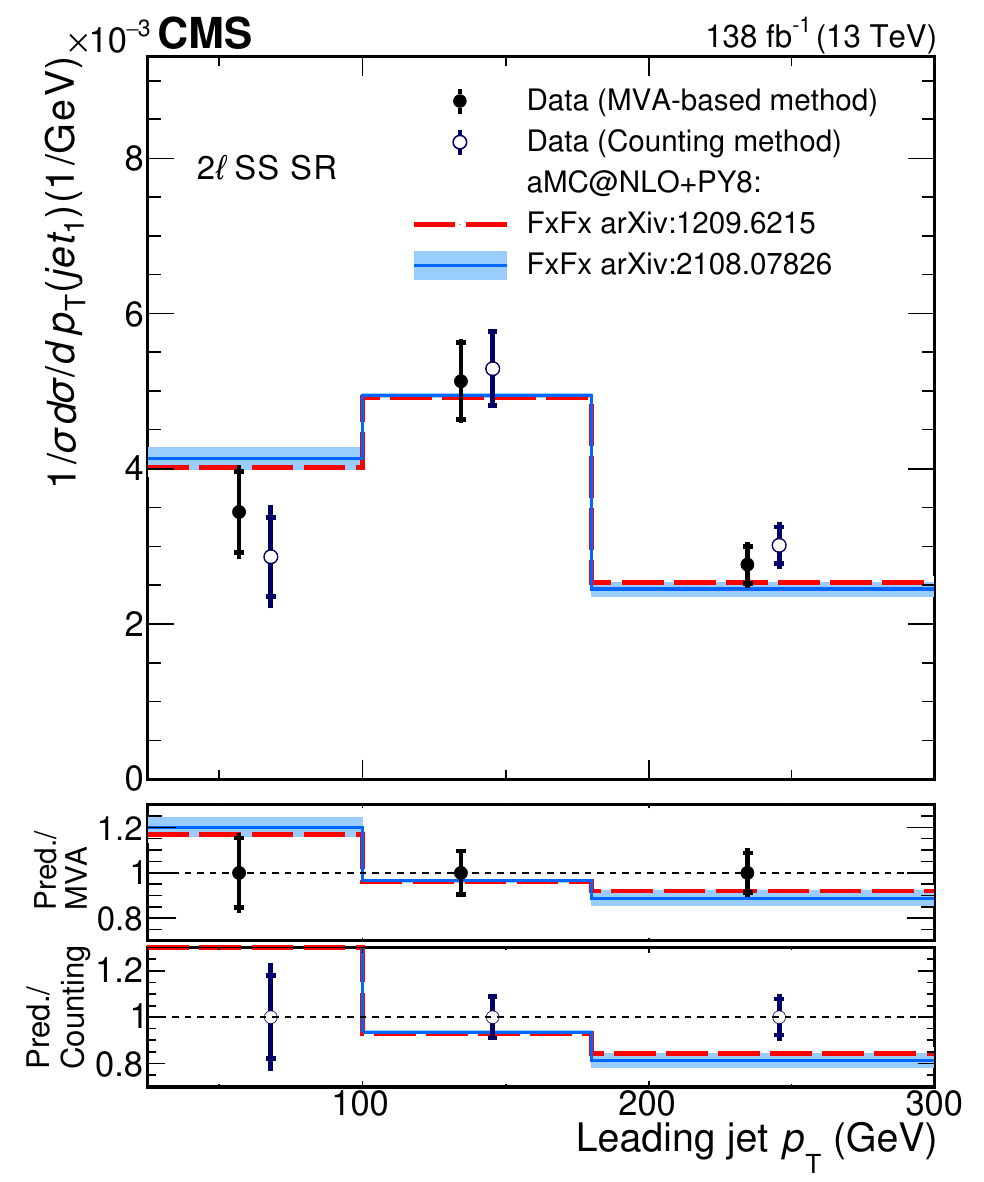}%
\hfill%
\includegraphics[width=0.475\textwidth]{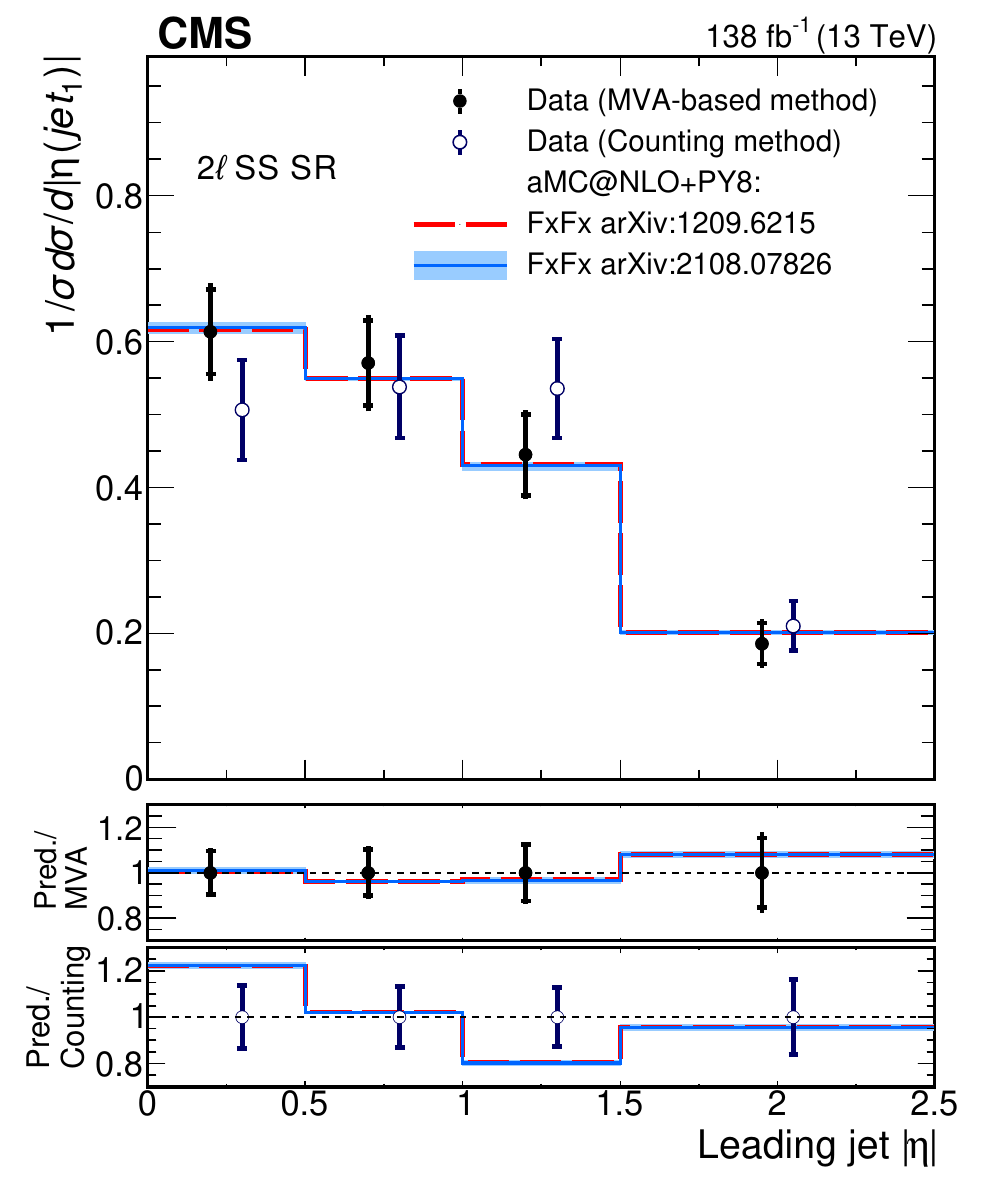}
\caption{%
    Absolute (upper row) and normalized (lower row) differential cross section measured as a function of the leading jet \pt (left) and leading jet \abseta (right), using the MVA-based and the counting method.
    \diffextracaptionTogether
}
\label{fig:differential_xsectogether_2}
\end{figure}

\begin{figure}[!ptt]
\centering
\includegraphics[width=0.475\textwidth]{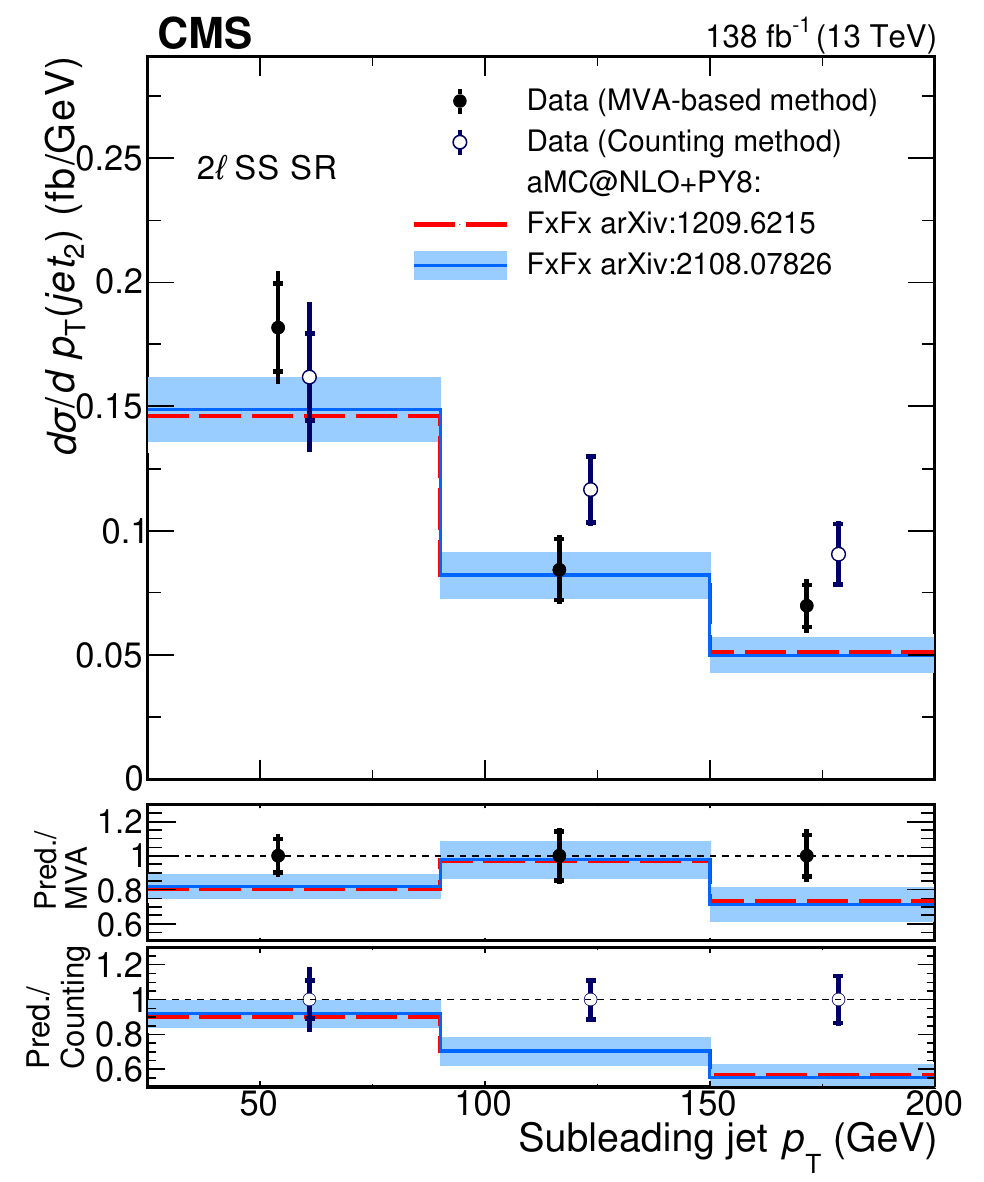}%
\hfill%
\includegraphics[width=0.475\textwidth]{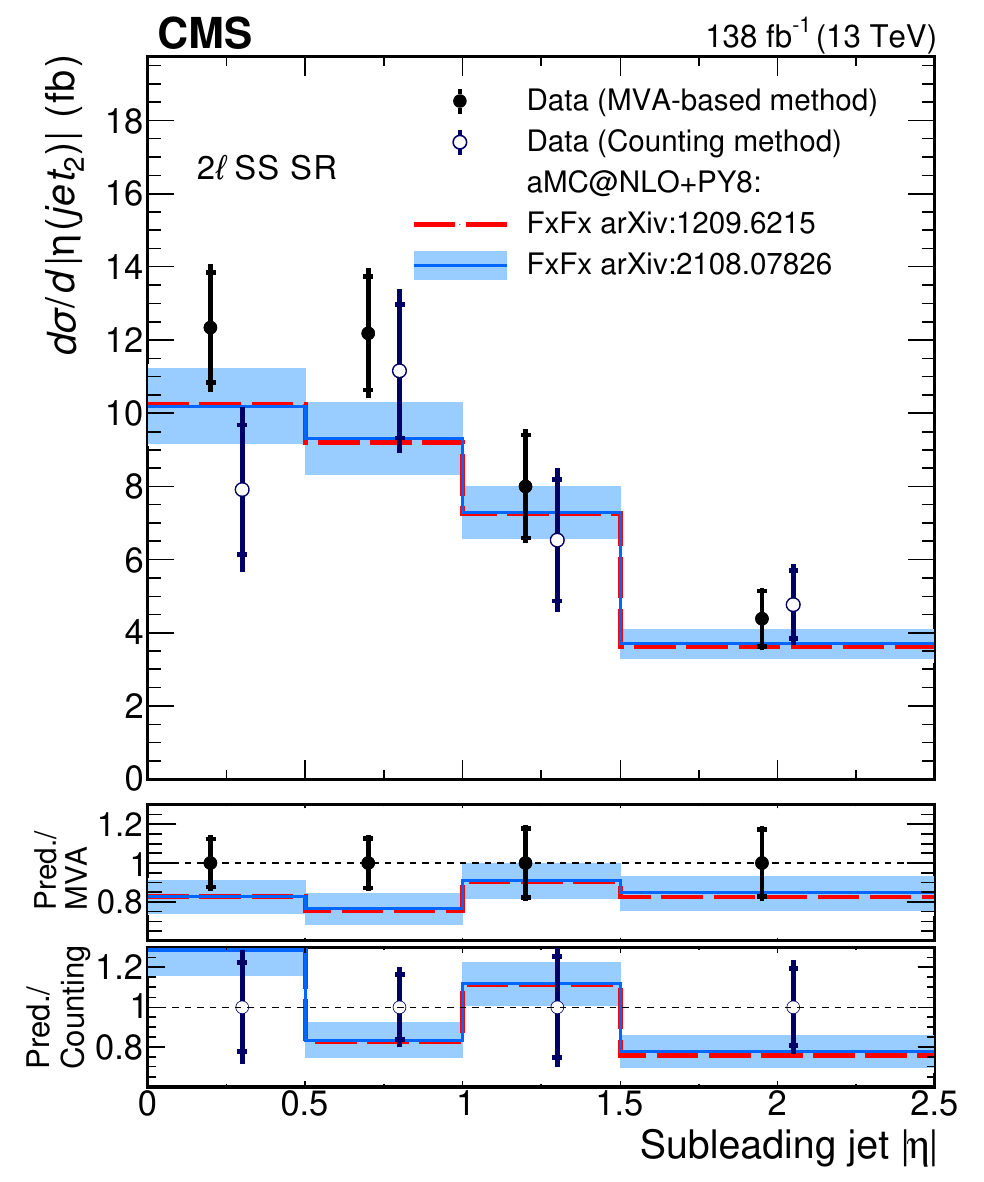} \\[\cmsTabSkip]
\includegraphics[width=0.475\textwidth]{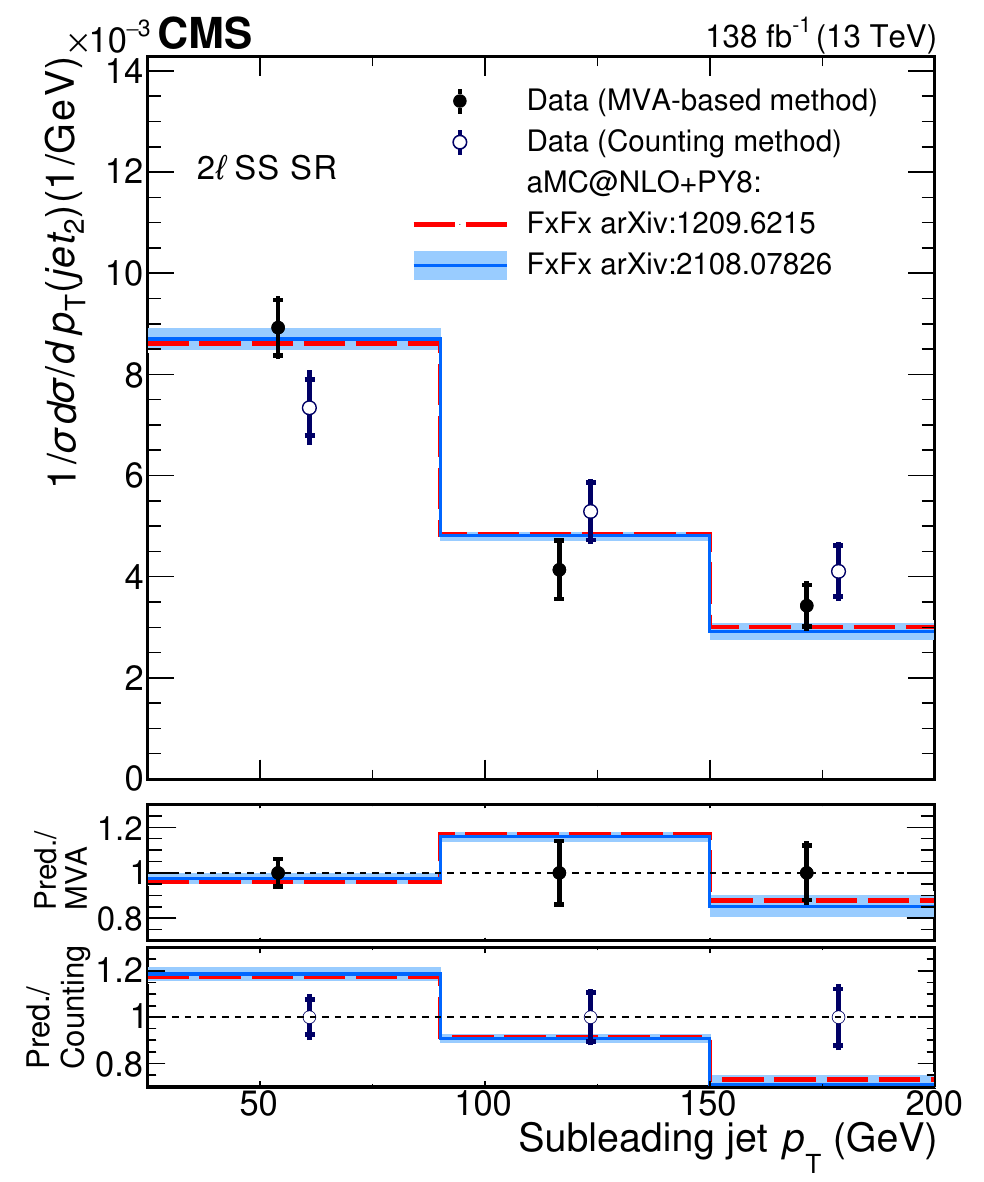}%
\hfill%
\includegraphics[width=0.475\textwidth]{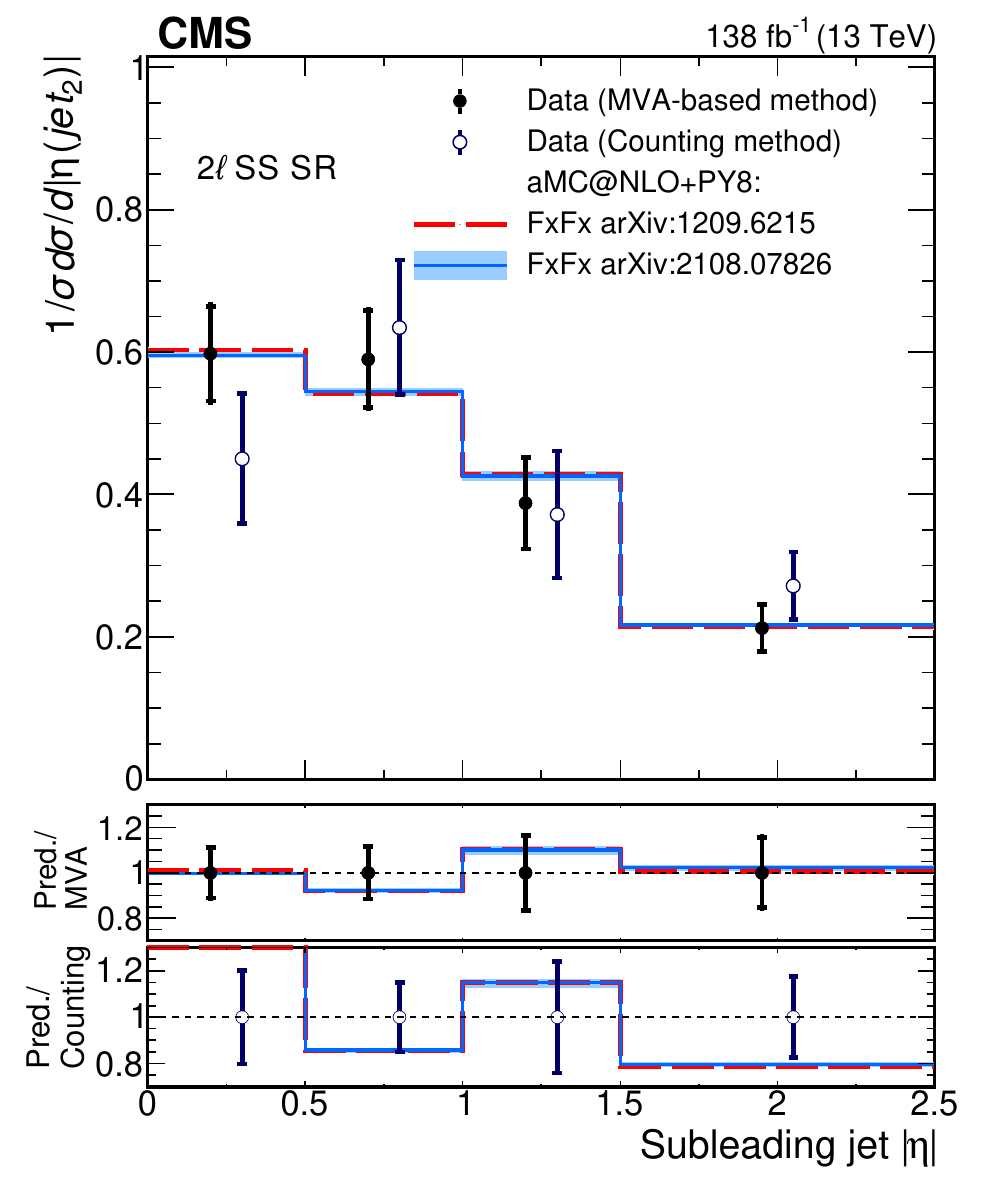}
\caption{%
    Absolute (upper row) and normalized (lower row) differential cross section measured as a function of the subleading jet \pt (left), and subleading jet \abseta (right), using the MVA-based and the counting method.
    \diffextracaptionTogether
}
\label{fig:differential_xsectogether_3}
\end{figure}

\begin{figure}[!ptt]
\centering
\includegraphics[width=0.475\textwidth]{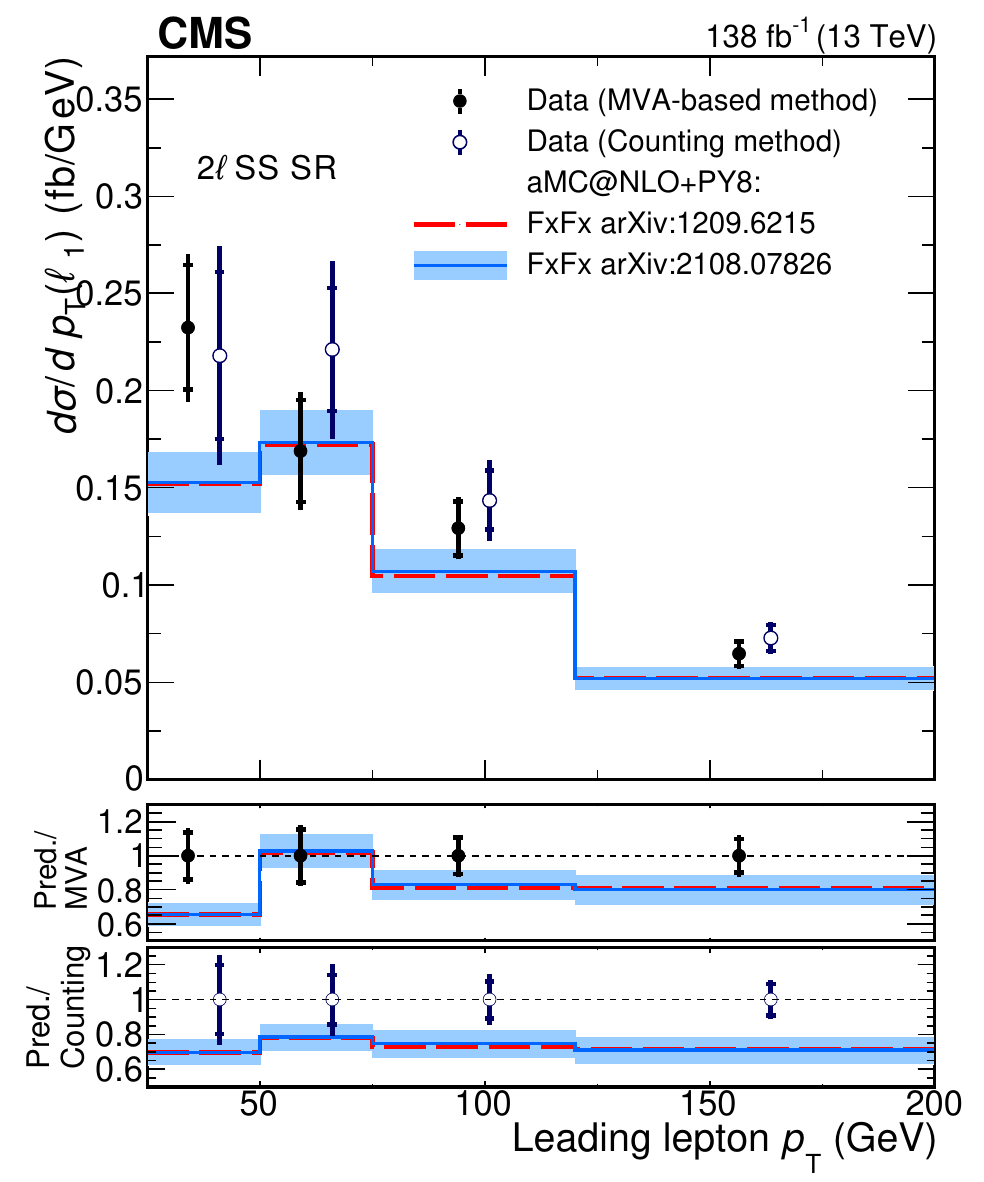}%
\hfill%
\includegraphics[width=0.475\textwidth]{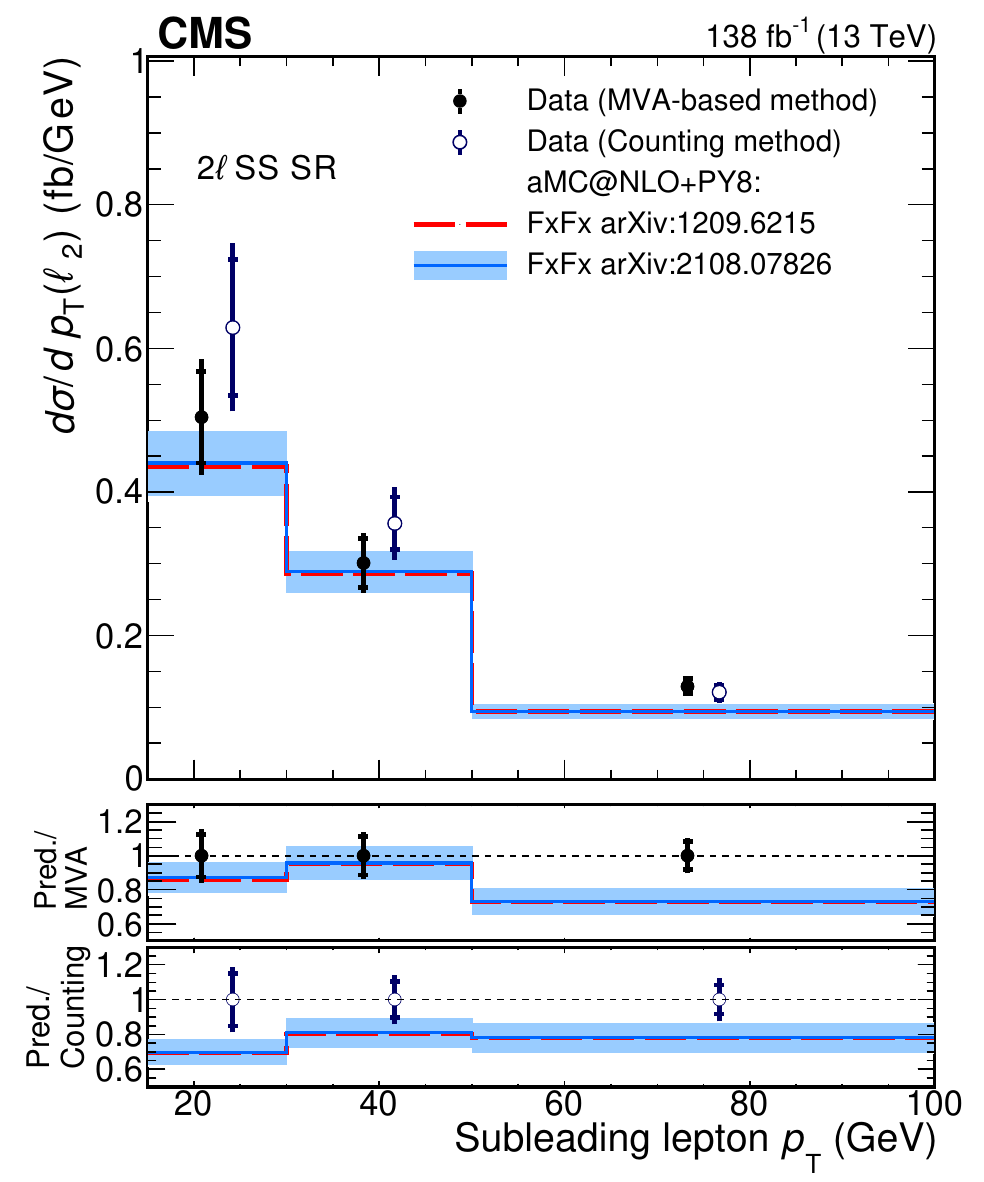} \\[\cmsTabSkip]
\includegraphics[width=0.475\textwidth]{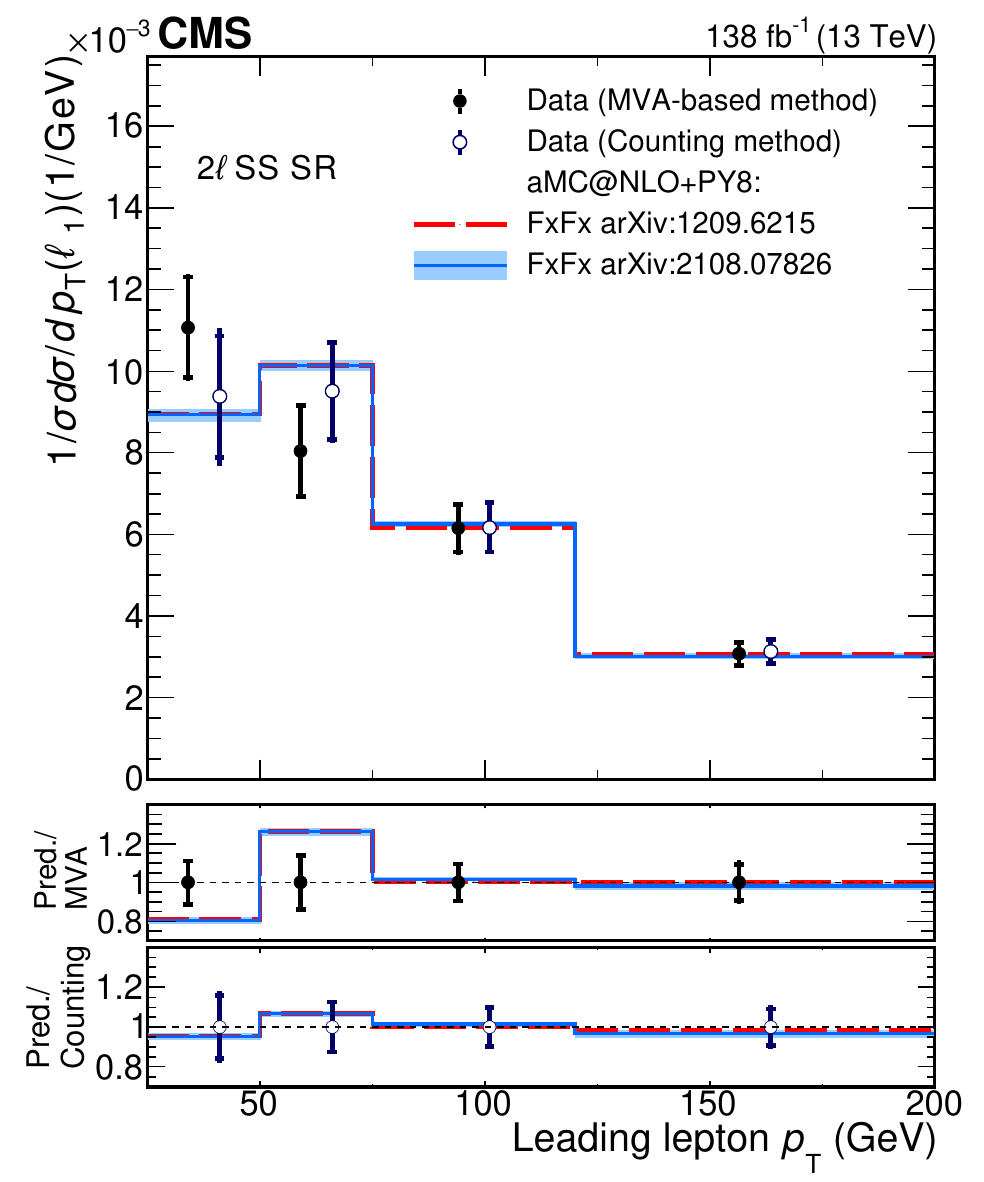}%
\hfill%
\includegraphics[width=0.475\textwidth]{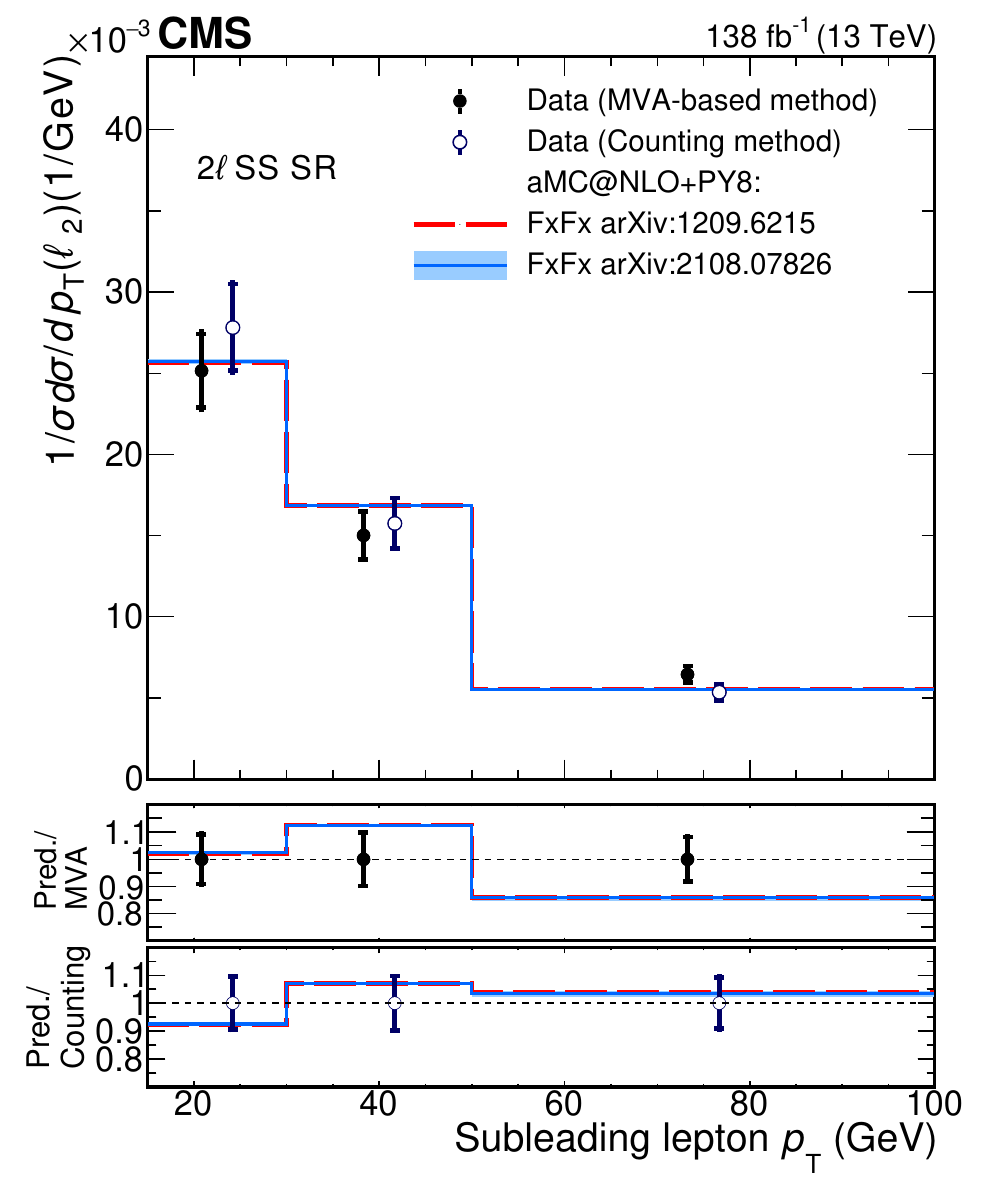}
\caption{%
    Absolute (upper row) and normalized (lower row) differential cross section measured as a function of the leading (left) and subleading (right) lepton \pt, using the MVA-based and the counting method.
    \diffextracaptionTogether
}
\label{fig:differential_xsectogether_4}
\end{figure}

\begin{figure}[!pt]
\centering
\includegraphics[width=0.475\textwidth]{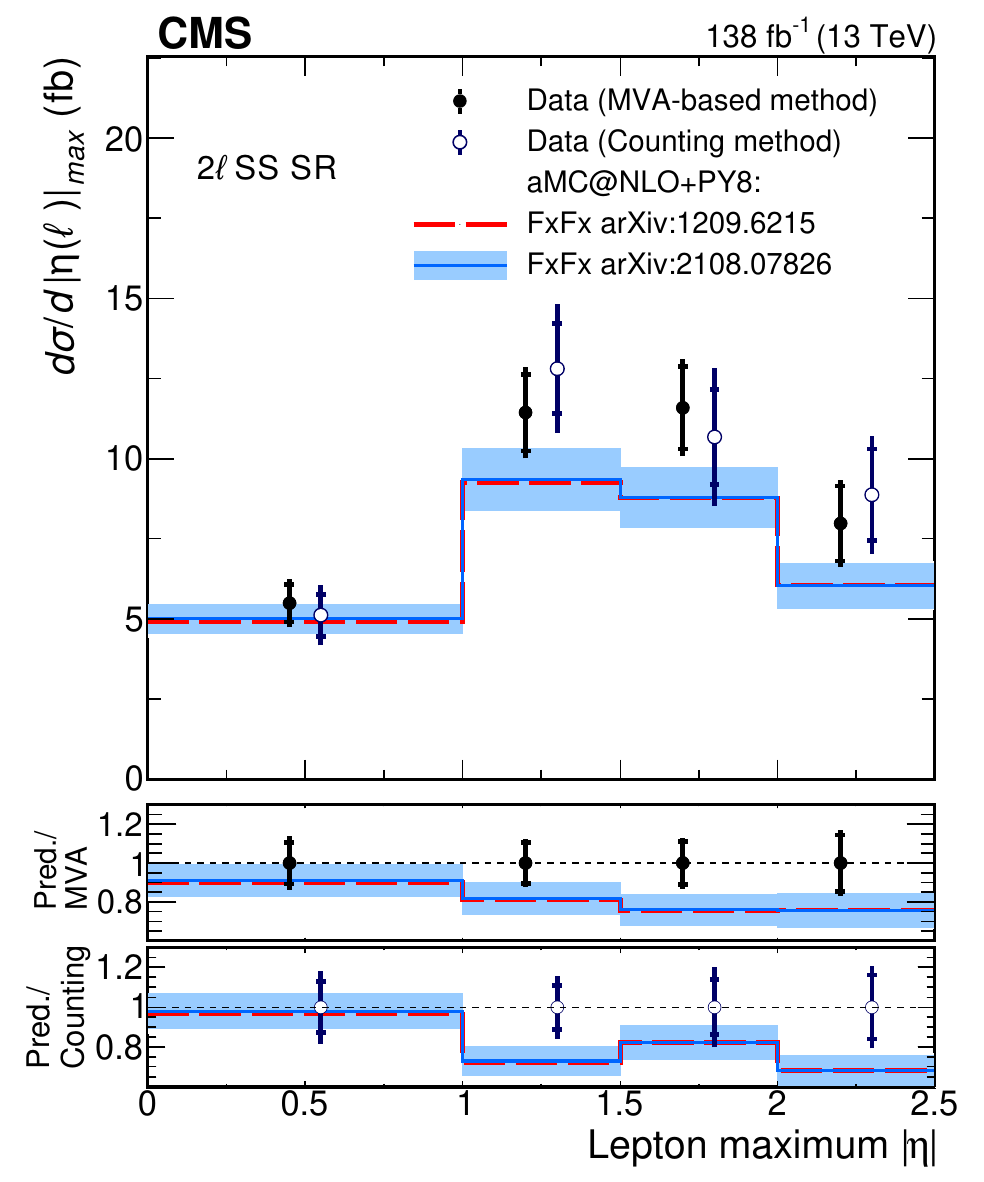}%
\hfill%
\includegraphics[width=0.475\textwidth]{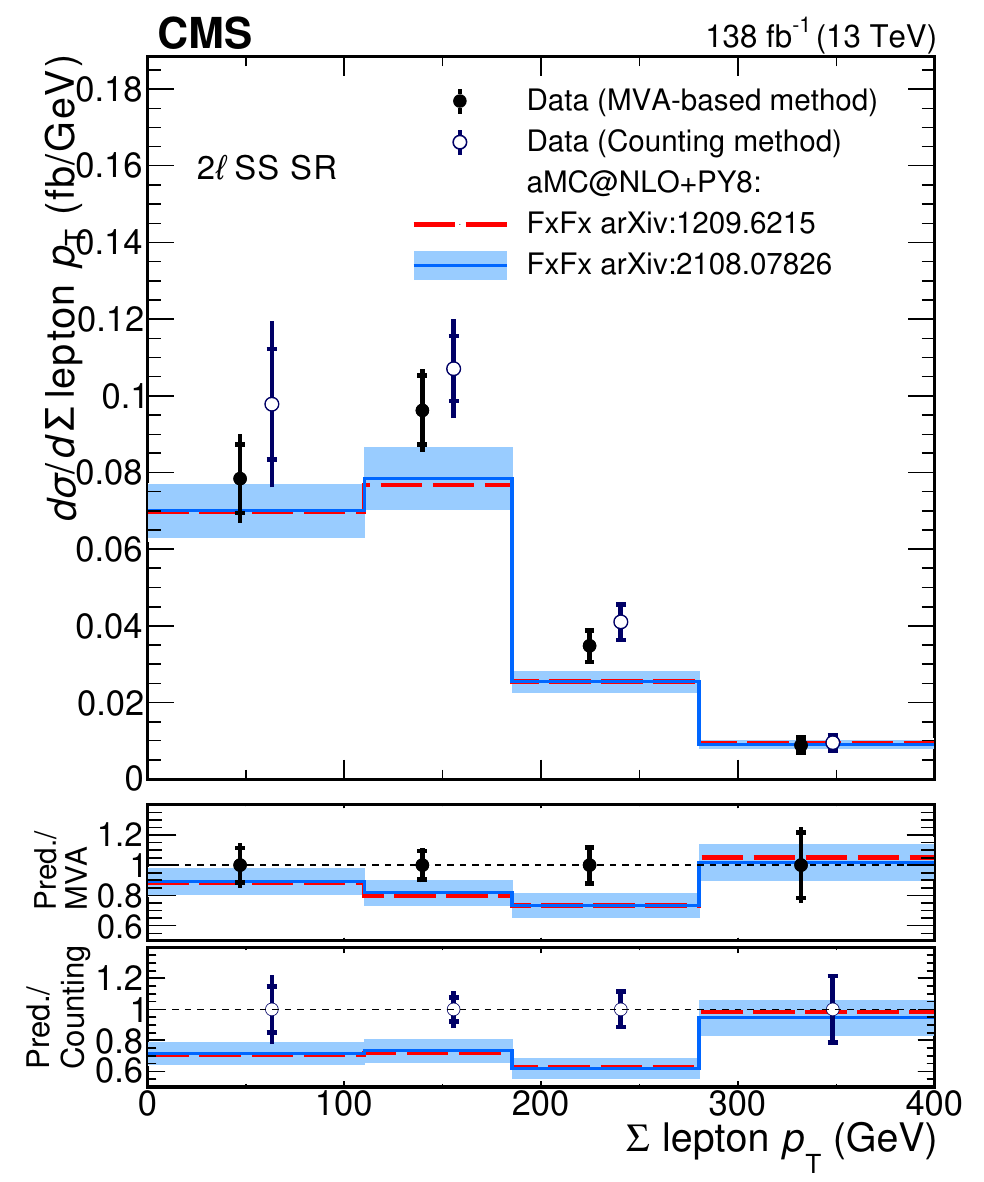} \\[\cmsTabSkip]
\includegraphics[width=0.475\textwidth]{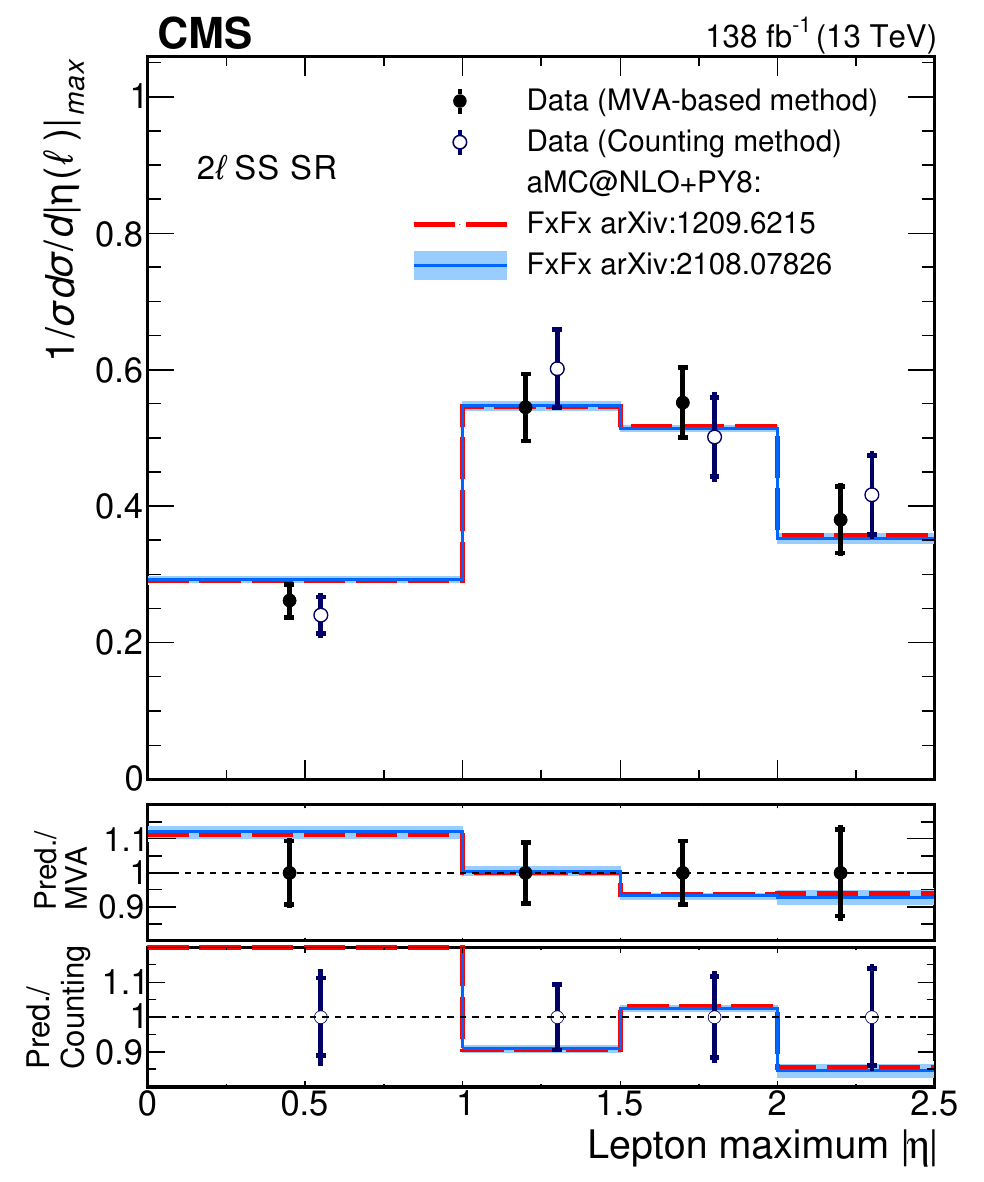}%
\hfill%
\includegraphics[width=0.475\textwidth]{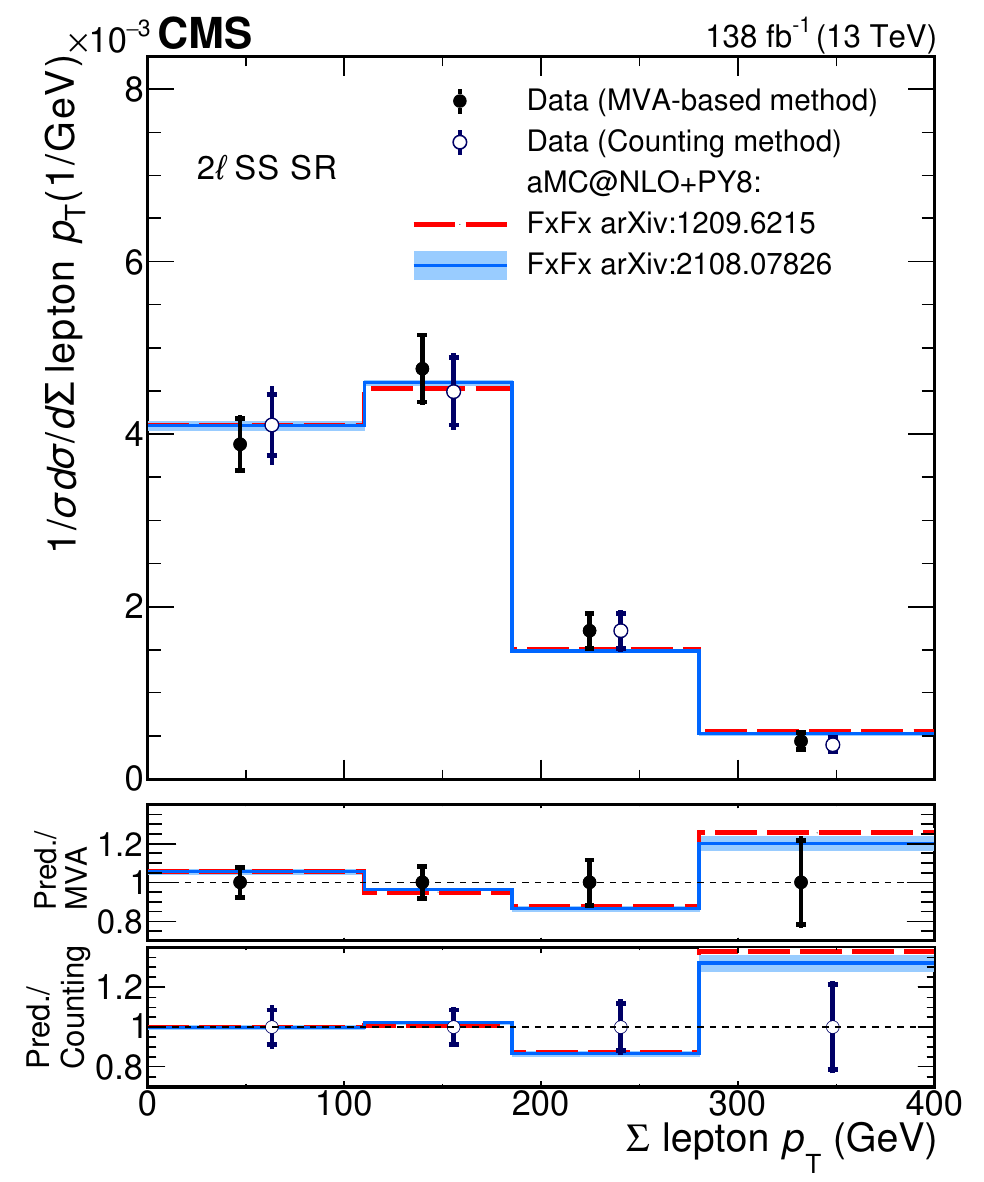}
\caption{%
    Absolute (upper row) and normalized (lower row) differential cross section measured as a function of the maximum \abseta of the selected leptons (left) and the sum of their \pt, using the MVA-based and the counting method.
    \diffextracaptionTogether
}
\label{fig:differential_xsectogether_5}
\end{figure}

\begin{figure}[!pt]
\centering
\includegraphics[width=0.475\textwidth]{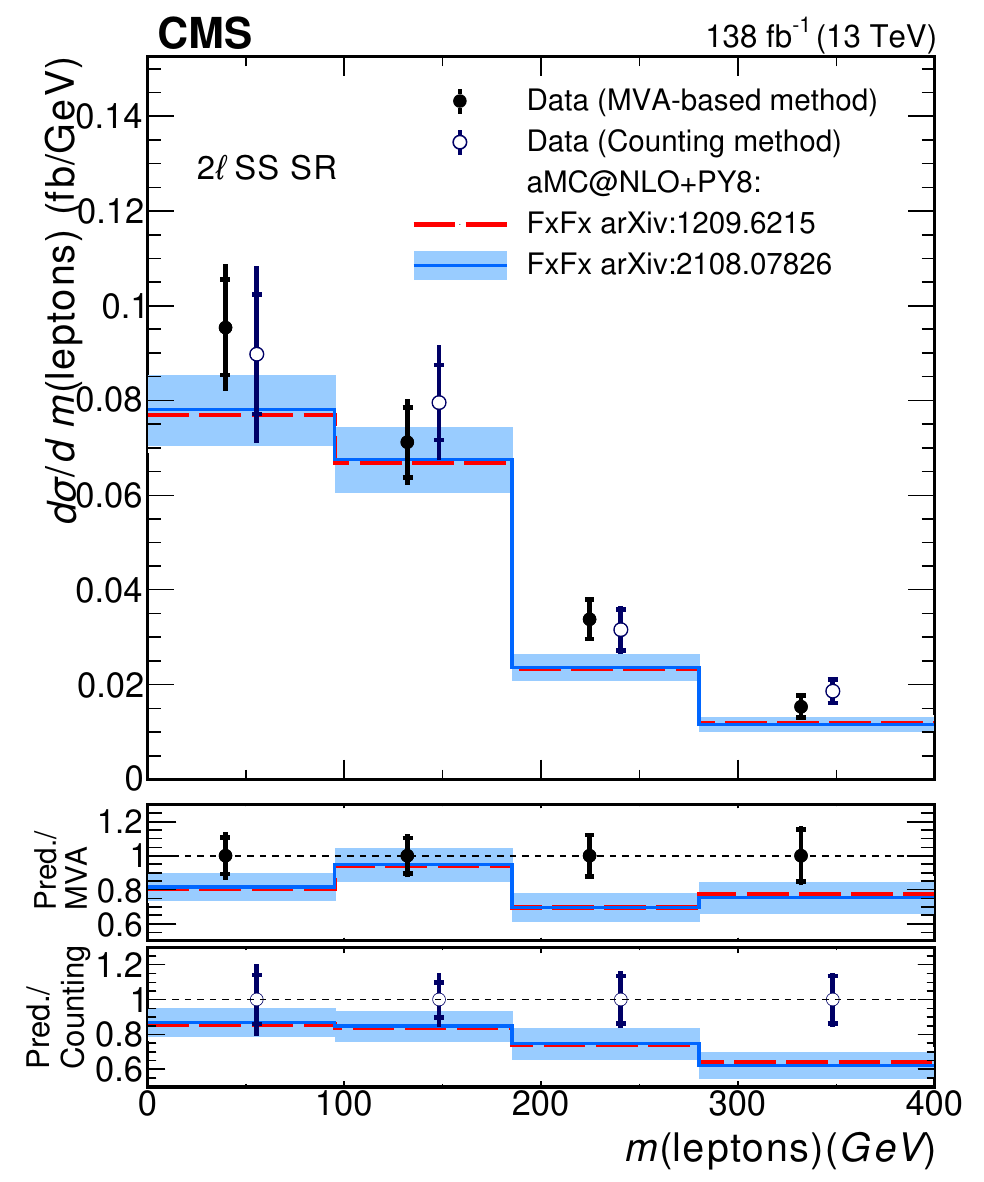}%
\hfill%
\includegraphics[width=0.475\textwidth]{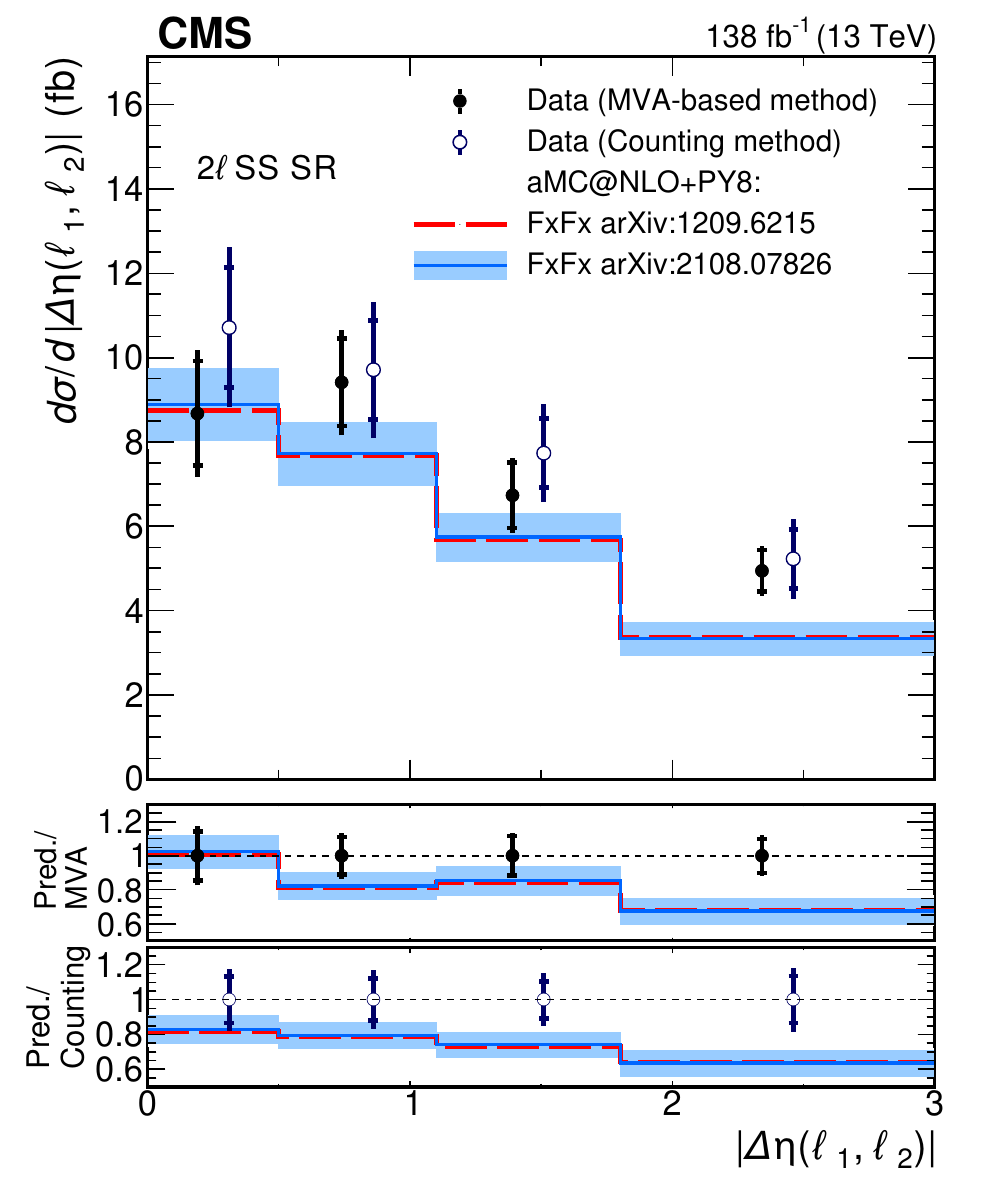} \\[\cmsTabSkip]
\includegraphics[width=0.475\textwidth]{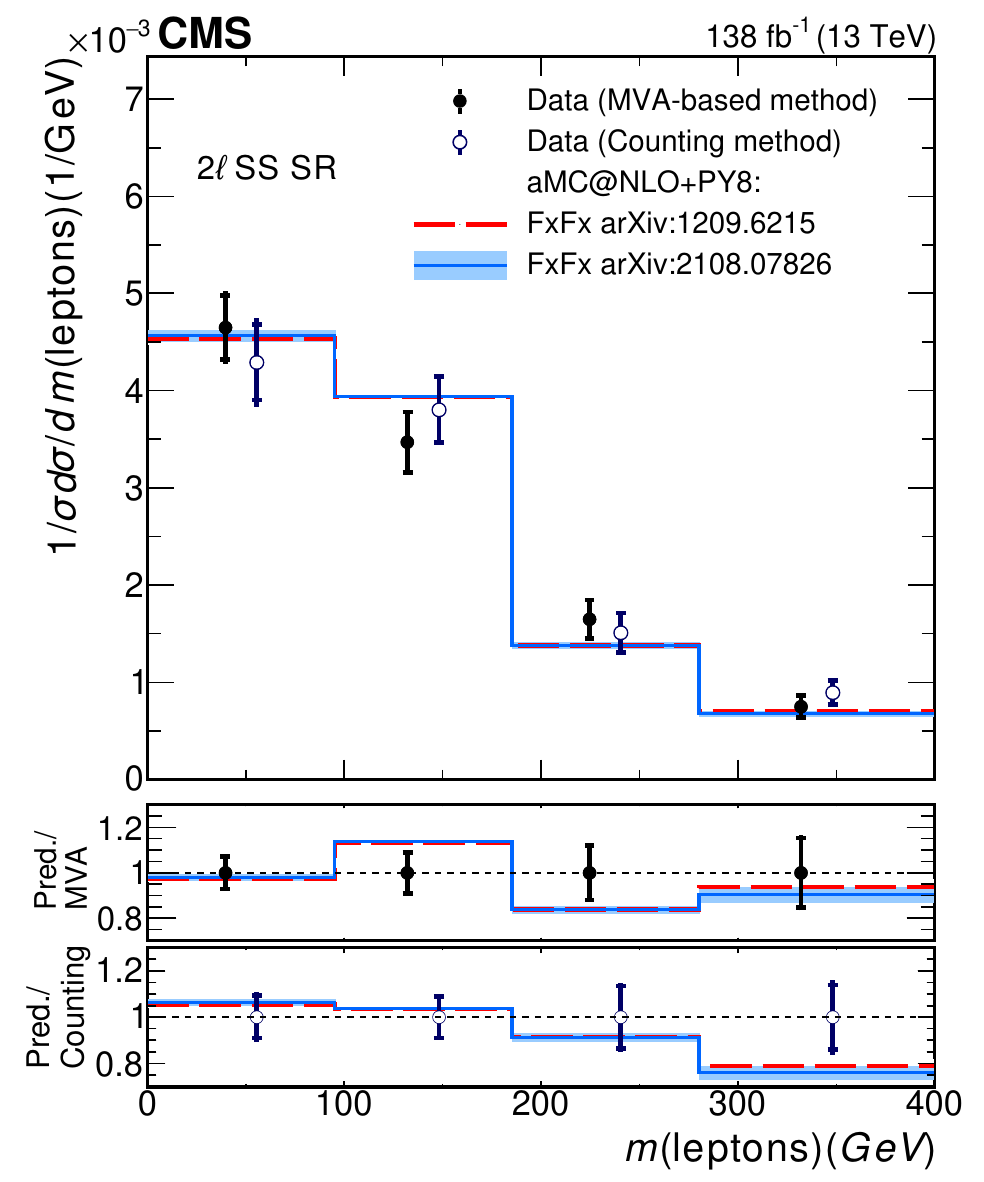}%
\hfill%
\includegraphics[width=0.475\textwidth]{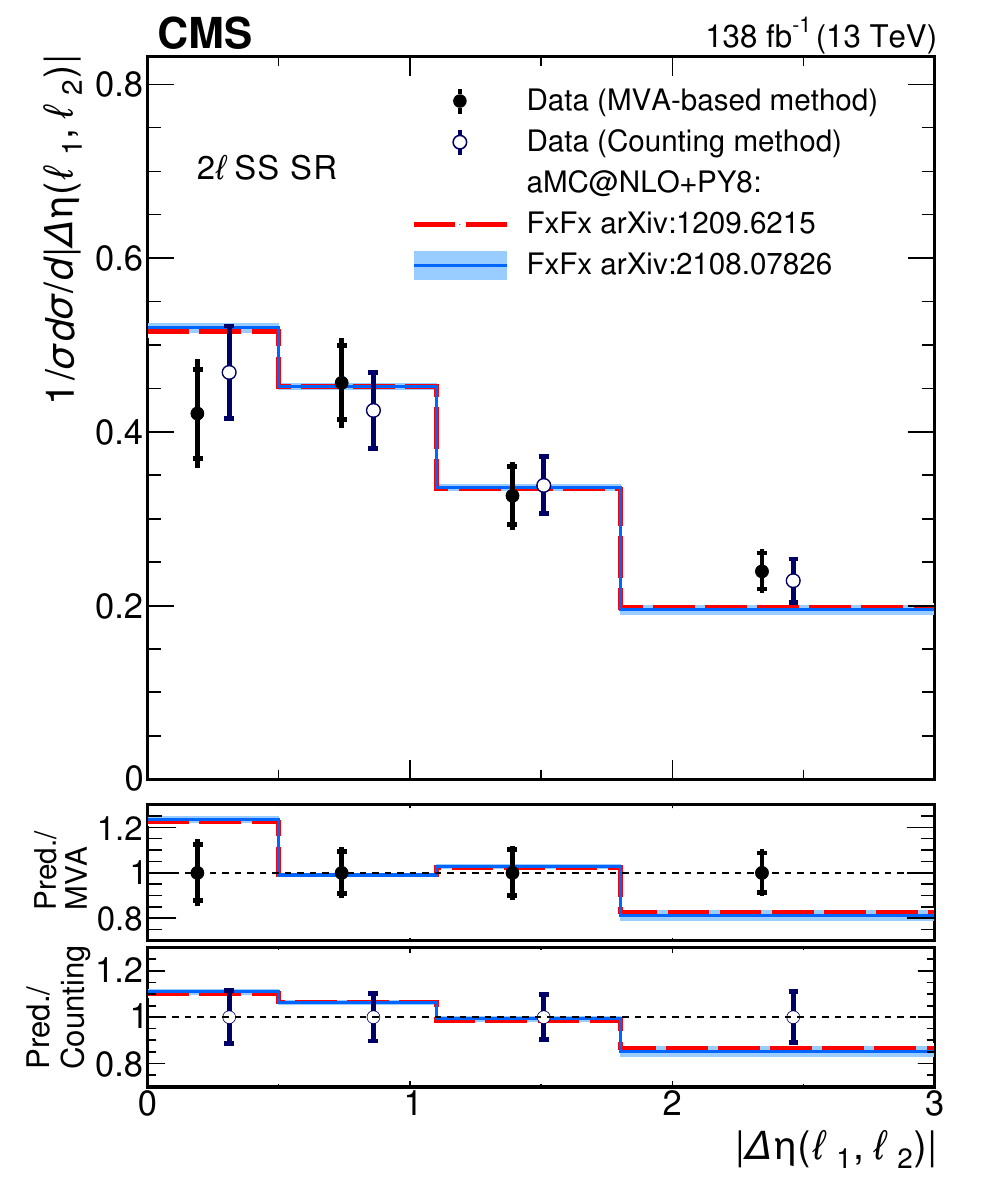}
\caption{%
    Absolute (upper row) and normalized (lower row) differential cross section measured as a function of the invariant mass of the leptons (left) and the \Detaloneltwo, using the MVA-based and the counting method.
    \diffextracaptionTogether
}
\label{fig:differential_xsectogether_6}
\end{figure}

\begin{figure}[!pt]
\centering
\includegraphics[width=0.475\textwidth]{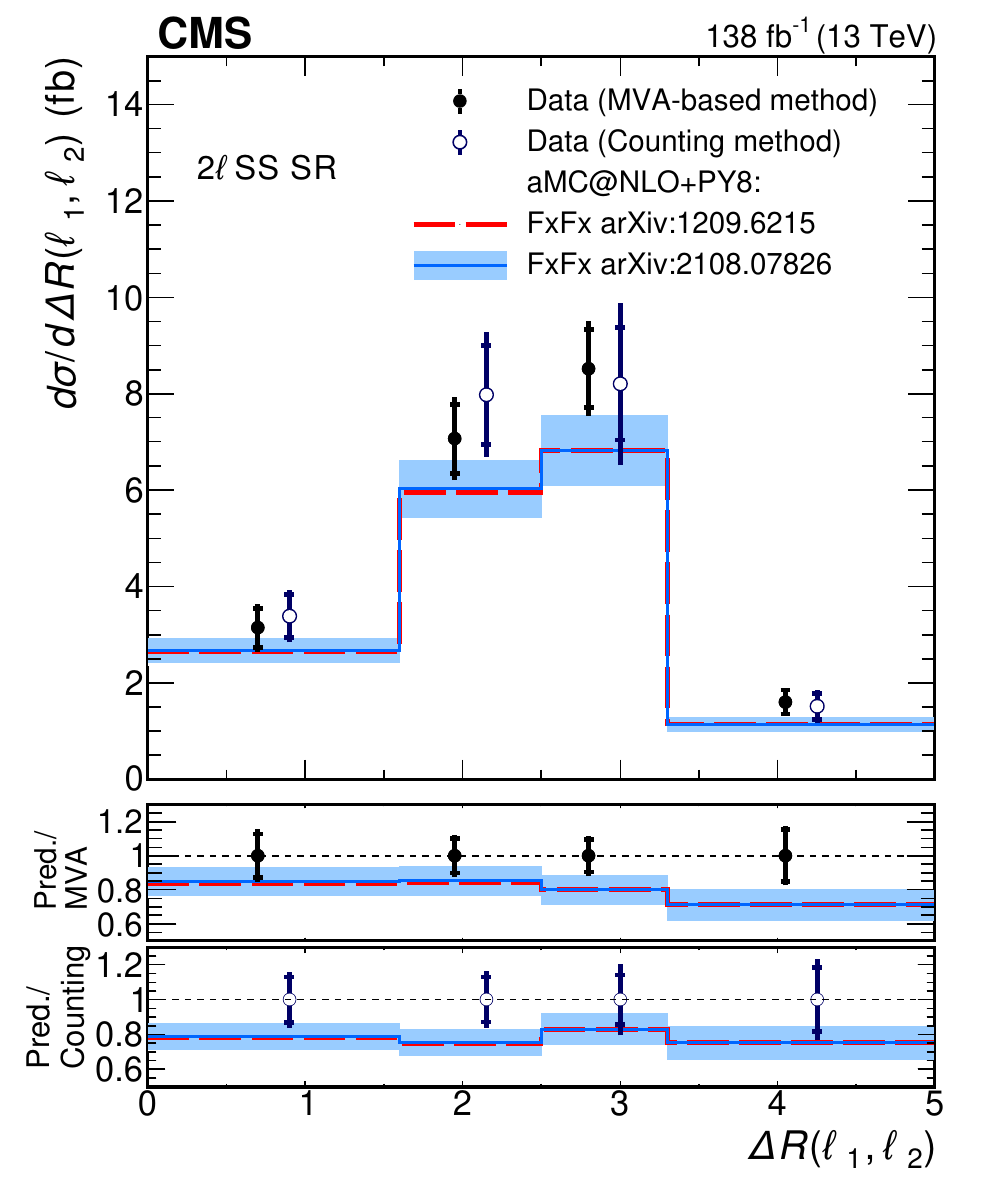}%
\hfill%
\includegraphics[width=0.475\textwidth]{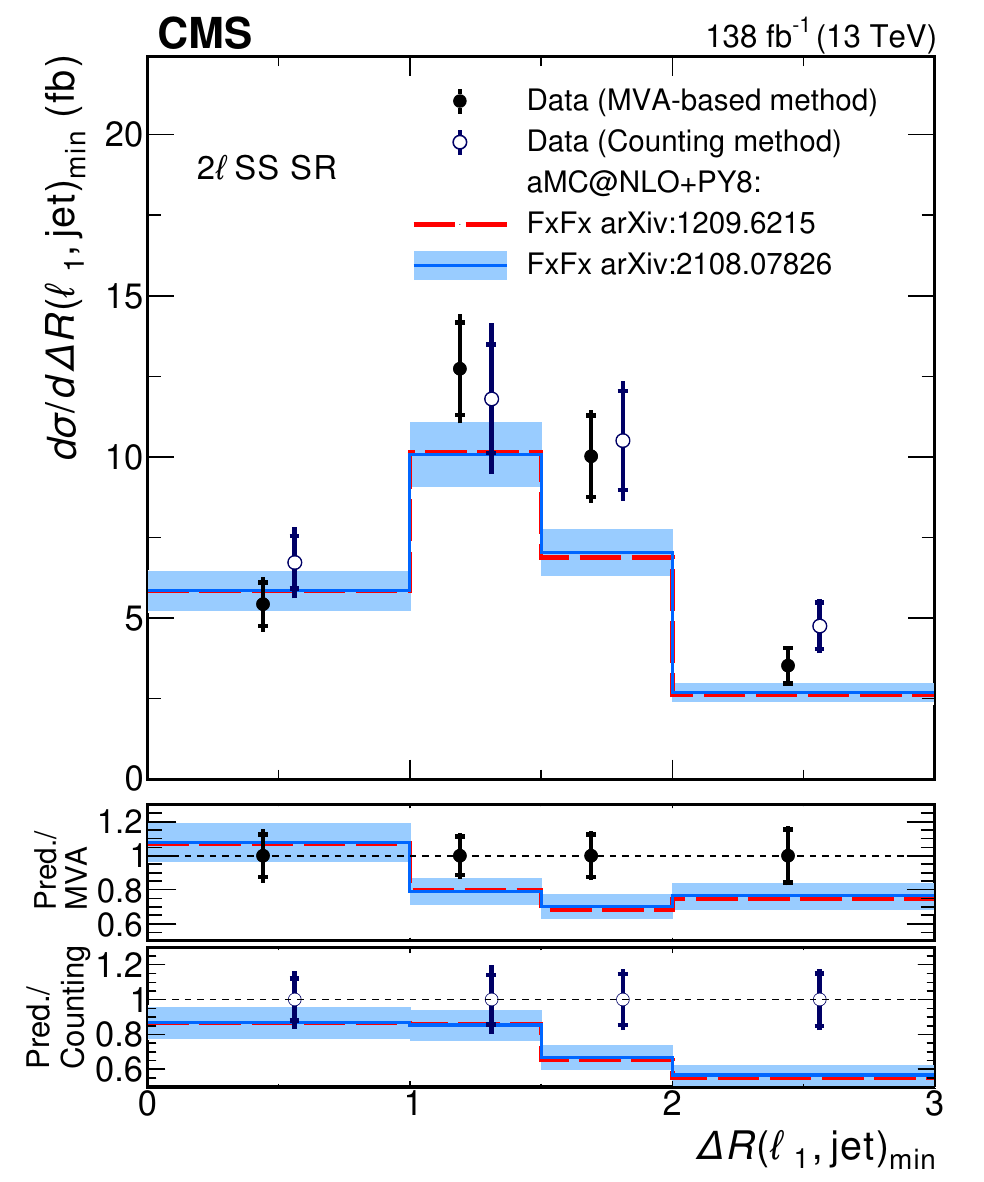} \\[\cmsTabSkip]
\includegraphics[width=0.475\textwidth]{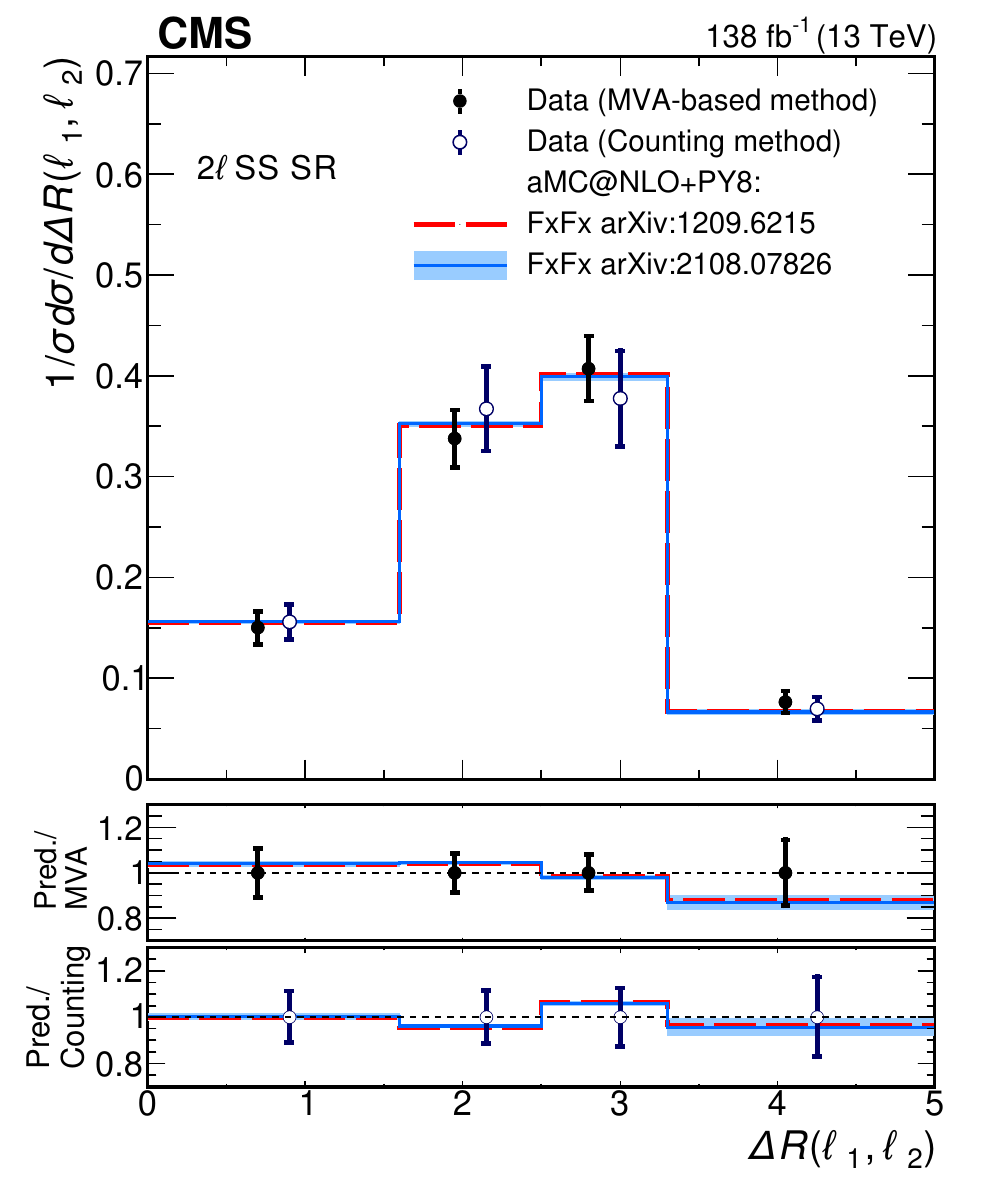}%
\hfill%
\includegraphics[width=0.475\textwidth]{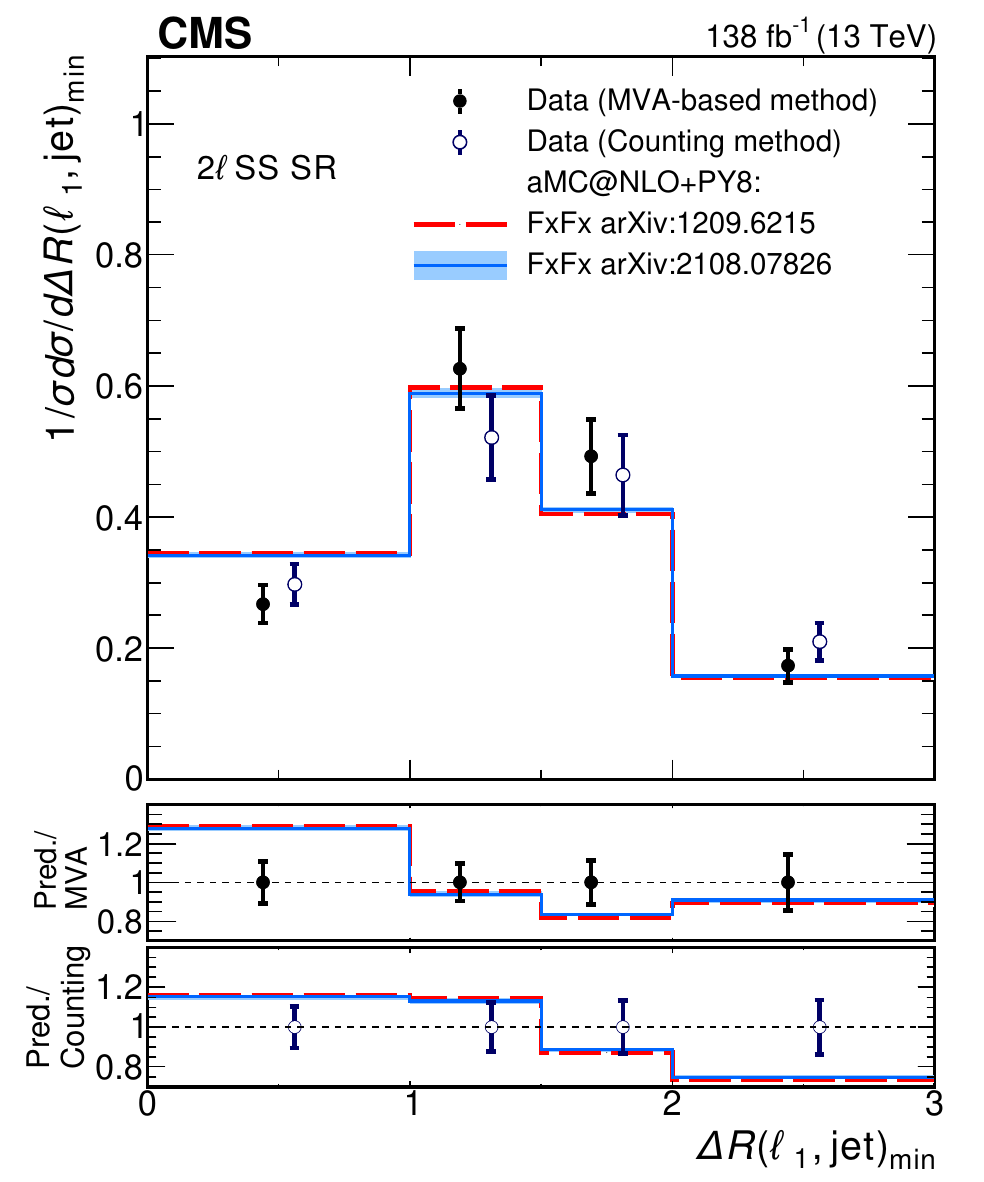}
\caption{%
    Absolute (upper row) and normalized (lower row) differential cross section measured as a function of the $\DR(\Pell_1, \Pell_2)$ and the $\DR(\Pell_1, jet)_{\text{min}}$, using the MVA-based and the counting method.
    \diffextracaptionTogether
}
\label{fig:differential_xsectogether_7}
\end{figure}
\cleardoublepage \section{The CMS Collaboration \label{app:collab}}\begin{sloppypar}\hyphenpenalty=5000\widowpenalty=500\clubpenalty=5000\cmsinstitute{Yerevan Physics Institute, Yerevan, Armenia}
{\tolerance=6000
A.~Hayrapetyan, V.~Makarenko\cmsorcid{0000-0002-8406-8605}, A.~Tumasyan\cmsAuthorMark{1}\cmsorcid{0009-0000-0684-6742}
\par}
\cmsinstitute{Institut f\"{u}r Hochenergiephysik, Vienna, Austria}
{\tolerance=6000
W.~Adam\cmsorcid{0000-0001-9099-4341}, J.W.~Andrejkovic, L.~Benato\cmsorcid{0000-0001-5135-7489}, T.~Bergauer\cmsorcid{0000-0002-5786-0293}, M.~Dragicevic\cmsorcid{0000-0003-1967-6783}, C.~Giordano, P.S.~Hussain\cmsorcid{0000-0002-4825-5278}, M.~Jeitler\cmsAuthorMark{2}\cmsorcid{0000-0002-5141-9560}, N.~Krammer\cmsorcid{0000-0002-0548-0985}, A.~Li\cmsorcid{0000-0002-4547-116X}, D.~Liko\cmsorcid{0000-0002-3380-473X}, I.~Mikulec\cmsorcid{0000-0003-0385-2746}, J.~Schieck\cmsAuthorMark{2}\cmsorcid{0000-0002-1058-8093}, R.~Sch\"{o}fbeck\cmsAuthorMark{2}\cmsorcid{0000-0002-2332-8784}, D.~Schwarz\cmsorcid{0000-0002-3821-7331}, M.~Shooshtari, M.~Sonawane\cmsorcid{0000-0003-0510-7010}, W.~Waltenberger\cmsorcid{0000-0002-6215-7228}, C.-E.~Wulz\cmsAuthorMark{2}\cmsorcid{0000-0001-9226-5812}
% ...
\par}
\vskip\cmsinstskip
\dag:~Deceased\\
$^{1}$Also at Yerevan State University, Yerevan, Armenia\\
$^{2}$Also at TU Wien, Vienna, Austria\\
% ...
\end{sloppypar}
\end{document}